\documentclass{jfm}
\usepackage{graphicx}
\usepackage{epstopdf}
\usepackage{amssymb}
\usepackage{enumerate}
\usepackage{amsmath}
\usepackage{psfrag}
\usepackage{amsbsy}
\usepackage{amsfonts}
\usepackage{setspace}
\usepackage[usenames, dvipsnames]{color}
\usepackage{float}
\usepackage{tabularx}
\usepackage{epsfig}
\usepackage{bm}
\usepackage{booktabs,array,dcolumn}
\usepackage{arydshln}
\usepackage{dashrule}
\usepackage{wasysym}
\usepackage{multicol}
\usepackage{multirow}
\usepackage{ulem}
\usepackage{nomencl}
\usepackage{multicol}
\usepackage{amsmath}
\usepackage{subfig}
\usepackage{epstopdf}
\usepackage{color, soul}
\usepackage{framed}
\usepackage{nomencl}
\makenomenclature
\usepackage{longtable,tabularx}
\sethlcolor{SkyBlue}

\newcommand{\reffig}[1]{figure \ref{#1}}
\newcommand{\reffigs}[1]{figures \ref{#1}}
\newcommand{\refFig}[1]{Figure \ref{#1}}
\newcommand{\refFigs}[1]{Figures \ref{#1}}
\newcommand{\reftab}[1]{table \ref{#1}}
\newcommand{\refTab}[1]{Table \ref{#1}}
\newcommand{\refeq}[1]{equation \eqref{#1}}
\newcommand{\refeqs}[1]{equations \eqref{#1}}
\newcommand{\refEq}[1]{Equation \eqref{#1}}

\graphicspath{{figures/}}
\begin{document}
	
\newtheorem{lemma}{Lemma}
\newtheorem{corollary}{Corollary}

\shorttitle{Flow-excited membrane instability} 
\shortauthor{G. Li, R. K. Jaiman and B. C. Khoo} 

\title{Flow-excited membrane instability at moderate Reynolds numbers}

\author
{
	G. Li\aff{1}
    \corresp{\email{li.guojun@u.nus.edu}},
	R. K. Jaiman\aff{2}
	\and 
	B. C. Khoo\aff{1}
}

\affiliation
{
	\aff{1}
	Department of Mechanical Engineering, National University of Singapore, Singapore 119077
	\aff{2}
	Department of Mechanical Engineering, University of British Columbia, Vancouver, BC  Canada V6T 1Z4
}

\maketitle

\begin{abstract}
In this paper, we study the fluid-structure interaction of a three-dimensional (3D) flexible membrane immersed in an unsteady separated flow at moderate Reynolds numbers. We employ a body-conforming variational fluid-structure interaction solver based on the recently developed partitioned iterative scheme for the coupling of turbulent fluid flow with nonlinear structural dynamics. Of particular interest is to understand the flow-excited instability of a 3D flexible membrane as a function of the non-dimensional mass ratio ($m^*$), Reynolds number ($Re$) and aeroelastic number ($Ae$). For a wide range of the parameters, we examine two distinctive stability regimes of fluid-membrane interaction: deformed-steady state (DSS) and dynamic balance state (DBS). We propose stability phase diagrams to demarcate the DSS and DBS regimes for the parameter space of mass ratio vs. Reynolds number ($m^*$-$Re$) and mass ratio vs. aeroelastic number ($m^*$-$Ae$). With the aid of the global Fourier mode decomposition technique, the distinctive dominant vibrational modes are identified from the intertwined membrane responses in the parameter space of $m^*$-$Re$ and $m^*$-$Ae$. Compared to the deformed-steady membrane, the flow-excited vibration produces relatively longer-attached leading edge vortices which improve the aerodynamic performance when the coupled system is near the flow-excited instability boundary. The optimal aerodynamic performance is achieved for lighter membranes with higher $Re$ and larger flexibility. Based on the global aeroelastic mode analysis, we observe a frequency lock-in phenomenon between the vortex shedding frequency and the membrane vibration frequency causing self-sustained vibrations in the dynamic balance state. 
To characterize the origin of the frequency lock-in, we propose an approximate analytical formula for the nonlinear natural frequency by considering the added mass effect and employing a large deflection theory for a simply supported rectangular membrane.
Through our systematic high-fidelity numerical investigation, we find that the onset of the membrane vibration and the mode transition has a direct dependence on the frequency lock-in between the natural frequency of the tensioned membrane and the vortex shedding frequency or its harmonics. These findings on the fluid-elastic instability of membranes have implications for the design and development of control strategies for membrane wing-based unmanned systems and drones.

\textbf{Key words:} flow-structure interactions, vortex dynamics
\end{abstract}

\section{Introduction}
Fluid-membrane interaction between a flexible membrane structure and a surrounding unsteady flow is ubiquitous in nature and engineering systems. In the past decades, morphing fins/wings with flexible membrane components have received substantial attention from the aerospace engineering community in the context of bio-inspired swimming/flying vehicles at moderate Reynolds numbers \citep{webster1962role,shyy1999flapping,bahlman2013design}. During the biological movement with the aid of the morphing structures, a flexible membrane can deform up passively and display complex spatial-temporal vibrational dynamics through the flow-excited instability as functions of wing shapes, physical parameters and the boundary conditions \citep{rojratsirikul2011flow,sun2017nonlinear}. Flow-excited instability of a flexible membrane can lead to self-sustained vibrations, the so-called flow-induced vibration, when the fluctuating flows are strongly coupled with the elastic thin membrane structures. Particularly, a frequency synchronization of the vortex shedding frequency ($f^{vs}$) and the frequency of the membrane vibrations ($f^s$) may be established in the coupled fluid-membrane systems under specific conditions, resulting in the well-known frequency lock-in phenomenon \citep{rojratsirikul2009unsteady,tregidgo2012frequency}. By proper harnessing of such flow-excited instability, one can regulate the membrane kinematics and the vortex shedding features through the fluid-membrane coupling effect \citep{song2008aeromechanics,gursul2014control,buoso2017demand}. The understanding of the coupled fluid-membrane dynamics can facilitate the development of effective control strategies to explore the potential for improved aerodynamic performance and smooth adaptability in a gusty flow. 

\begin{figure}
	\centering
	\subfloat[][]{\includegraphics[width=0.5\textwidth]{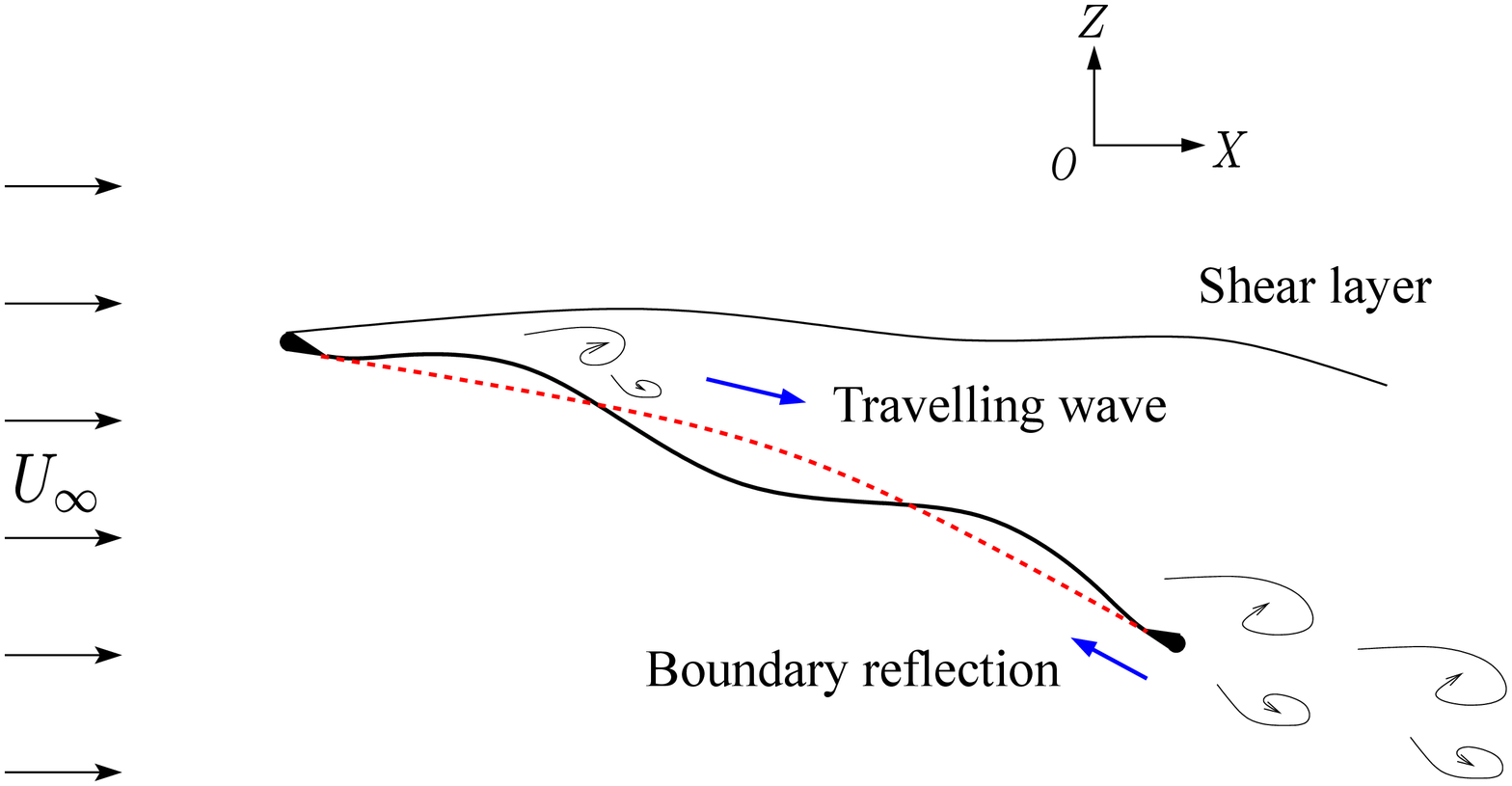}\label{membranea}}
	\subfloat[][]{\includegraphics[width=0.4\textwidth]{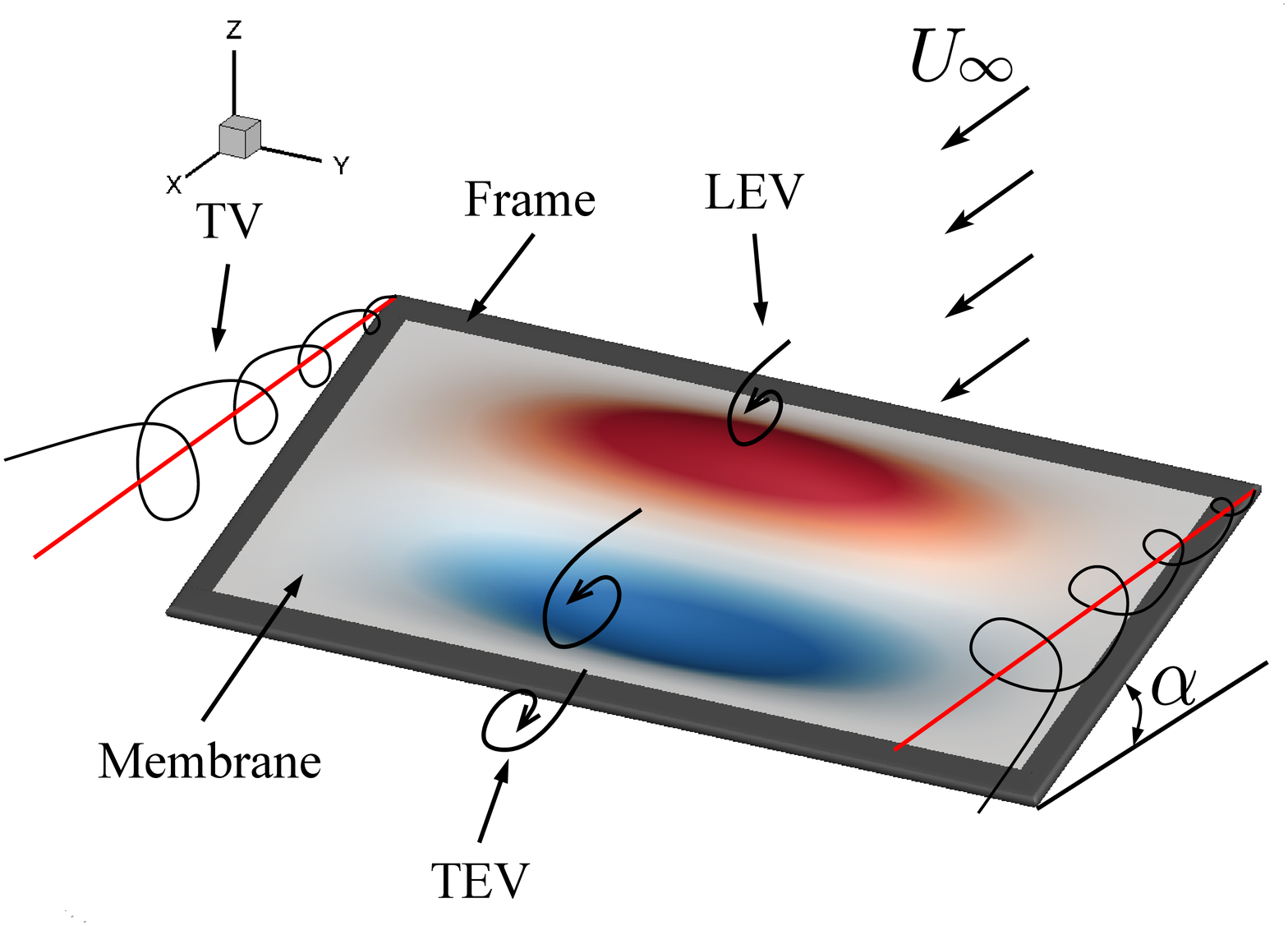}\label{membraneb}}
	\caption{Flow past a flexible membrane wing: (a) schematic of fluid-membrane interaction, (b) illustration of three-dimensional membrane wing with  free-stream velocity $U_{\infty}$ and a given angle of attack $\alpha$. In (a), the red dashed line ({\textbf{\color{red}- - -}}) indicates the time-averaged membrane shape, and the black solid line ({\textbf{\color{black}---}}) represents the instantaneous membrane shape. In (b), the red and blue contours are the positive and negative membrane displacements, respectively.}
	\label{membrane}
\end{figure}

During fluid-membrane interaction, the self-sustained vibration associated with the flow-excited instability is found to appear under certain conditions \citep{song2008aeromechanics,rojratsirikul2011flow,bleischwitz2017fluid}. As illustrated in \reffig{membrane} \subref{membranea}, the perturbation caused by the pressure fluctuations in the shear layer propagates backward and interacts with the flexible membrane to vibrate. The induced travelling waves and their boundary reflections are interacted to form intertwined vibrational modes. These multiple vibrational modes further increase the complexity of the flow-induced vibrations. \refFig{membrane} \subref{membraneb} shows a representative schematic of the unsteady separated flow past a 3D rectangular flexible membrane wing vibrating in a chord-wise second and span-wise first mode. Different from the two-dimensional (2D) membrane, the vibrations and the unsteady flows along the span-wise direction of a 3D membrane become important in the coupled system. These 3D effects further make the coupled membrane dynamics more complex and rich. From the perspective of the unsteady separated flow, the leading edge vortex (LEV), the trailing edge vortex (TEV) and the tip vortex (TV) are highly coupled with the membrane vibrations. The vibrational energy and the flow kinetic energy are redistributed through the fluid-membrane coupling effect. In the observed frequency lock-in phenomenon, these redistributed energies were mainly concentrated in several specific frequencies of the coupled fluid-membrane system, compared to its rigid counterpart \citep{serrano2018fluid}.

As the governing fluid-structure parameters (e.g., Reynolds number, mass ratio and aeroelastic number) vary, the transition of the dominant aeroelastic modes associated with the vibration and the vortex shedding process may be observed \citep{rojratsirikul2011flow,bleischwitz2015aspect,sun2016nonlinear,bleischwitz2017fluid,sun2017nonlinear}. Through numerous investigations, the natural frequency of the flexible membrane was found to play an important role in the flow-induced vibration \citep{gordnier2009high,bleischwitz2016aeromechanics}. However, the processes of flow-induced vibrations of the membrane and the aeroelastic mode transition are not fully understood. The highly nonlinear dependence of the coupled dynamics on the fluid-structure parameters restricts us to gain a deeper insight into these phenomena. This paper aims to explain how the flow-excited instability governs the 3D fluid-membrane interaction characteristics and can be linked with the mode transition. Specifically, we examine the coupled fluid-membrane dynamics and the dominant dynamic modes during fluid-membrane interaction. The variation of the natural frequency of the flexible membrane relative to the vortex shedding frequency is monitored by properly changing the physical parameters.

Recent studies based on experiments and numerical simulations suggested that the characteristics of the flow-induced vibration and the underlying dynamic modes were strongly related to the membrane performance. Some of these investigations primarily focused on the mean dynamic responses of the flexible membrane exposed to an unsteady flow. However, the unsteady dynamic behaviour induced by the flow-excited instability depicted in \reffig{membrane} \subref{membranea} plays an important role in the membrane performance \citep{rojratsirikul2009unsteady,serrano2018fluid}. \cite{song2008aeromechanics} examined the effects of aspect ratio, compliance and pre-strain values on the unsteady membrane dynamics for a low aspect ratio wing in wind tunnel experiments. The results indicated that the compliant wing can produce larger lift forces and delay stall, compared to its rigid counterpart. Secondary vortices induced by local separation at higher vibrational modes weakened the aerodynamic performance. Moreover, the flow-induced vibrations in the coupled system were found to exhibit different characteristics for membranes with different geometries of the leading and trailing edges \citep{arbos2013leading,serrano2016effect}, different excess lengths and pre-strain values \citep{rojratsirikul2010effect} and at various flow conditions (e.g., angle of attack and Reynolds number) \citep{rojratsirikul2009unsteady}. A series of wind tunnel experiments were conducted to study the aeromechanics and near-wake characteristics of flexible membrane wings in and out of ground effect \citep{bleischwitz2016aeromechanics,bleischwitz2017fluid,bleischwitz2018near}. In the experiments, the flexible membrane wings were found to produce a variety of flow features related to different types of flow-induced vibrations. In nature, for example, bats are capable of dynamically changing the material properties via miniature muscles and the high degree-of-freedom skeleton system. Therefore, the material properties also play an important role in the flow-induced vibration and the membrane performance. However, a few investigations on the effects of membrane material properties have been performed to understand the flow-induced vibration in the coupled system.

Concerning the flow-induced vibration phenomenon, various vibrational modes and wake patterns are excited in the coupled system via the flow-excited instability. Mode transition between different aeroelastic modes was observed as the angle of attack (AOA) changes \citep{song2008aeromechanics,bleischwitz2015aspect} or under forced pitching motion \citep{tregidgo2013unsteady}. The reason for the mode transition phenomenon during fluid-membrane interaction remains unclear and has not been systematically studied. Once the mode transition is triggered, the altered dominant modes might be fixed at a certain mode within a range of physical parameter space. For example, a typical chord-wise second mode with varied vibrational frequencies was always observed for a 2D membrane wing with the fixed leading and trailing edges when the AOA exceeded the stall angle at different free-stream velocities \citep{rojratsirikul2009unsteady}. Similar conclusions have been drawn in wind tunnel experiments for 3D membrane wings at high incidences \citep{rojratsirikul2010unsteady,tregidgo2011fluid}. Moreover, the occurrence of the chord-wise second mode was independent of the geometries at the leading and trailing edges \citep{arbos2013leading}. \cite{tregidgo2011fluid} found that the membrane vibrating in the chord-wise second mode was related to the frequency lock-in behaviour of a rigid airfoil. The possible reason was attributed to the strong coupling of the membrane oscillations and the particular vortex shedding process in the wake.

In parallel to experimental research, some numerical studies have been carried out to gain further insight into the fluid-membrane interaction phenomenon. One of the earliest relevant works has focused on the 2D linear elastic membrane model with the small deformation assumption in a potential flow \citep{newman1987aerodynamic} or in a laminar flow \citep{smith1995computational}. The recently rapid development in computational fluid dynamics, computational structural dynamics and fluid-structure coupling algorithms provided a practical and reliable way to investigate various nonlinear membrane dynamics for a wide range of physical parameters. \cite{gordnier2009high} developed a computational simulation solver by coupling a high-order Cartesian Navier-Stokes solver and a nonlinear membrane solver to study the 2D fluid-membrane interaction at low Reynolds numbers. The effects of AOA, the membrane rigidity, the membrane pretension and Reynolds number on the membrane dynamic responses have been investigated. The results confirmed the tight coupling between the membrane oscillations and the vortical structures in the wake, which was similar to the phenomena observed in the experiments \citep{song2008aeromechanics,rojratsirikul2009unsteady}. The evolutions of the nonlinear dynamic behaviour of a 2D membrane under periodic load \citep{sun2016nonlinear} and the membrane with the perimeter reinforcement \citep{sun2017nonlinear,sun2017effect} have been studied systematically. The results indicated that the membrane dynamics were closely dependent on the initial conditions applied to the membrane wing and the material properties. Besides, the membrane inertia enlarged the AOA range of a 2D flexible membrane in a stable state in a laminar flow \citep{tiomkin2019on}. However, the study on the effect of membrane inertia focused on the fluid-membrane interaction at small AOAs without obvious vortex structures. Only a handful of publications on fully coupled 3D fluid-membrane interaction analysis of flexible membrane dynamics can be found in the literature. Recently, \cite{yang2018aeroelasticity} developed a fluid-membrane interaction solver based on the coupling between a fluid solver and a finite element structural solver and conducted a systematic validation for 3D membrane dynamics. The numerical results achieved a good agreement with the experimental data \citep{rojratsirikul2009unsteady}.

In previous studies on the membrane dynamics, the relationship between the natural frequency of the tensioned membrane and the vortex shedding frequency has been found to play an important role in the coupled system \citep{song2008aeromechanics,rojratsirikul2009unsteady,bleischwitz2015aspect,waldman2017camber}. Different from the fluid-structure interaction problems for freely vibrating rigid bodies with the fixed natural frequency, the natural frequency of the flexible membrane can be changed when the aerodynamic loads are applied to a stretchable membrane surface. A simple linear natural frequency model of a 3D rectangular flexible membrane with all fixed edges can be given as \citep{kinsler1999fundamentals}:
\begin{equation}
f^n_{ij}=\frac{1}{2} \sqrt{\frac{N^s}{\rho^s h}} \left[\left(\frac{i}{c} \right)^2 + \left( \frac{j}{b} \right)^2 \right]^{\frac{1}{2}},
\label{linear_frequency}
\end{equation}
where $f^n_{ij}$ is the linear natural frequency of the membrane with the chord-wise $i$ and span-wise $j$ mode. The membrane chord length and the span length are $c$ and $b$, respectively. $N^s=E^s h \varepsilon^s$ denotes the membrane tension per unit length. $E^s$, $h$ and $\varepsilon^s$ represent the Young's modulus, the membrane thickness and the strain of the stretched membrane. $\rho^s$ is the membrane density. For a membrane with fixed geometries, $f^n_{ij}$ is only dependent on the structural properties ($\rho^s$ and $E^s$) and the strain $\varepsilon^s$ caused by the membrane deformation under aerodynamic loads. The variation of the natural frequency estimated from the linear natural frequency equation was examined for flexible membranes with different membrane rigidities, AOAs, Reynolds numbers \citep{gordnier2009high} and undergoing pitching motion \citep{tregidgo2013unsteady}. The dominant structural mode of the vibrating membrane and the vortex shedding pattern were also changed, a phenomenon known as mode transition. \cite{song2008aeromechanics} reported a series of wind tunnel experiments on membrane wings with different aspect ratios, compliance and prestrain values. The mode transition from the first mode to the higher-order mode was observed. They argued that this mode transition phenomenon was possibly caused by the nonlinear resonance mechanism between the constant vortex shedding frequency and the reduced natural frequency of the membrane or influenced by the increasing forcing frequency due to the vortex shedding process as the Reynolds number increased. However, the employed linear natural frequency model in \refeq{linear_frequency} neglects the dynamic stress caused by the nonlinear vibration and the added mass of the vibrating membrane immersed in uniform flow. Thus, this simplified model is not suitable for the analysis of the nonlinear dynamics and the mode transition in a coupled fluid-membrane system with moderate and large amplitude vibrations. A nonlinear natural frequency model that considers the added mass effect is required for further analysis. While some studies have been done to examine the physics of the flow-induced vibration and the mode transition, the underlying relationship between these phenomena and the role of the natural frequency are not yet fully understood in these coupled systems. The complexity and nonlinearity of the 3D coupled fluid-structure dynamics pose a serious challenge in the analysis. In addition, a large physical parameter space that governs the flow-induced vibration increases the difficulty to explore the coupled dynamics in a comprehensive manner.

In this study, we present novel physical insights into the coupled fluid-membrane dynamics of an extensible 3D membrane immersed in unsteady separated flows at moderate Reynolds numbers. Of particular interest is to examine the role of the flow-excited instability and the natural frequency of the 3D flexible membrane in the coupled fluid-membrane characteristics and the mode transition phenomenon. To shed light on the variation of the flow-induced vibration, we establish the stability phase diagrams in the $m^*$-$Re$ space and the $m^*$-$Ae$ space through a series of coupled numerical simulations. New empirical equations of the flow-excited instability boundary demarcated between the DSS and DBS regimes are determined based on our numerical simulations. We employ the global Fourier mode decomposition (FMD) technique \citep{li2020aeroelastic} to identify the dominant aeroelastic modes and classify the membrane vibrational states from the vibrating membrane responses. To isolate the impact of the relevant physical parameters on the flow-induced vibration and the membrane dynamics, we investigate the coupled fluid-membrane dynamics as a function of mass ratio $m^*$, Reynolds number $Re$ and aeroelastic number $Ae$. We further compare the mode frequency spectra and the energy distributions in the fluid and structural domains based on the FMD analysis. The natural frequency of the tensioned membrane and the vortex structures are important to understand the onset of the flow-induced vibration and the mode transition. An approximate analytical formula of the nonlinear natural frequency of a simply supported 3D rectangular membrane in uniform flow considering the added mass is derived using the nonlinear structural dynamic equation based on large deflection theory. The vortex shedding frequencies are measured at the monitoring line in the wake via the Fourier-based signal analysis. The natural frequency of the membrane and the vortex shedding frequency are compared as a function of $m^*$, $Re$ and $Ae$. We systematically examine the onset of the flow-induced vibration and the mode transition by monitoring the variation of the natural frequency. The findings and conclusions can be used to promote the development of active/passive control strategies in producing superior aerodynamic performance.

The remainder of this manuscript is organized as follows. In $\S$2, we briefly introduce the coupled fluid-structure formulation and the global Fourier mode decomposition technique. The description of the fluid-membrane interaction problem and the validation of a 3D membrane with a supporting rigid frame is conducted in $\S$3. We present the membrane dynamics as functions of three physical parameters in $\S$4. The dependence of the flow-induced vibration and the mode transition on the frequency lock-in and the structural natural frequency is further explored. In $\S$5, the main conclusions of the flow-excited instability are summarized.

\section{Numerical methodology}
\subsection{Coupled fluid-structure formulation}
The governing equations for the incompressible unsteady viscous flow with an arbitrary Lagrangian-Eulerian (ALE) reference frame are discretized using the stabilized Petrov-Galerkin variational formulation. The generalized-$\alpha$ method is utilized to integrate the ALE flow solution in the time domain and it can ensure unconditionally stable and second-order accuracy for linear dynamics problems. The motion equations for a flexible structure are discretized using the standard Galerkin finite element method. The kinematic joints are considered as constraints on the displacement field. A hybrid RANS/LES model based on the delayed detached eddy simulation treatment is employed to simulate the separated turbulent flow \citep{joshi2017variationally}. A partitioned iterative coupling algorithm is adopted to integrate the fluid equations and the multibody structural equations. A typical predictor-corrector scheme is used to solve the coupled fluid-structure governing equations. The velocity and traction continuity along the coupling interface with non-matching meshes is satisfied in the fluid-structure coupling procedure. An efficient interpolation scheme via the compactly-supported radial-basis function (RBF) with a third-order of convergence is implemented to exchange coupling data through the fluid-structure interface and to update the mesh in the fluid domain \citep{joshi2020variational}. To avoid the numerical instability caused by significant added mass effect, a recently developed nonlinear interface force correction scheme \citep{jaiman2016stable} is utilized to correct and stabilize the fluid forces at each iterative step. The fluid-multibody structure interaction equations and the variational partitioned formulations for the fluid-structure interaction framework have been described in \cite{li2018novel} in detail. This fluid-structure interaction solver has been validated for an anisotropic flexible wing with supporting battens and covered membrane components \citep{li2018novel}.

\subsection{Global Fourier mode decomposition in fluid-membrane interaction}
The flexible membrane wing is coupled with the unsteady flow fluctuations to vibrate in a variety of mode shapes. The multi-mode mixed dynamic responses of the membrane vibrations and the vortices with varied scales become an obstacle to identify the dominant aeroelastic modes from the complex coupled system. The global Fourier mode decomposition technique provides an effective way to isolate the most influential modes of interest from the membrane dynamic responses and the surrounding flow field. The global FMD method transfers the spatial-time physical variables to the spatial-frequency data at each spatial position for the whole physical field via the discrete Fourier transform. The phase, the amplitude and the mode energy of each dynamic mode are calculated quantitatively with the aid of FMD. Given the inherent coupling effect between the unsteady flow and the flexible structures in fluid-structure interaction problems, the dynamic modes in the fluid and structure domains should be isolated from the coupled system together. Thus, a direct connection can be established between these aeroelastic modes when exploring the coupling mechanism. The detailed algorithm of the FMD method and its application to fluid-membrane interaction is presented in \cite{li2020aeroelastic}.

\section{Numerical set-up and validation}
\begin{figure}
	\centering
	\subfloat[][]{\includegraphics[width=0.45\textwidth]{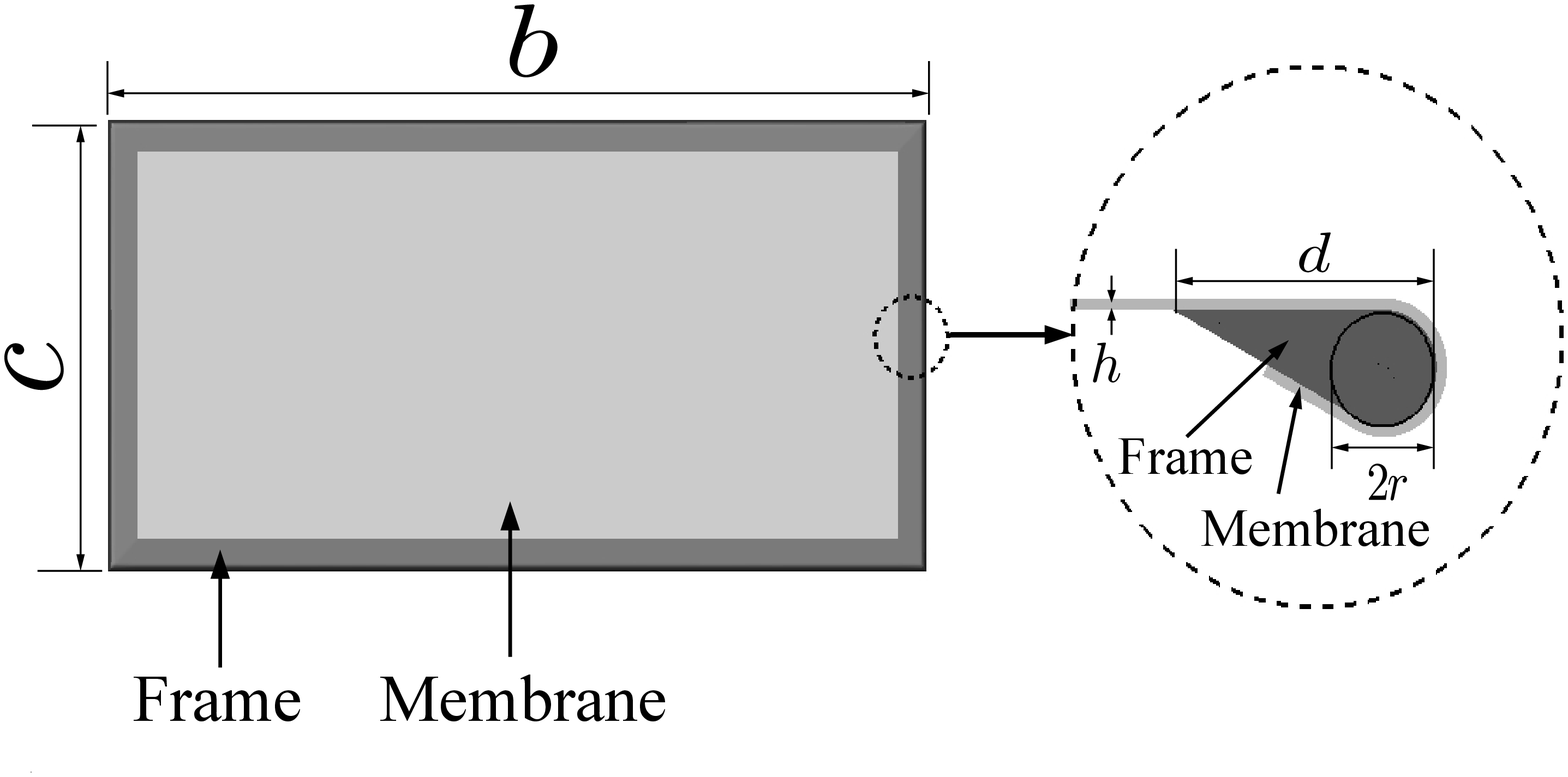}\label{membranegeo}}
	\quad
	\subfloat[][]{\includegraphics[width=0.5\textwidth]{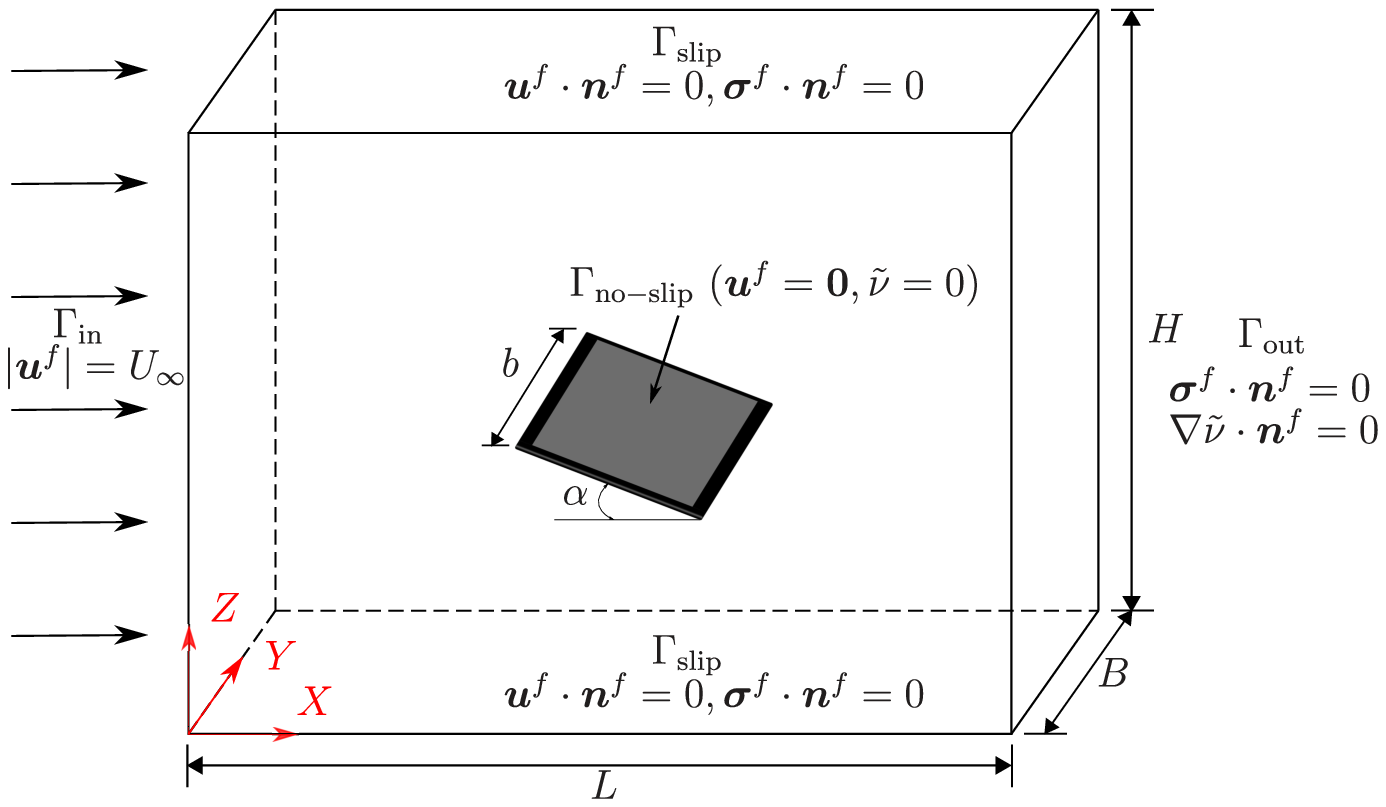}\label{membrane_domaina}}
	\caption{Three-dimensional computational set-up for fluid-membrane interaction: (a) membrane wing geometry and support frame section, (b) schematic diagram of computational domain and boundary conditions.}
	\label{membrane_domain}
\end{figure}

In this section, we provide an overview of the fluid-membrane interaction simulation set-up for a 3D rectangular flexible membrane wing immersed in a viscous incompressible fluid flow. The 3D wing consists of a latex rubber membrane sheet and a rectangular stainless steel frame. The thin membrane component has a thickness of $h$=0.2 mm. The Young's modulus of the latex rubber is $E^s$=2.2 MPa and its material density is $\rho^s$=1000 kg/m$^3$. The wing with an aspect ratio of 2 has a chord length of $c$=68.75 mm. The membrane wing geometry and the information of the support frame section are displayed in \reffig{membrane_domain} \subref{membranegeo}. The width of the supporting frame section is $d$=5 mm and the diameter of the support rob is $2r$=2 mm. Three key non-dimensional physical parameters, namely aeroelastic number $Ae$, mass ratio $m^*$ and Reynolds number $Re$, govern the nonlinear dynamic responses of the flexible membrane aerofoils, which are given as
\begin{equation}
Ae=\frac{E^s h}{\frac{1}{2} \rho^f U_{\infty}^2 c},\quad m^*=\frac{\rho^s h}{\rho^f c}, \quad Re=\frac{\rho^f U_{\infty } c}{\mu^f},
\end{equation}
where $\rho^f$ represents the air density, $U_{\infty}$ is the free-stream velocity and $\mu^f$ denotes the dynamic viscosity of the air.

The air load is calculated by integrating the surface traction taking into account the first layer of elements on the membrane surface. The instantaneous lift, drag and normal force coefficients are defined as
\begin{equation}
C_L = \frac{1}{\frac{1}{2} \rho^f U_{\infty}^2 S} \int_{\Gamma} (\boldsymbol{\bar{\sigma}}^f \cdot \boldsymbol{n}) \cdot \boldsymbol{n}_z \rm{d \Gamma},
\label{CL} 
\end{equation}
\begin{equation}
C_D = \frac{1}{\frac{1}{2} \rho^f U_{\infty}^2 S} \int_{\Gamma} (\boldsymbol{\bar{\sigma}}^f \cdot \boldsymbol{n}) \cdot \boldsymbol{n}_x \rm{d \Gamma},
\label{CD} 
\end{equation}
\begin{equation}
C_N = \frac{1}{\frac{1}{2} \rho^f U_{\infty}^2 S} \int_{\Gamma} (\boldsymbol{\bar{\sigma}}^f \cdot \boldsymbol{n}) \cdot \boldsymbol{n}_c \rm{d \Gamma},
\label{CN} 
\end{equation}
where $\boldsymbol{n}_x$ and $\boldsymbol{n}_z$ are the Cartesian components of the unit normal vector $\boldsymbol{n}$ to the membrane surface and $\boldsymbol{n}_c$ is the unit normal vector to the chord line. $\boldsymbol{\bar{\sigma}}^f$ is the fluid stress tensor with $\Gamma$ being the surface boundary of the membrane. The pressure coefficient is defined as
\begin{equation}
C_p = \frac{p-p_\infty}{\frac{1}{2} \rho^f U_{\infty}^2},
\label{Cp} 
\end{equation}
where $p$ and $p_\infty$ are the pressure at the local point and the pressure at the far-field, respectively. 

A sketch of the 3D computational domain used for the fluid-membrane interaction problem is shown in \reffig{membrane_domain}\subref{membrane_domaina}. We set the same length of $50c$ for the distances between the side walls ($\Gamma_{\text{slip}}$) on top and bottom as well as these on both sides and the distance between inlet ($\Gamma_{\text{in}}$) and outlet ($\Gamma_{\text{out}}$) boundaries. A uniform flow velocity $\boldsymbol{\bar{u}}^f=(u^f, v^f, w^f)$ is set at the inlet boundary $\Gamma_{\text{in}}$ for the computational domain. The boundary condition on the top, bottom and both sides boundaries ($\Gamma_{\text{slip}}$) is defined as the slip-wall boundary condition. We apply a traction-free boundary condition for the outlet boundary $\Gamma_{\text{out}}$, where $\sigma_{xx}=\sigma_{yx}=\sigma_{zx}=0$. A no-slip boundary condition is applied on the membrane wing surface. The structural model of this rectangular membrane wing includes the rigid body elements to model the rigid frame and the geometrically exact co-rotational shell elements to simulate the elastic membrane. A clamped boundary condition with all fixed degrees of freedom is imposed on the rigid frame.

Before we proceed to the validation of the employed fluid-structure interaction solver for this 3D flexible membrane, a systematic convergence study is performed for the adequacy of mesh resolution and time step size. The convergence results are summarized in Appendix A. At different AOAs, we next compare the membrane displacements, the aerodynamic forces, the membrane vibration frequency and the unsteady flow features with the experimental data \citep{rojratsirikul2010unsteady,rojratsirikul2011flow} in Appendix B. Overall, the coupled fluid-membrane dynamics are well captured by our high-fidelity aeroelasticity solver and the results match reasonably well with the experimental data. A detailed validation study and the comparisons with other numerical simulations can be found in \cite{li2020computational}. Notably, the coupled fluid-membrane dynamics of 2D flexible membrane predicted by our high-fidelity solver and the numerical solver of \cite{gordnier2009high} were found to have an excellent agreement. A validation study for a 3D rectangular membrane wing was also presented in \cite{li2020computational} along with the comparison with the simulation results of \cite{gordnier2014impact}. Owing to the differences in the underlying fluid-structure interaction formulations and the choice of turbulence modelling in our present solver and that of \cite{gordnier2014impact}, some discrepancies were observed in the unsteady flow features for the 3D set-up of flexible membrane wing. Besides the solver details, these discrepancies can be also attributed to differences in the post-processing such as the time window selected for the time average calculation and the time instant chosen for plotting the instantaneous membrane dynamics. Further investigation will be helpful to consolidate some of these numerical solvers for 3D unsteady flow-membrane interaction with turbulence effects.

\section{Results and discussion}
The current study aims to investigate the role of the flow-excited instability in a coupled system for a 3D membrane immersed in an unsteady separated flow. To fully understand the onset of the flow-induced membrane vibration and the mode transition phenomenon, we systematically simulate the membrane dynamics of the 3D rectangular membrane wing over a wide range of parameter space at a representative AOA of $\alpha$=$15^\circ$ with an apparent separated flow. Different stability regimes can be classified from the phase diagrams and the new empirical solutions of the flow-excited instability boundary are determined via our high-fidelity numerical simulations. The dominant aeroelastic mode is further identified through the FMD analysis when the flow-induced vibration occurs. To examine the connection between the natural frequency of the tensioned membrane and the vortex shedding frequency, we systematically study the effect of mass ratio, Reynolds number and aeroelastic number on the membrane dynamics. We further discuss the onset of the flow-induced vibration and the mode transition in the coupled system by correlating with the variation of the natural frequency of the flexible membrane.

\subsection{Flow-excited membrane instability}
\subsubsection{Classification of membrane states}
\begin{figure}
	\centering 
	\includegraphics[width=0.9\textwidth]{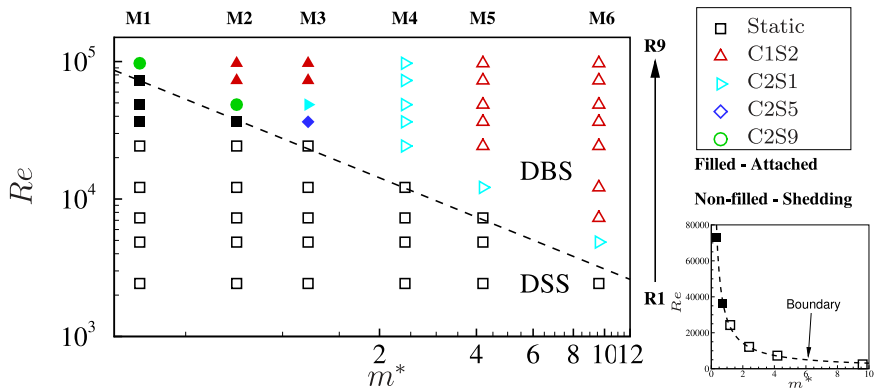}
	\caption{Stability phase diagram: non-dimensional Reynolds number $Re$ (R1$\to$9) versus mass ratio $m^*$ (M1$\to$6) for the 3D flexible membrane at $Ae=423.14$ for $\alpha$=$15^\circ$. Here, the dashed line ($\textcolor[RGB]{1,1,1}{---}$) is plotted to distinguish the flow-excited instability boundary. $\square$ denotes the simulation results corresponding to the deformed-steady state. $\triangle$, $\triangleright$, $\Diamond$ and $\circ$ represent the chord-wise first and span-wise second mode (C1S2), the chord-wise second and span-wise first mode (C2S1), the chord-wise second and span-wise fifth mode (C2S5) and the chord-wise second and span-wise ninth mode (C2S9) in the dynamic balance state. The label with or without filled colour is the flexible membrane with attached vortices or vortex shedding, respectively. }
	\label{mode_map}
\end{figure}

Before we proceed to exploring the flow-induced membrane vibration and the mode transition, we simulate the coupled 3D rectangular membrane dynamics as functions of mass ratio and Reynolds number. Six groups of mass ratios (M1$\to$6) within [0.36, 9.6] and nine sets of Reynolds numbers (R1$\to$9) ranging from 2430 to 97200 are chosen to form the parameter space to ensure that the flow-excited instability can be characterized. \refFig{mode_map} shows the stability phase diagram and the classification of the membrane states over the full parameter space. Two distinctive stability regimes are observed from the stability phase diagram termed a deformed-steady state (DSS) and a dynamic balance state (DBS). These two stability states are termed from the perspective of the tension force involved in the nonlinear natural frequency model presented in \refeq{mem32} in Appendix C. The membrane within the DSS regime refers to the membrane that is statically deformed under the aerodynamic loads and finally reaches a steady state. Meanwhile, the inherent tension force $N^s=E^sh \varepsilon^s$ caused by membrane deformations is statically balanced with the aerodynamic loads $\Delta p$. The dynamic balance state is achieved when the dynamically changing tension force is balanced with the unsteady aerodynamic forces and the inertial forces. The membrane vibrates with a limited amplitude within the DBS regime. It can be seen from \reffig{mode_map} that the flexible membrane maintains a static equilibrium state at low $Re$ for light membrane structures. For brevity, the Young-Laplace equation can be employed here to describe the deformed-steady state of the membrane \citep{song2008aeromechanics,waldman2017camber}, which can be written as
\begin{equation}
E^s h + \frac{\Delta p}{\varepsilon^s  \kappa^s} = 0,
\label{membrane_balance}
\end{equation}
where $\kappa^s$ is the curvature of the deformed membrane. In the current simulation, the first term in \refeq{membrane_balance} remains a constant value over the full $m^*$-$Re$ parameter space since the Young's modulus $E^s$ and the thickness $h$ are fixed. The values of the aerodynamic loads $\Delta p$, the membrane strain $\varepsilon^s$ and the curvature $\kappa^s$ of the deformed membrane are dependent on Reynolds number $Re$ and mass ratio $m^*$. Thus, the ratio $R$ in the second term can be written as a function of $Re$ and $m^*$, expressed as $R=\frac{\Delta p}{\varepsilon^s  \kappa^s}=f(Re,m^*)$. In other words, the ratio $R$ remains constant as $Re$ and $m^*$ vary. Through our high-fidelity numerical simulations, the flow-excited instability boundary between the DSS regime and the DBS regime can be observed in \reffig{mode_map}, which is plotted as a dashed curve line ($\textcolor[RGB]{1,1,1}{---}$). The flow-excited instability boundary described by the function of $Re$ and $m^*$ should satisfy the relationship shown above. Based on the simulation results, a new empirical relation that governs the flow-excited instability boundary curve can be expressed as
\begin{equation}
Re_{cr}=c_0+c_1(m^*)^n,
\label{recr}
\end{equation}
The above relationship is constructed using the numerical simulation results for a fixed $\alpha$=$15^\circ$, where $c_0$=0, $c_1$=27550 and $n$=-0.9495. 
The onset of the membrane vibration related to the flow-excited instability will be discussed further in $\S$\ref{onset}.

\begin{figure}
	\centering 
	\subfloat[][]{\includegraphics[width=0.237\textwidth]{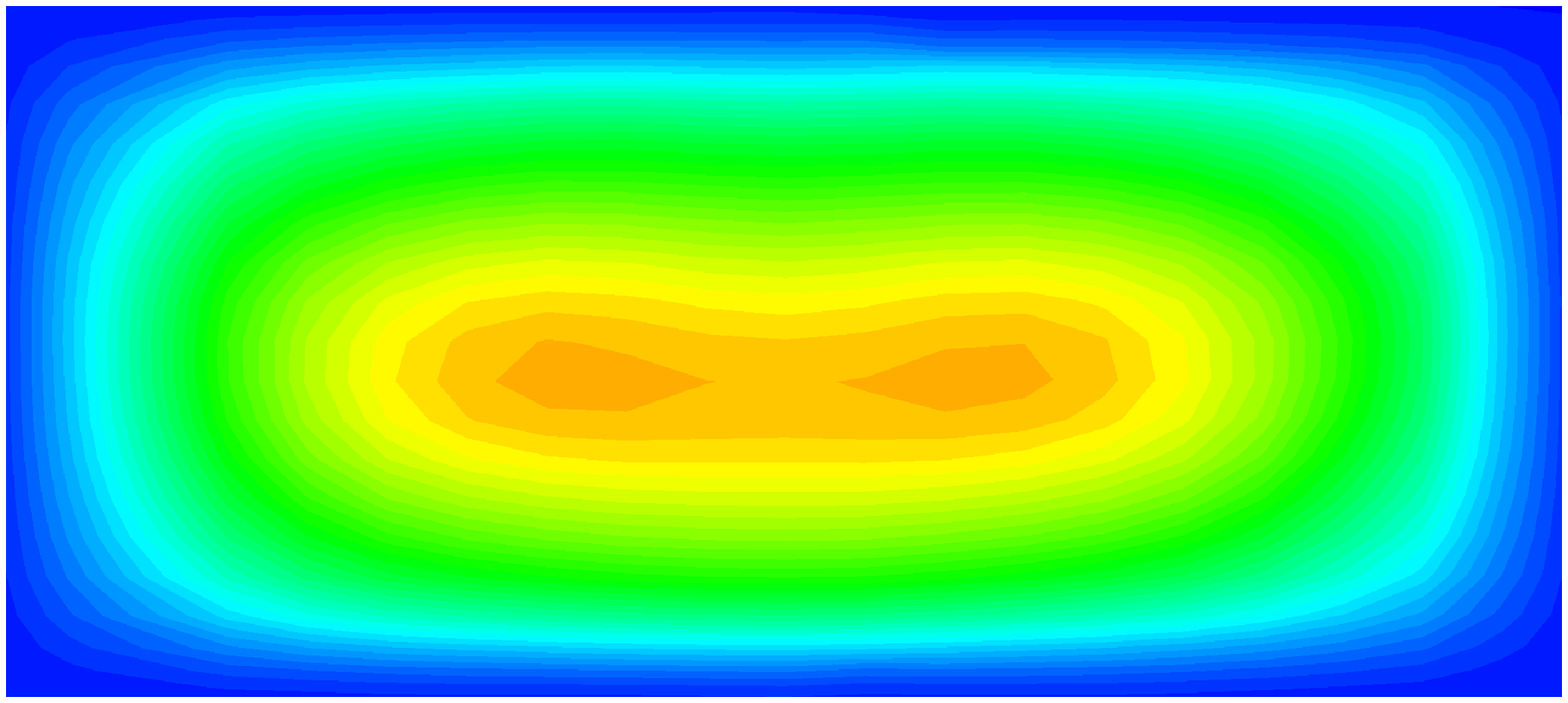}\label{modea}}
	\
	\subfloat[][]{\includegraphics[width=0.237\textwidth]{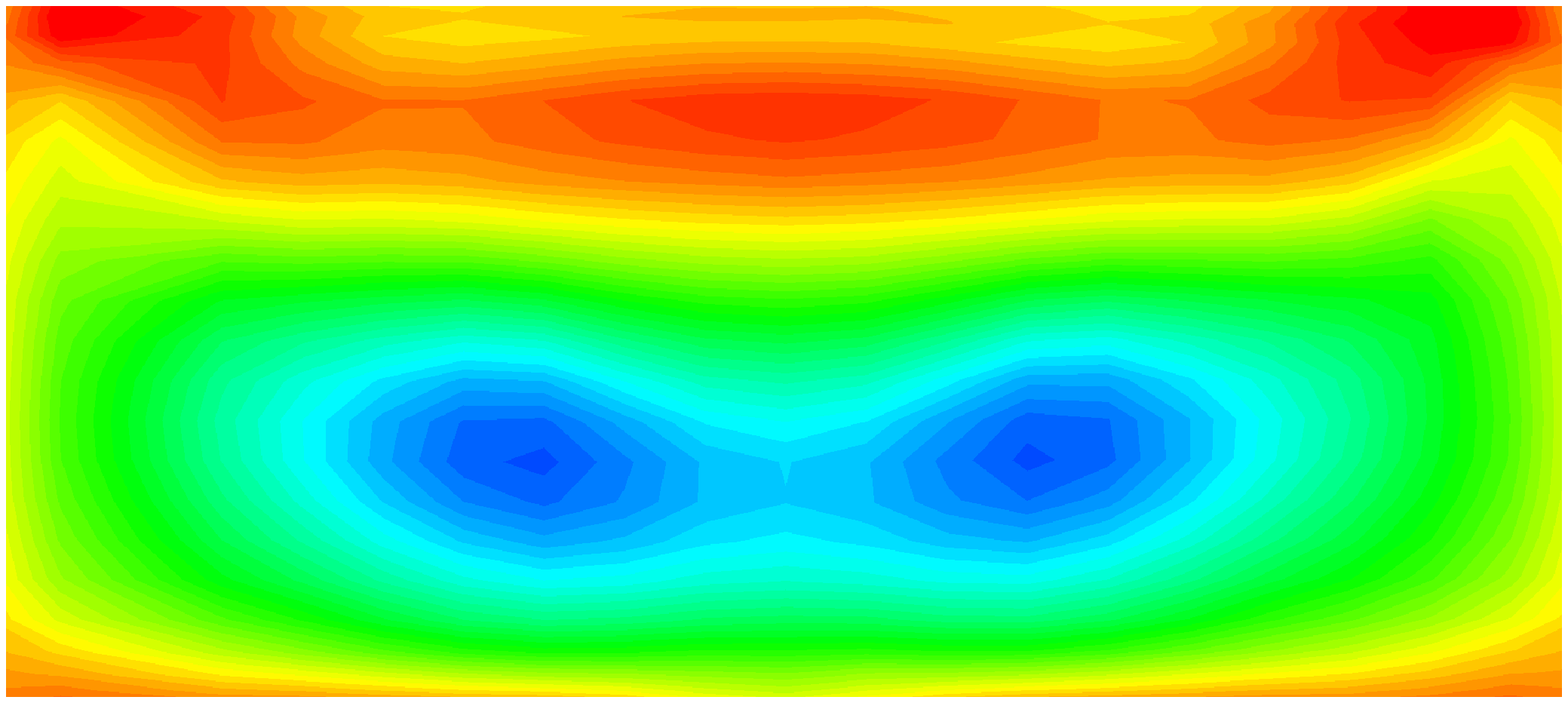}\label{modeb}}
	\
	\subfloat[][]{\includegraphics[width=0.237\textwidth]{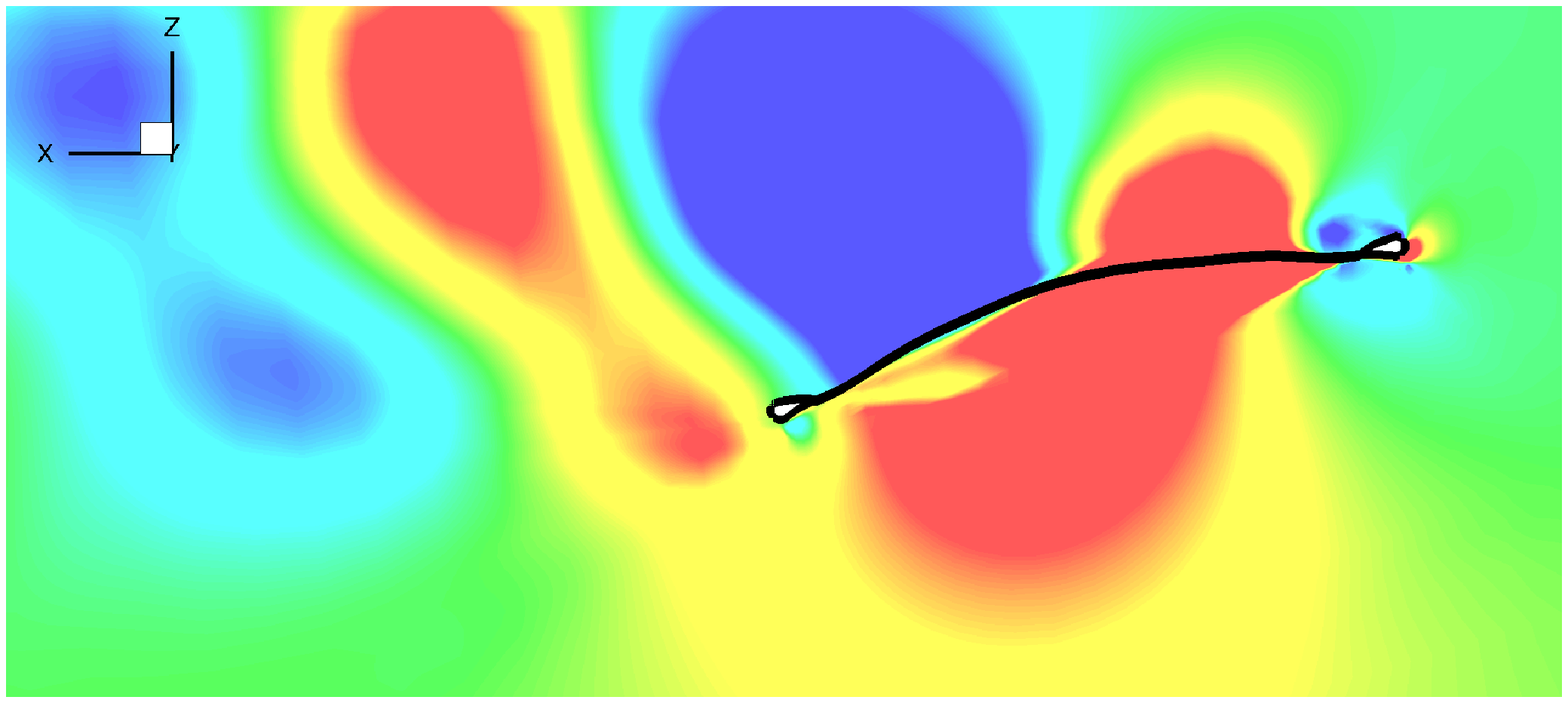}\label{modec}}
	\
	\subfloat[][]{\includegraphics[width=0.237\textwidth]{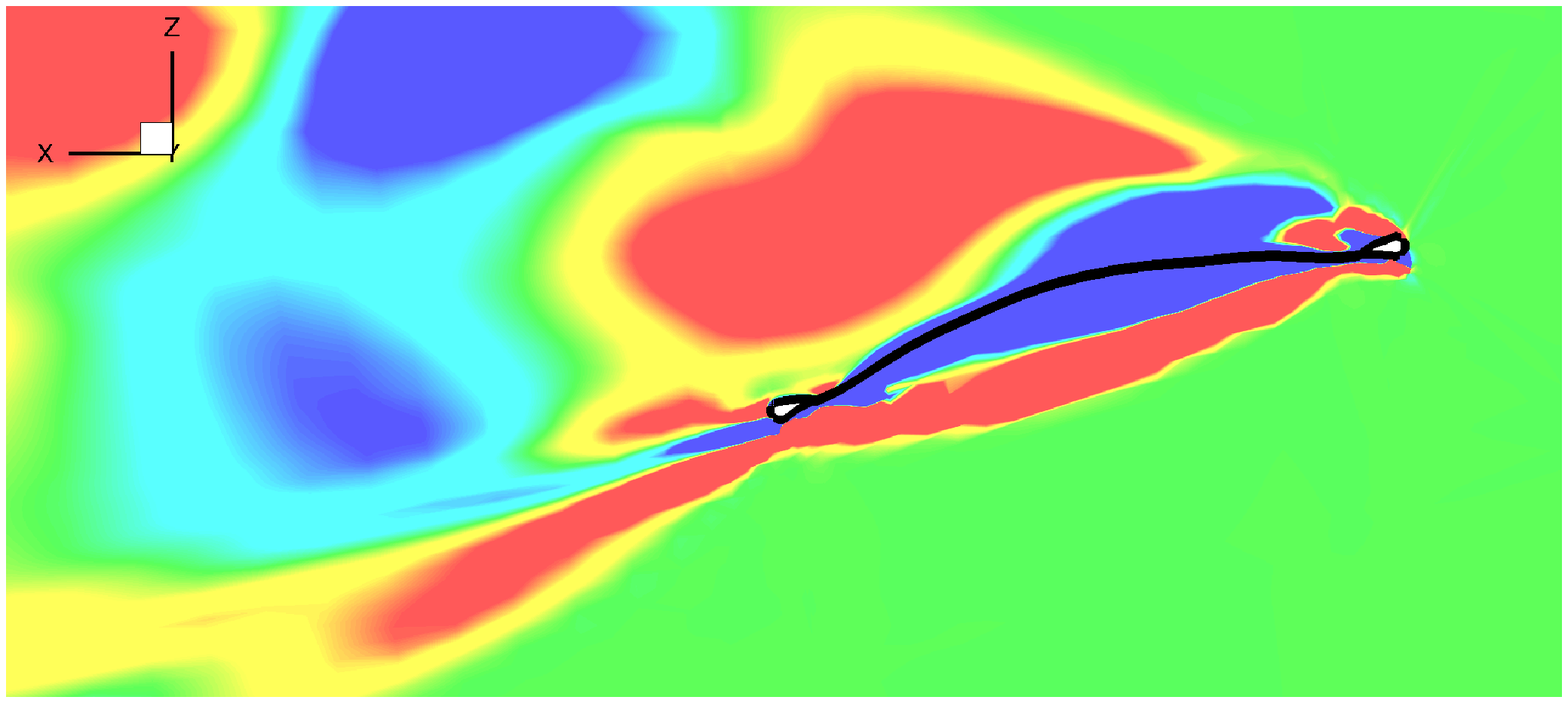}\label{moded}}
	\\
	\subfloat[][]{\includegraphics[width=0.24\textwidth]{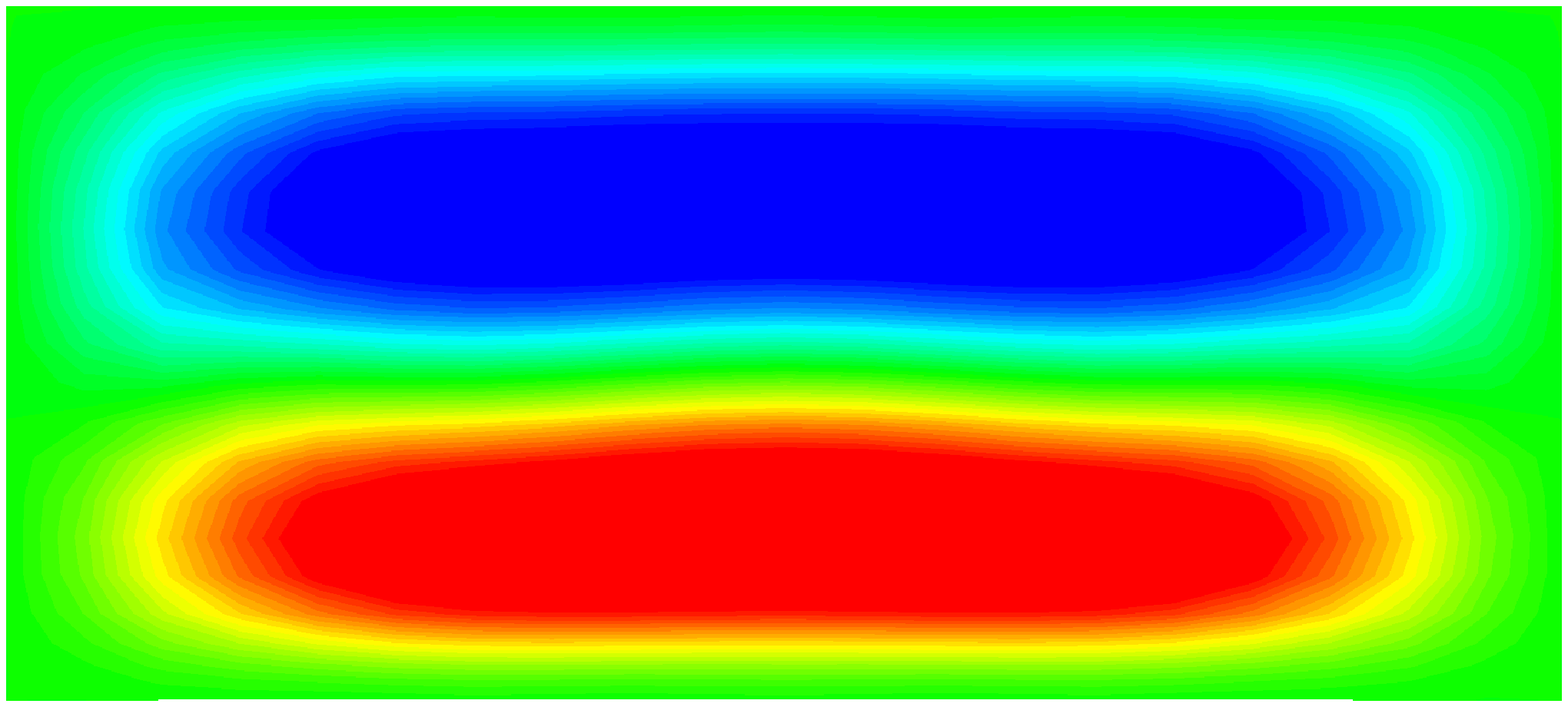}\label{modee}}
	\subfloat[][]{\includegraphics[width=0.24\textwidth]{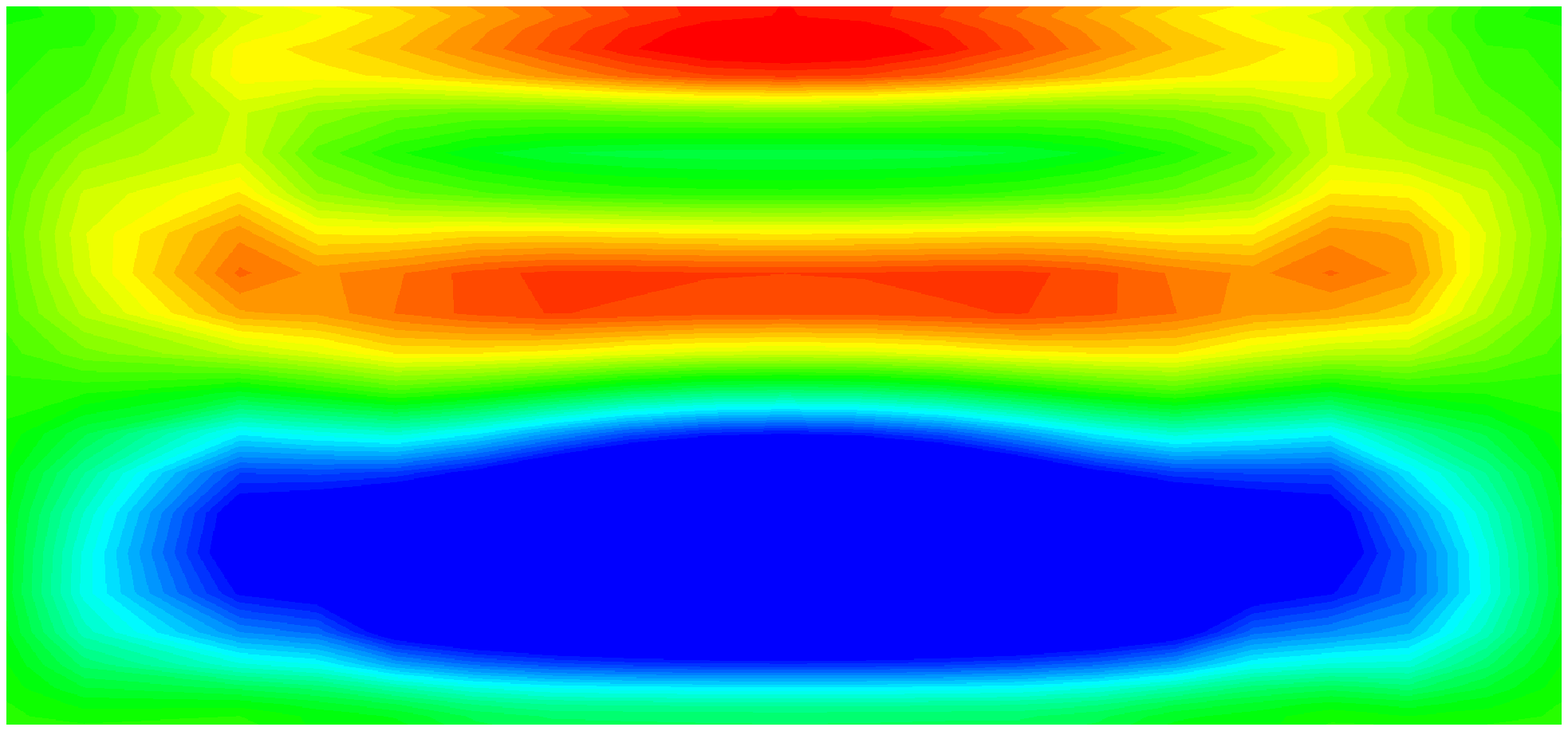}\label{modef}}
	\subfloat[][]{\includegraphics[width=0.24\textwidth]{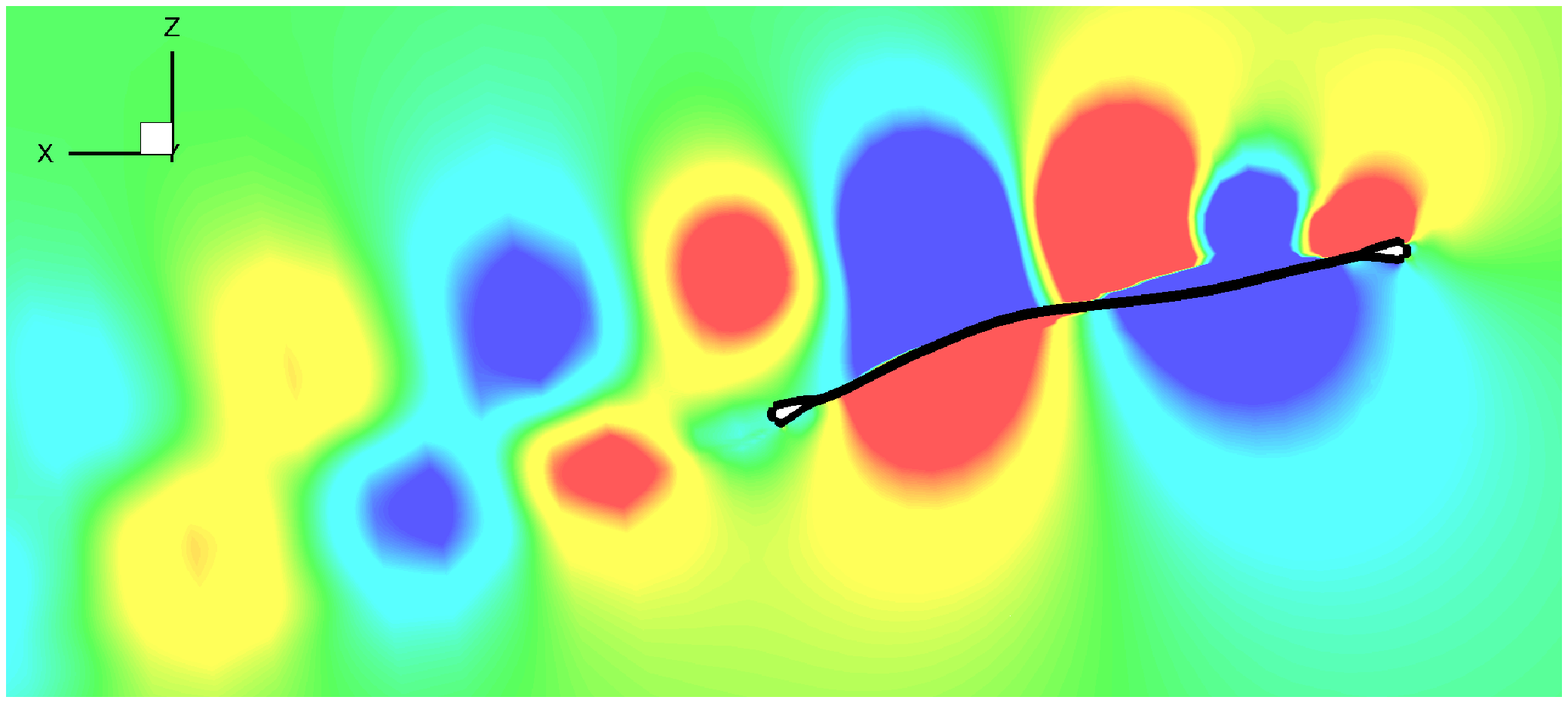}\label{modeg}}
	\
	\subfloat[][]{\includegraphics[width=0.24\textwidth]{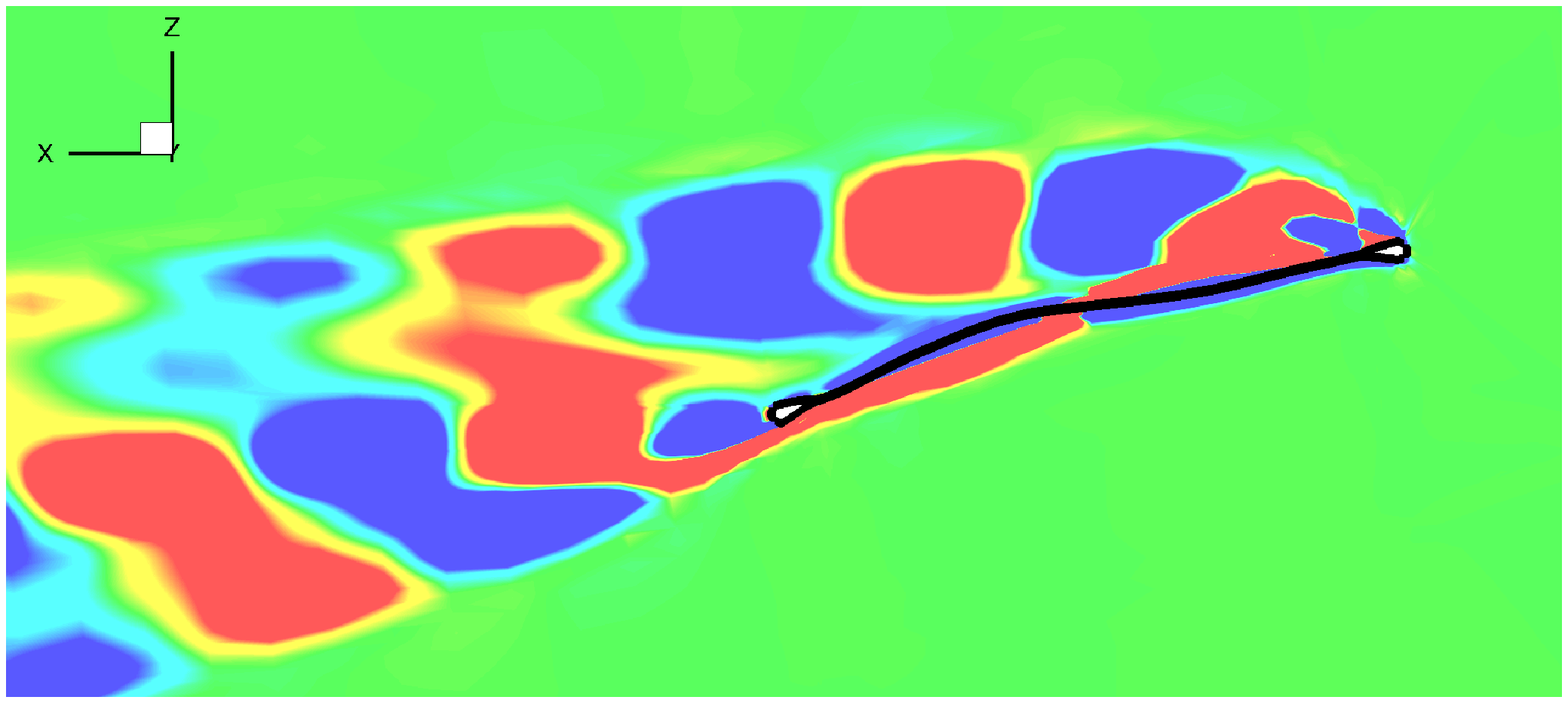}\label{modeh}}
	\\
	\caption{Dominant aeroelastic modes of 3D flexible membrane in DBS regime: (a,b,c,d) the chord-wise first and span-wise second mode at (M6,R5) and (e,f,g,h) the chord-wise second and span-wise first mode at (M4,R5). Contours are coloured by the real part of the Fourier transform coefficients of (a,e) membrane displacement on the whole surface, (b,f) membrane surface pressure difference (c,g) pressure coefficient fluctuation on the mid-span plane and (d,h) $Y$-vorticity fluctuation on the mid-span plane.}
	\label{modes}
\end{figure}

As Reynolds number and mass ratio exceed the boundary, it can be inferred from \reffig{mode_map} that the flexible membrane has a greater tendency for self-excited vibration and a stronger inertia effect on the fluid-structure coupling. Consequently, the inertia effect caused by the membrane vibration should be taken into account. The flexible membrane exhibits complex vibrational behaviour with overlapping structural modes within the DBS regime. With the aid of the proposed FMD method, the most influential aeroelastic mode is identified from the completed membrane responses. To represent the structural mode shapes conveniently, the chord-wise $i$ and span-wise $j$ mode of the 3D membrane is termed as C$i$S$j$ in the following discussions. Four typical dominant membrane modes are further classified within the DBS regime. We observe almost attached vortices on the membrane surface within the parameter space (M1$\to$3,R6$\to$9), which is indicated by the filled labels in \reffig{mode_map}. The obvious vortex shedding process is noticed at the rest parameter space for labels without filled colour. The dominant mode gradually transitions from the chord-wise second mode to the chord-wise first mode as the membrane inertia becomes significant and the Reynolds number increases. In \reffig{modes}, we present the dominant chord-wise first and second aeroelastic modes obtained from the membrane dynamic responses based on the FMD method for two representative cases (M6,R5) and (M4,R5) within the DBS regime, respectively. These dominant modes are classified as the corresponding mode energy calculated based on the FMD method is greater than other modes.

\subsubsection{Coupled fluid-membrane dynamics}
\begin{figure}
	\centering 
	\subfloat[][]{\includegraphics[width=0.48\textwidth]{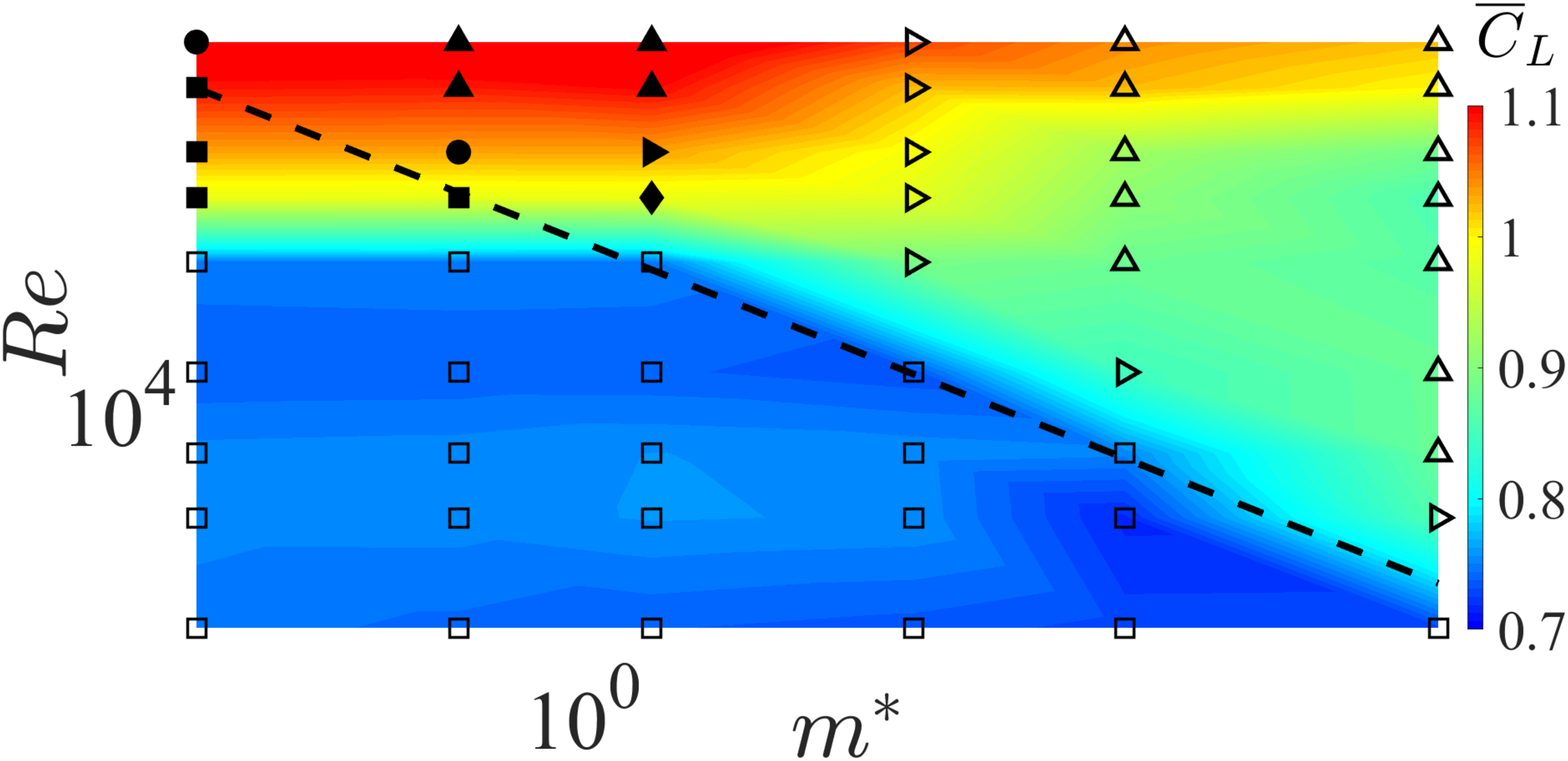}\label{phasea}}
	\quad
	\subfloat[][]{\includegraphics[width=0.48\textwidth]{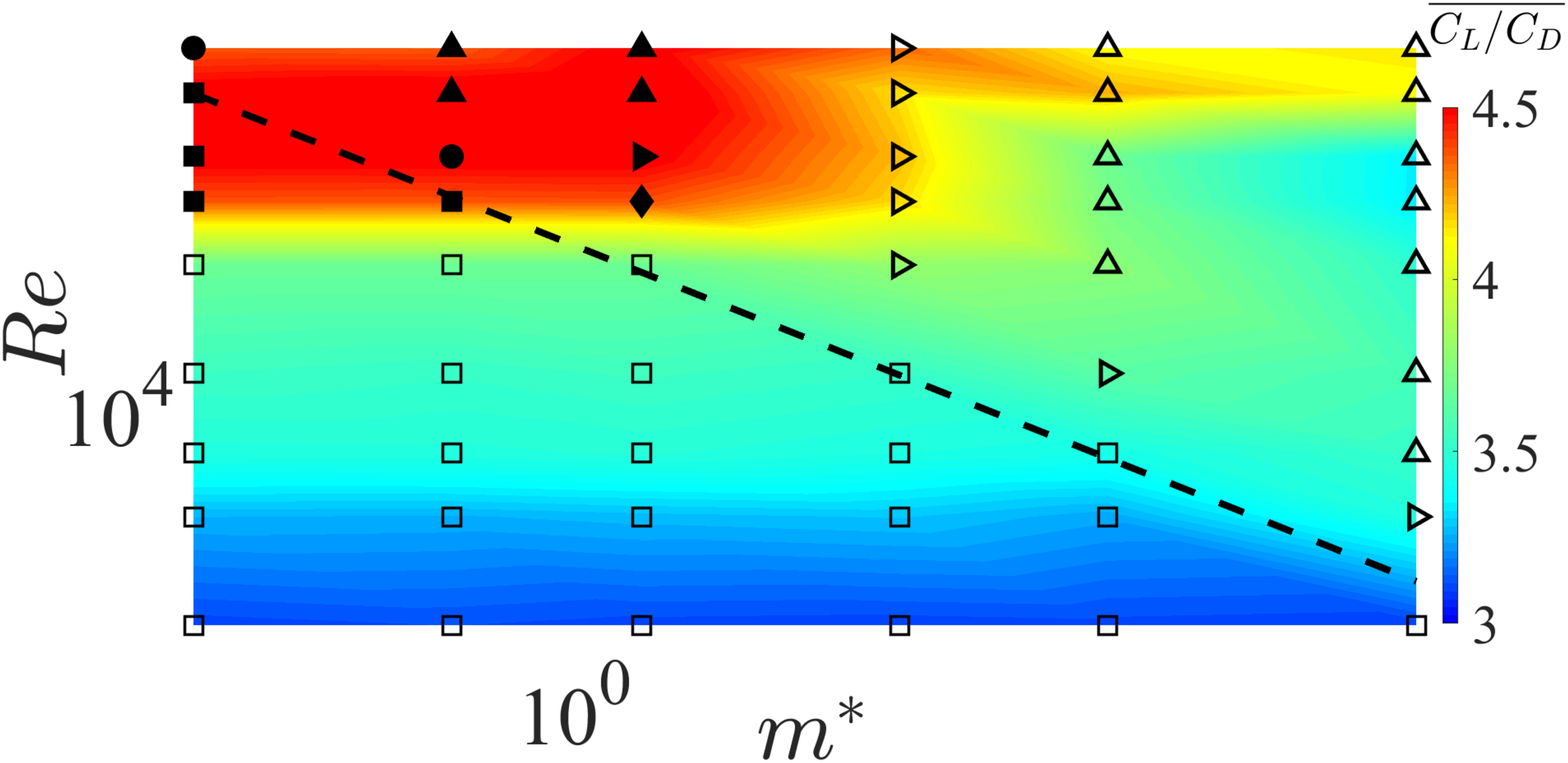}\label{phaseb}}
	\\
	\subfloat[][]{\includegraphics[width=0.48\textwidth]{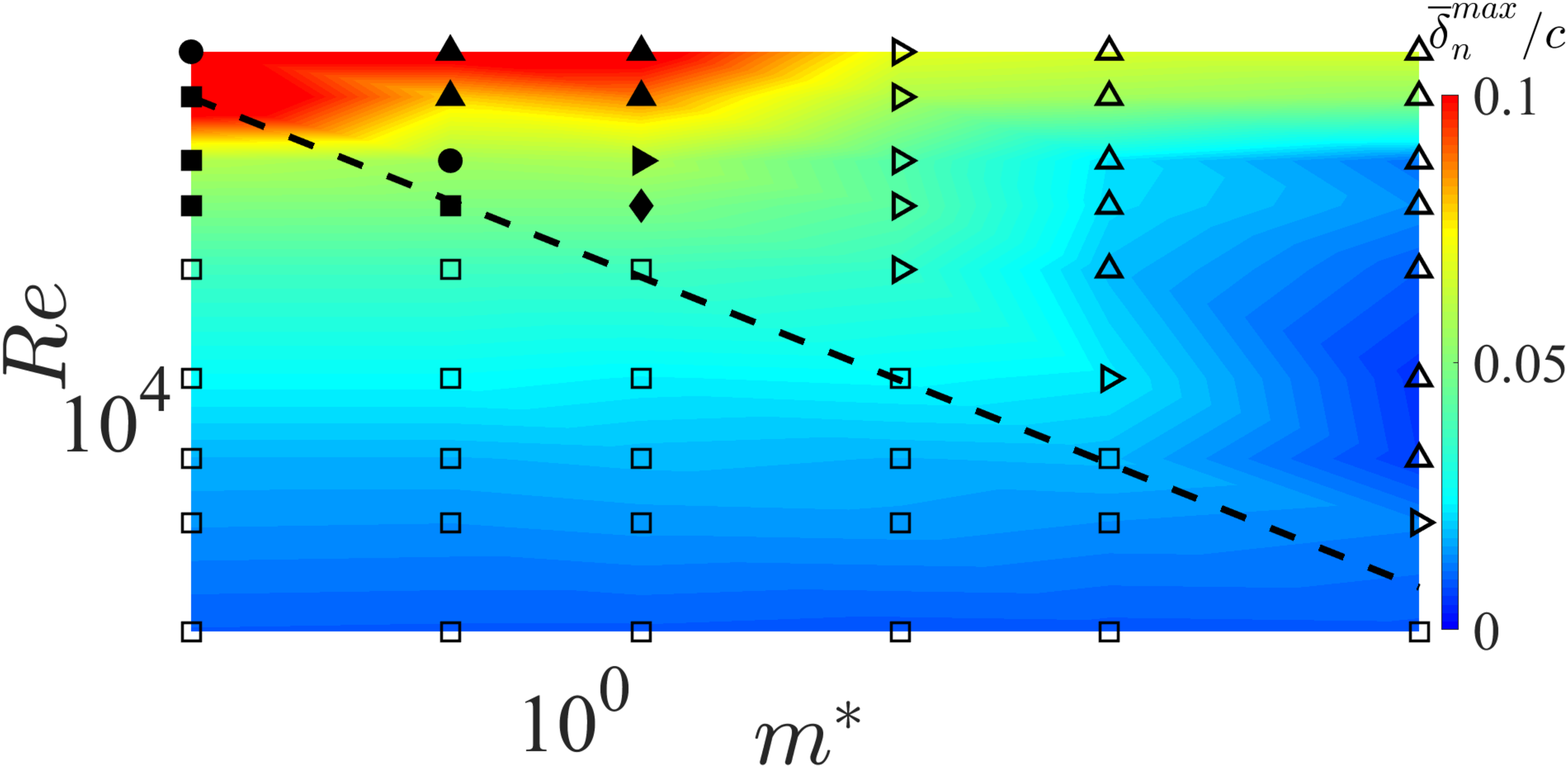}\label{phasec}}
	\quad
	\subfloat[][]{\includegraphics[width=0.48\textwidth]{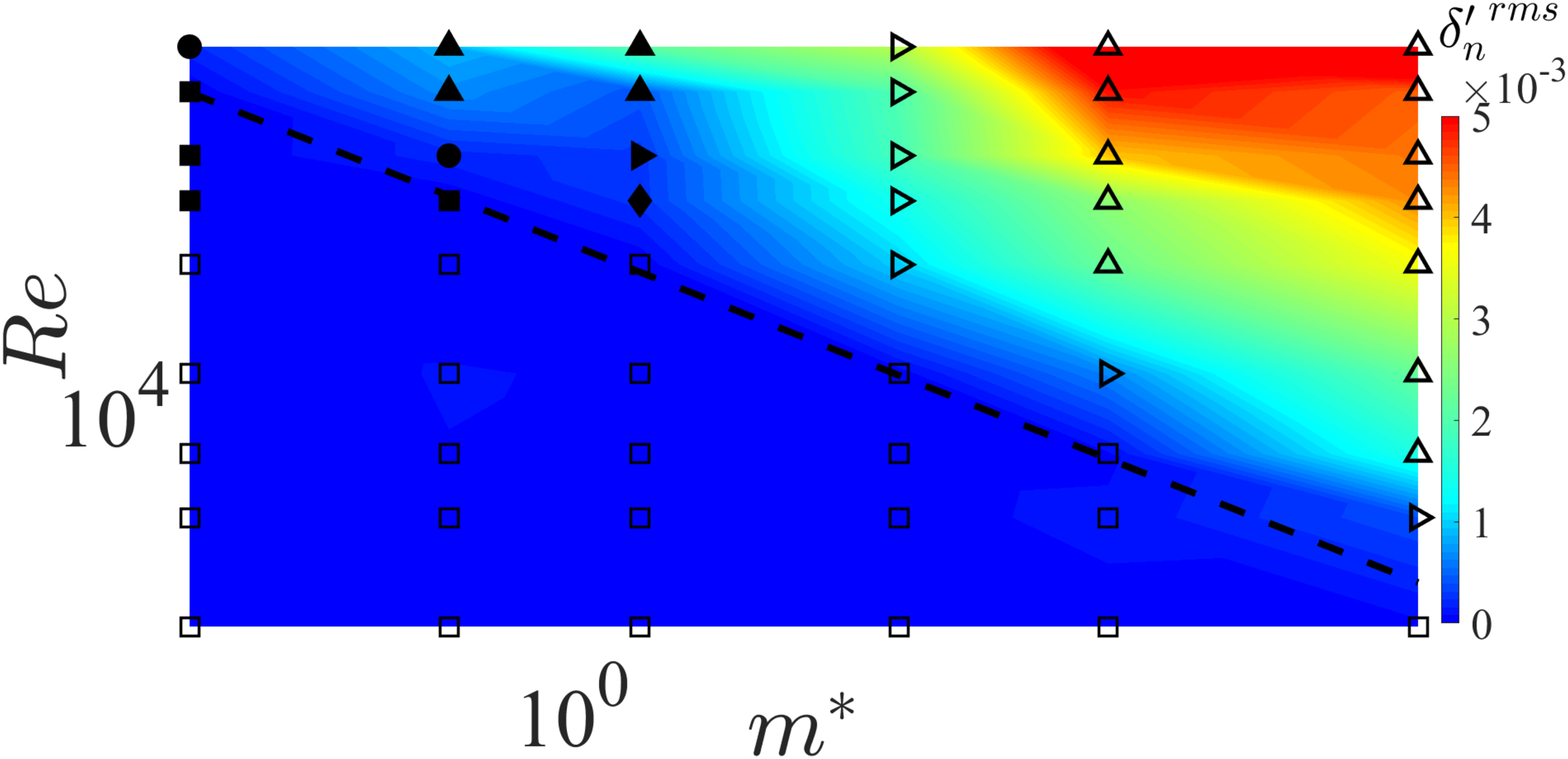}\label{phased}}
	\\
	\subfloat[][]{\includegraphics[width=0.48\textwidth]{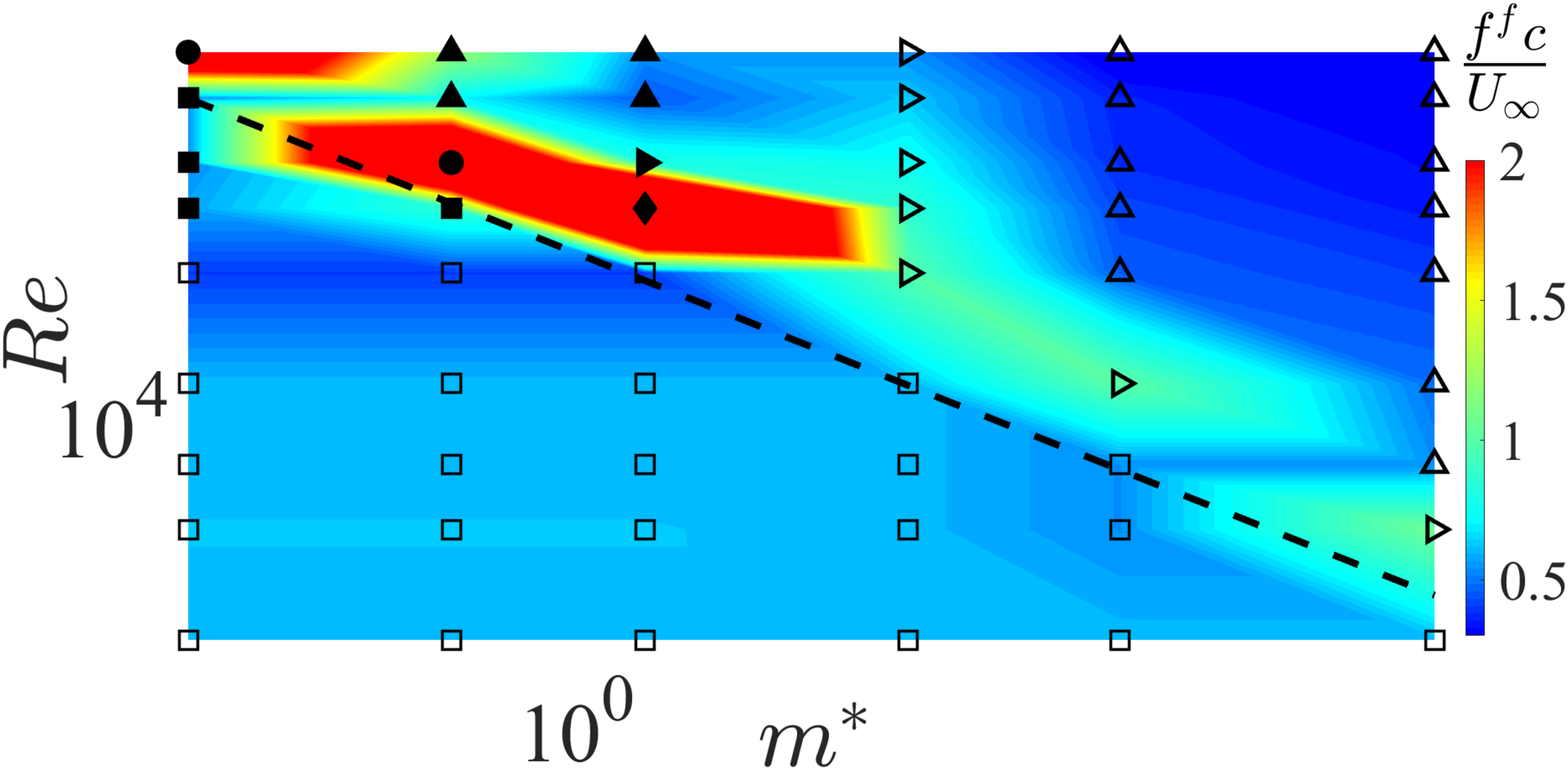}\label{phasee}}
	\quad
	\subfloat[][]{\includegraphics[width=0.48\textwidth]{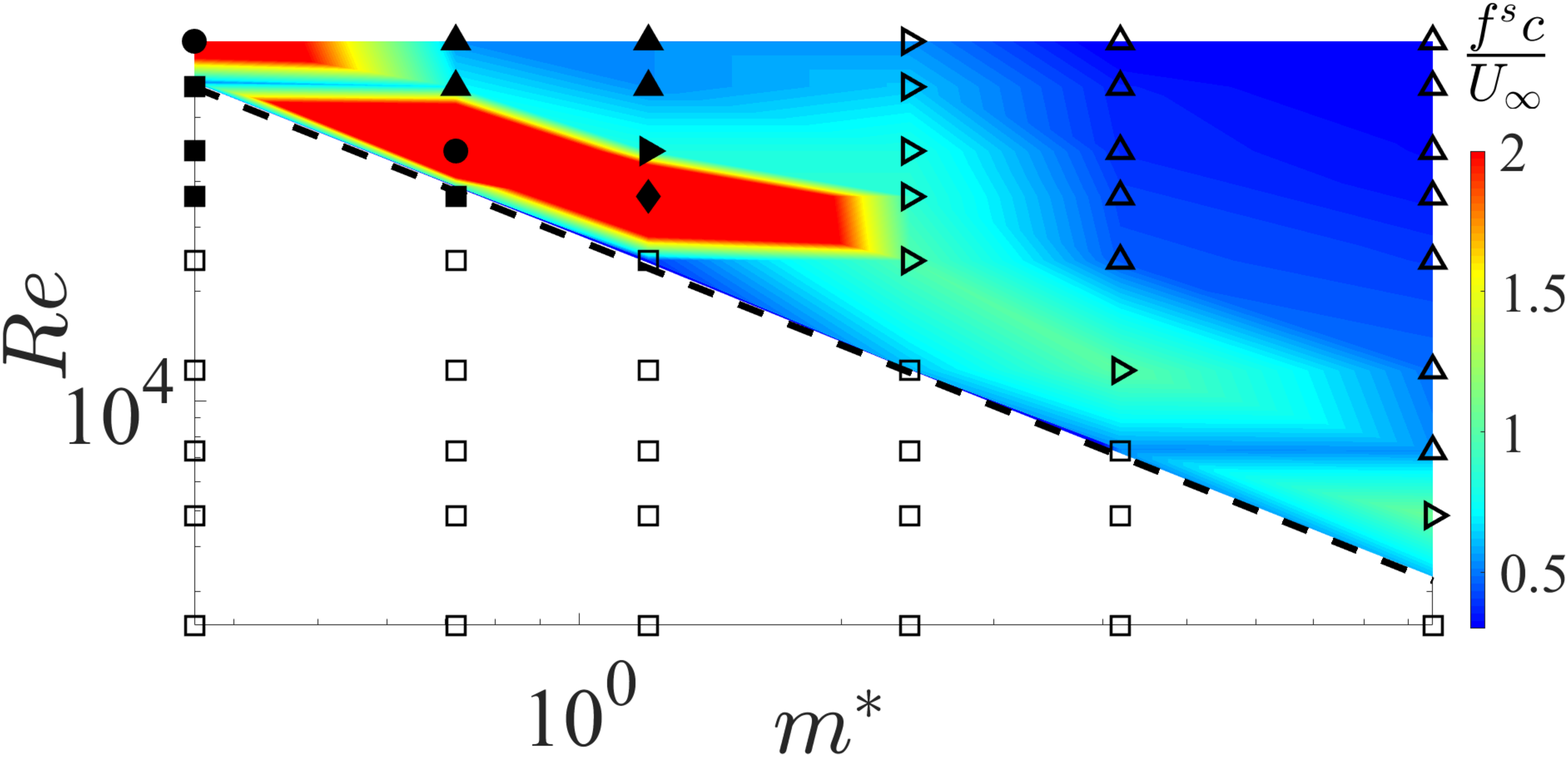}\label{phasef}}
	\caption{Coupled fluid-membrane dynamic characteristics of the 3D flexible membrane at $Ae$=423.14 for $\alpha$=$15^\circ$ over the full $m^*$-$Re$ parameter space: (a) time-averaged lift coefficient, (b) time-averaged lift-to-drag ratio, (c) maximum time-averaged non-dimensional membrane deflection, (d) maximum r.m.s. membrane deflection fluctuation, (e) non-dimensional dominant frequency in fluid domain and (f) non-dimensional dominant frequency in structural domain via FMD approach. The dashed line ($\textcolor[RGB]{1,1,1}{---}$) is the flow-excited instability boundary given by the empirical solution \refeq{recr}. $\square$ represents the membrane with a deformed-steady state. $\triangle$, $\triangleright$, $\Diamond$ and $\circ$ denote the dominant modes with the C1S2, C2S1, C2S5 and C2S9 mode shapes, respectively. The blank in (f) represents the deformed-steady membrane without vibration.}
	\label{phase}
\end{figure}

The coupled fluid-membrane dynamic characteristics of the flexible membrane over the given parameter space are summarized in \reffig{phase}. The contours are coloured by the time-averaged lift coefficient, the time-averaged lift-to-drag ratio, the normalized maximum time-averaged deflection, the maximum root-mean-squared value (r.m.s.) of the deflection fluctuation, the non-dimensional dominant frequency of the fluid domain and the structural domain via the FMD approach, respectively. Both the $X$ and $Y$ axes are displayed on a logarithmic scale. The dashed line is given by \refeq{recr} to distinguish the instability boundary between the DSS and DBS regimes.

In the DSS regime, it can be observed from \reffig{phase} \subref{phasea} that the mean lift coefficient maintains similar values. We notice from \reffigs{phase} \subref{phaseb} and \subref{phasec} that the mean lift-to-drag ratio and the mean membrane deflection is almost independent of $m^*$ at a fixed $Re$. As $Re$ increases, the aerodynamic efficiency improves substantially. The reason is attributed to the drag reduction caused by the cambered up membrane shape. A boundary is notice in \reffig{phase} \subref{phased} to separate the DSS and DBS regimes. Since the deformed-steady membrane shape almost keeps constant as $m^*$ changes, the dominant frequency of the unsteady flow presented in \reffig{phase} \subref{phasee} does not dependent on $m^*$. The dominant frequency of the structural vibration plotted in \reffig{phase} \subref{phasef} is set to blank within the DSS regime due to the steady state.

\begin{figure}
	\centering 
	\subfloat[][]{\includegraphics[width=0.47\textwidth]{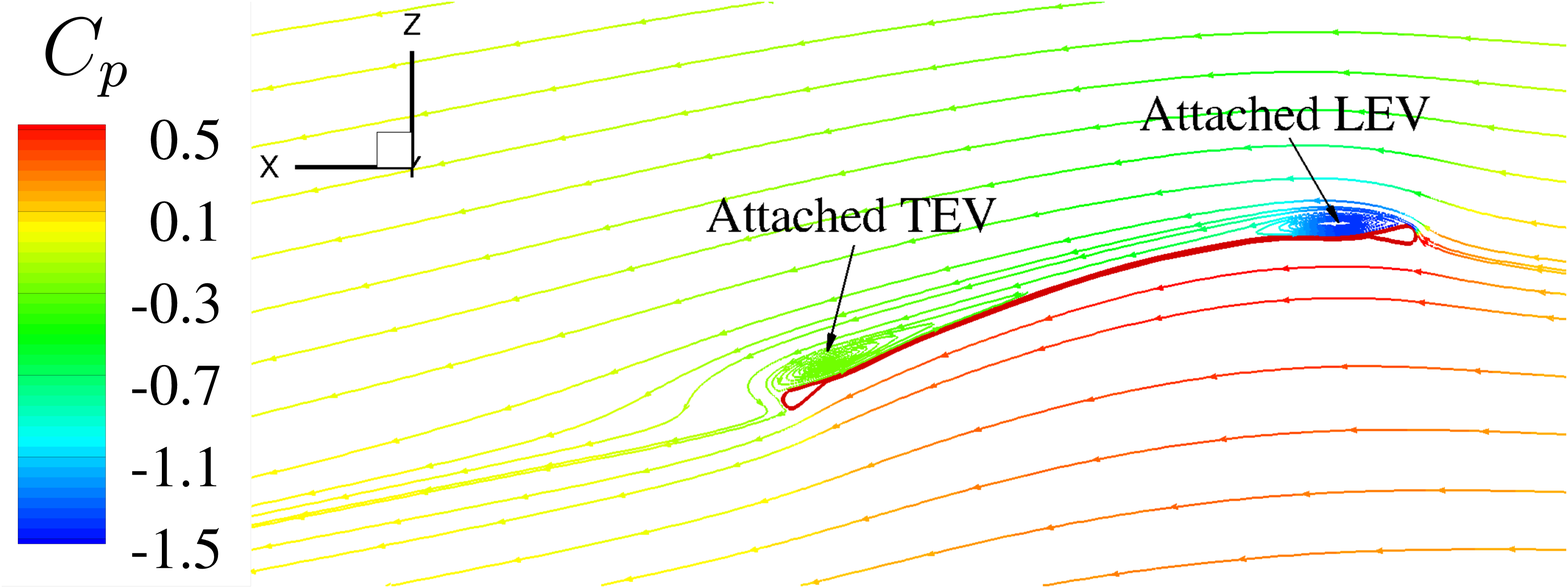}\label{attach_sheddinga}}
	\quad
	\subfloat[][]{\includegraphics[width=0.4\textwidth]{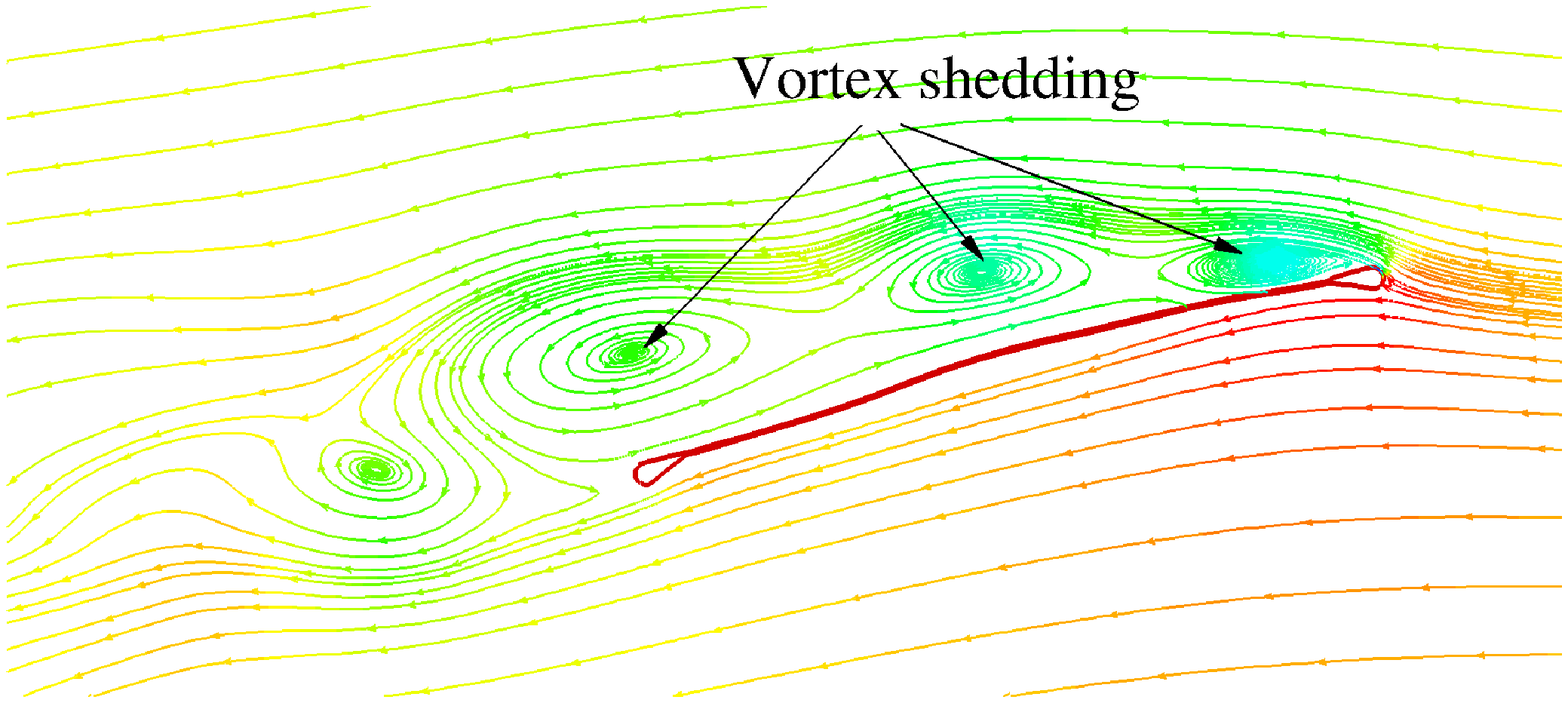}\label{attach_sheddingb}}
	\\
	\subfloat[][]{\includegraphics[width=0.47\textwidth]{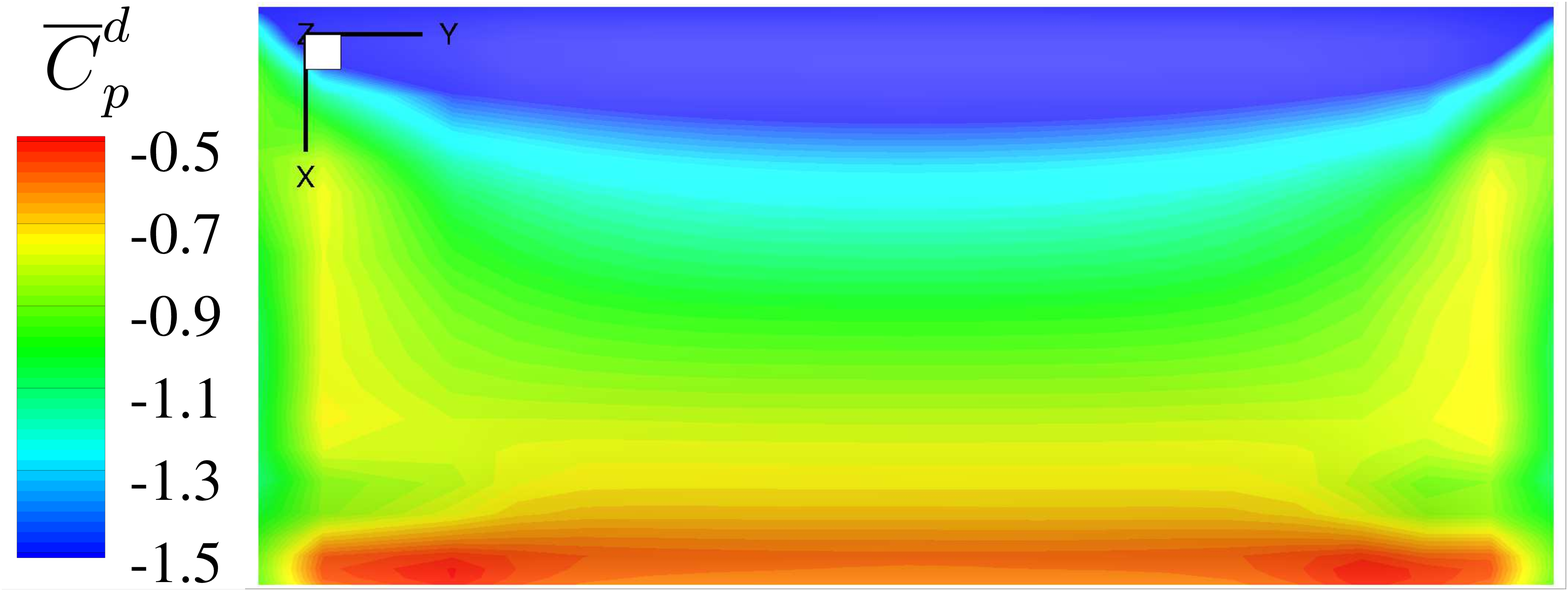}\label{attach_sheddingc}}
	\quad
	\subfloat[][]{\includegraphics[width=0.4\textwidth]{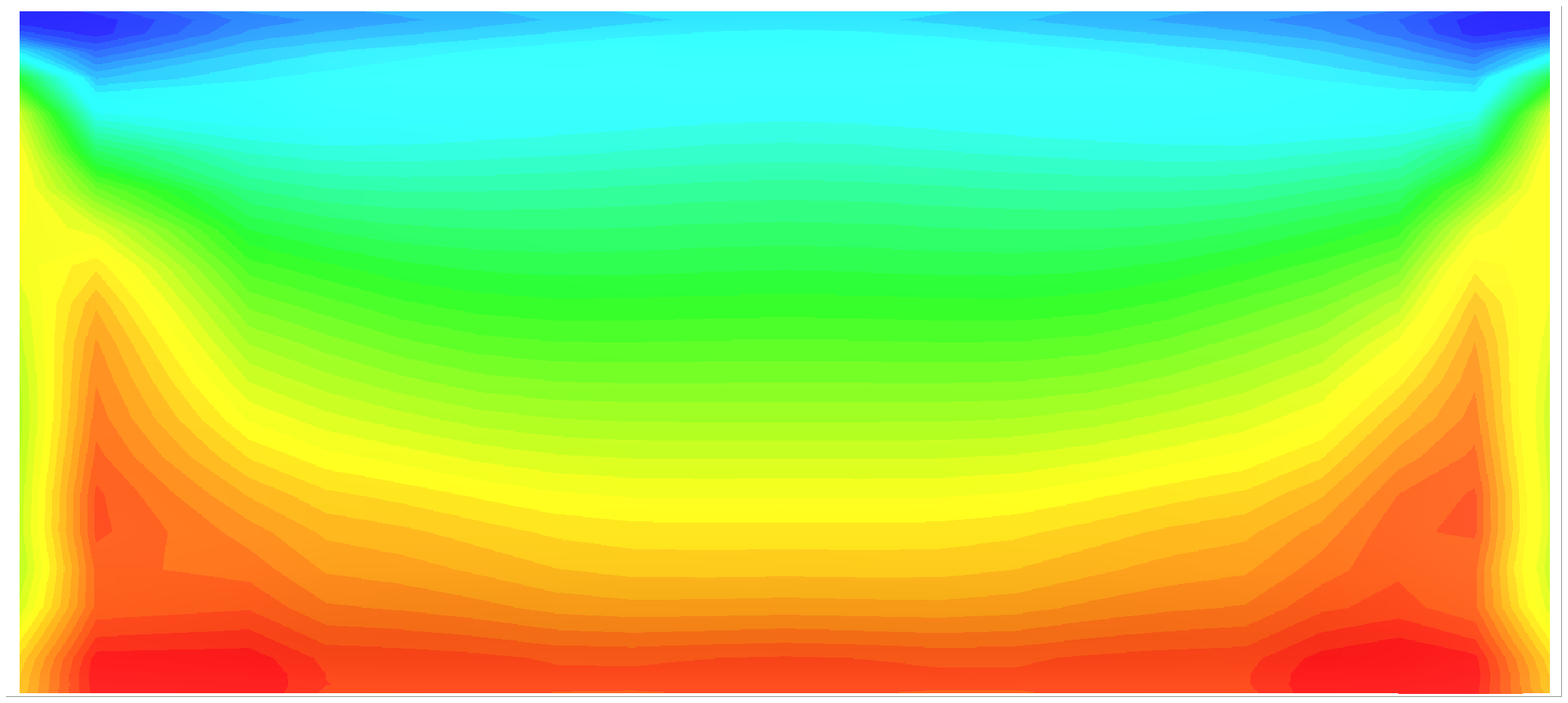}\label{attach_sheddingd}}
	\caption{Comparison of flow features of the flexible membrane with (a,c) attached vortices (M3,R6) and (b,d) vortex shedding (M5,R4) for (a,b) instantaneous streamlines coloured by pressure coefficient and (c,d) time-averaged pressure coefficient difference $\overline{C}_p^d$ between upper and lower surfaces.}
	\label{attach_shedding}
\end{figure}

Different from the deformed-steady membrane, the aerodynamic characteristics of the oscillating membrane are significantly affected by the interaction of the unsteady separated flow and the membrane vibration. The mean lift coefficient is further improved when the flexible membrane maintains the dynamic balance state. The optimal aerodynamic performance is observed in the parameter space of (M1$\to$3, R6$\to$9). Meanwhile, we notice the largest membrane deflections and mild membrane vibrations in this region. Through the comparison of the flow features of the flexible membrane for two selected cases (M3,R6) and (M5,R4), the attached vortices at the leading and trailing edges in \reffig{attach_shedding} \subref{attach_sheddinga} enlarge the suction area on the membrane surface shown in \reffig{attach_shedding} \subref{attach_sheddingc}. When the light membrane deforms up to some extent at a relatively high $Re$ without high-intensity oscillation, the large separated flow region is suppressed and the attached vortices are formed on the membrane surface. It can be seen from \reffig{attach_shedding} \subref{attach_sheddingb} that the vortices cannot keep attached to the membrane surface. These vortices are coupled with the vibrating membrane and shed into the wake alternatively. As a result, the excited separated flow and the vortex shedding process degrade the aerodynamic performance as compared to the coupled system with attached vortices. A similar distribution of the dominant frequency of the unsteady flow and the membrane vibration is observed in \reffigs{phase} \subref{phasee} and \subref{phasef}, resulting in the frequency lock-in phenomenon.

\subsection{Effect of mass ratio} \label{mass}
\begin{figure}
	\centering 
	\subfloat[][]{\includegraphics[width=0.48\textwidth]{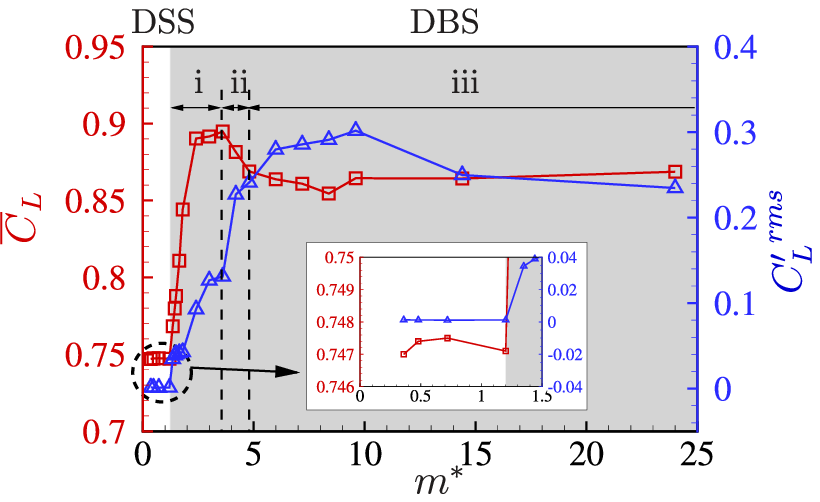}\label{clcdmassa}}
	\quad
	\subfloat[][]{\includegraphics[width=0.48\textwidth]{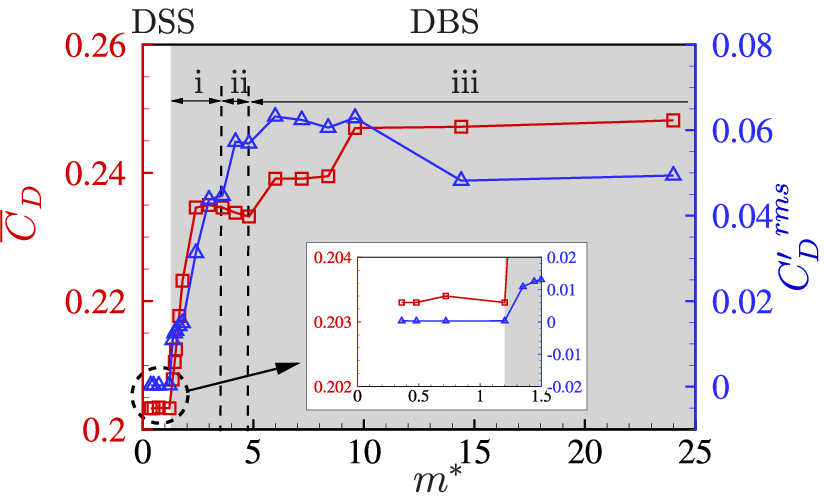}\label{clcdmassb}}
	\\
	\subfloat[][]{\includegraphics[width=0.45\textwidth]{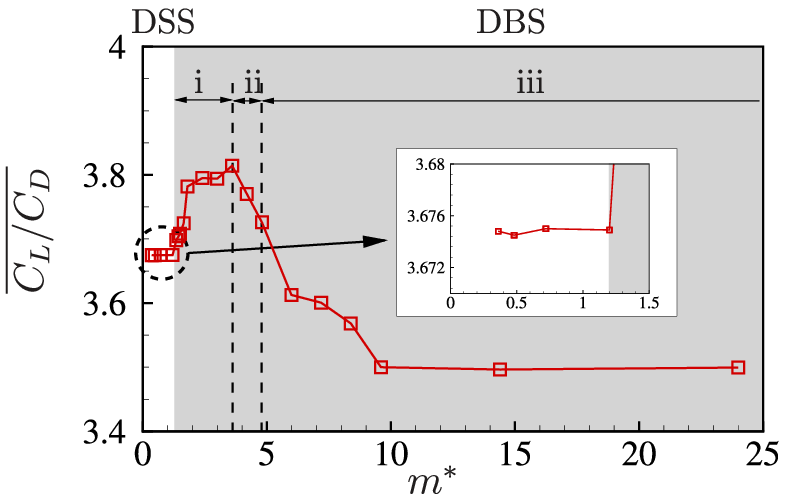}\label{clcdmassc}}
	\quad
	\subfloat[][]{\includegraphics[width=0.45\textwidth]{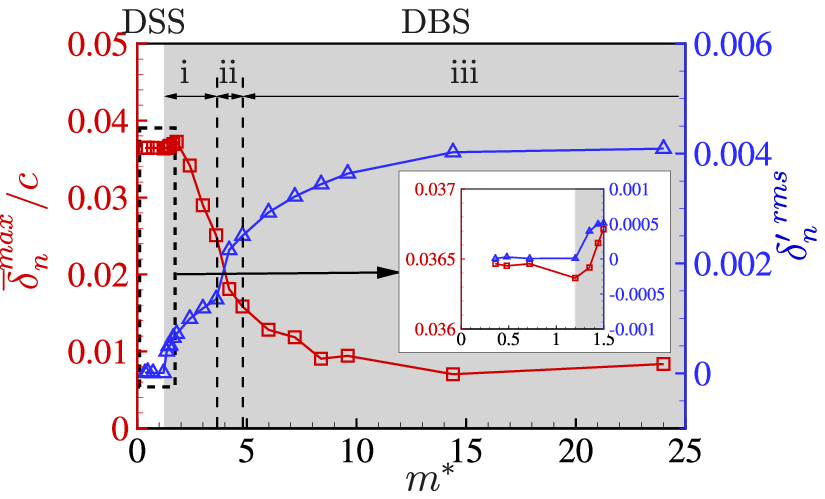}\label{clcdmassd}}
	\caption{Evolution of 3D membrane dynamics as a function of $m^*$: (a) time-averaged lift coefficient and r.m.s. lift coefficient fluctuation, (b) time-averaged drag coefficient and r.m.s. drag coefficient fluctuation, (c) time-averaged lift-to-drag ratio and (d) maximum time-averaged normalized membrane deflection and r.m.s. membrane deflection fluctuation.}
	\label{clcdmass}
\end{figure}

We observe from \reffig{mode_map} that the membrane exhibits various dynamic behaviours over the studied parameter space. It is found that the mass ratio plays an important role in the flow-induced vibration. To investigate the effect of mass ratio on the coupled fluid-membrane characteristics and to isolate the impact of Reynolds number, we present and compare the evolution of the membrane dynamics as a function of mass ratio at a fixed Reynolds number of 24300 (R5). This representative Reynolds number selected for this study covers all the typical dynamic states. To fully understand the flow-induced vibration and the mode transition phenomena, extra cases were run for the membrane dynamics near the flow-excited instability boundary and the mode transition region in the range of $m^* \in [0.36,24]$. \refFig{clcdmass} summarizes the evolution of the force characteristics, the membrane displacements and the membrane vibration intensities as a function of $m^*$. 

It can be seen from \reffig{clcdmass} that the aerodynamic force characteristics and the membrane deflections almost keep constant within the range of $m^* \in [0.36, 1.2]$, leading to no fluctuation in the membrane responses. Similar conclusions can be drawn for the membrane dynamics of the coupled system within the DSS regime at different Reynolds numbers in \reffig{phase}. We can infer that the inertia effect is neglected during the fluid-membrane interaction when $m^*$ is less than the critical value. Consequently, a static equilibrium state of the cambered-up membrane is achieved when the tension force is balanced with the aerodynamic loads acting on the membrane surface.

When the membrane vibration occurs, the flexible membrane enters a dynamic balance state. The inertia effect participates in the dynamics of the coupled system in this regime. With the aid of the global FMD method, three distinctive vibrational modes are further identified from the membrane dynamic response, namely (i) chord-wise first mode, (ii) transitional mode, and (iii) chord-wise second mode. The mean lift coefficient and the mean lift-to-drag ratio increase dramatically from the constant values at the deformed-steady state to the overall peak values when the membrane vibrates in the chord-wise second mode. As the dominant structural mode gradually transitions from the chord-wise second mode to the chord-wise first mode, the aerodynamic performance begins to degrade. A kink is noticed at the boundary between two types of vibrational modes. The mean lift coefficient and the mean lift-to-drag ratio keep decreasing until approximately constant values are reached at $m^*$=9.6 in the dominant first mode regime. The drag coefficient jumps to a plain when $m^*$ approaches the transitional region and more drag penalties are achieved as $m^*$ further increases. Finally, no obvious increment of the drag coefficient is noticed for a relatively heavy membrane. The mean camber of the membrane decreases continuously within the range of $m^* \in (1.2, 24]$. The root-mean-squared values of the aerodynamic forces and the membrane displacement fluctuations show an overall growing trend in the simulated parameter space, which indicates higher oscillation intensities for a heavier membrane.

\begin{figure}
	\centering 
	\subfloat[][]{\includegraphics[width=0.49\textwidth]{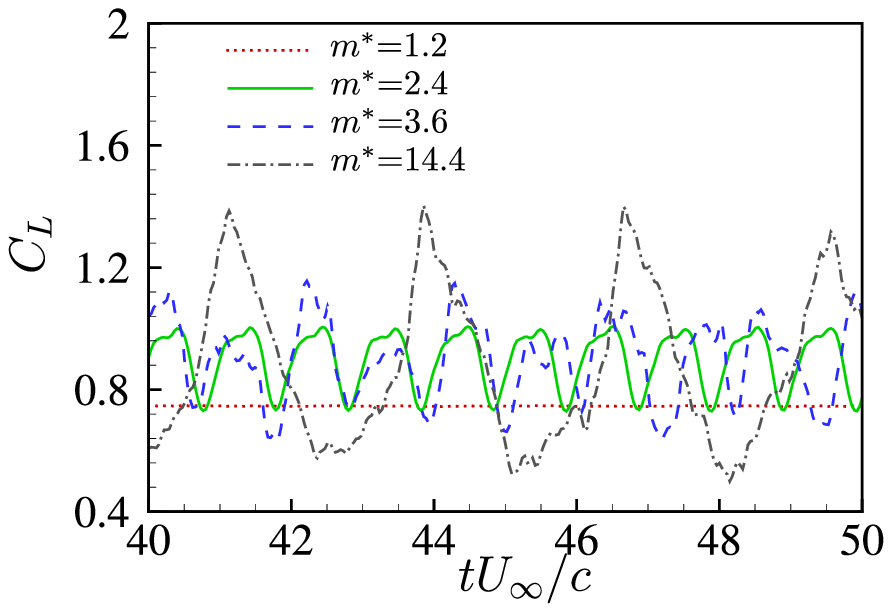}\label{response_massa}}
	\subfloat[][]{\includegraphics[width=0.49\textwidth]{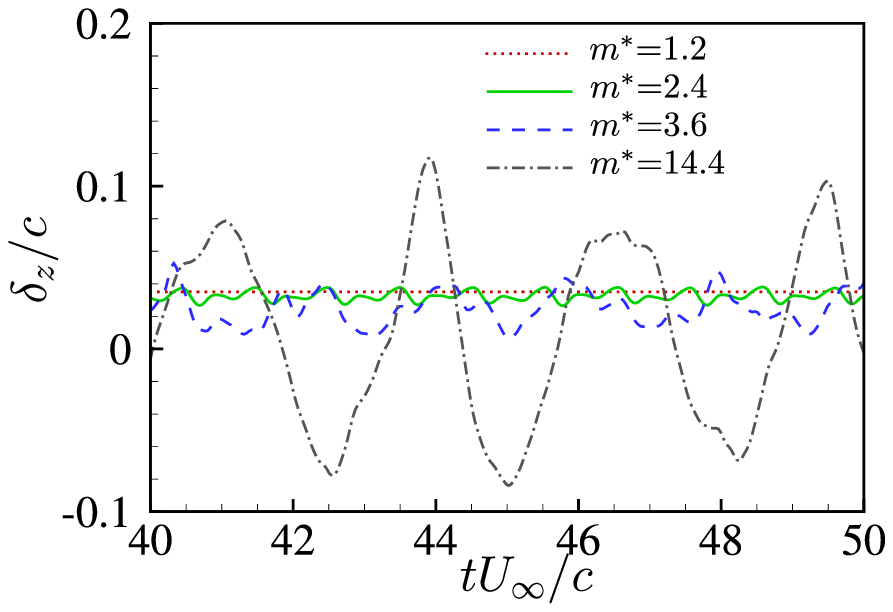}\label{response_massb}}
	\caption{Time histories of instantaneous: (a) lift coefficients and (b) normalized displacements at the membrane centre along the $Z$-direction for four selected cases with $m^*$=1.2, 2.4, 3.6 and 14.4 at a fixed $Re$=24300.}
	\label{response_mass}
\end{figure}

To further investigate the effect of mass ratio, we compare the time histories of the instantaneous lift coefficients and the instantaneous displacement at the membrane centre along the $Z$ direction in \reffig{response_mass}. We choose one case at $m^*$=1.2 corresponding to the DSS state and three cases at $m^*$=2.4, 3.6 and 14.4 for the second mode, the transitional mode and the first mode in the DBS regime, respectively. It can be seen that the time-varying lift coefficient and the time-varying membrane displacement show steady responses at $m^*$=1.2. As $m^*$ further increases, both the time-varying lift coefficient and displacement exhibit growing oscillation amplitude and decreasing oscillation frequency. Meanwhile, the dynamic responses gradually change from periodic oscillation to non-periodic oscillation.

\begin{figure}
	\centering 
	\subfloat[][]{
		\includegraphics[width=1.0\textwidth]{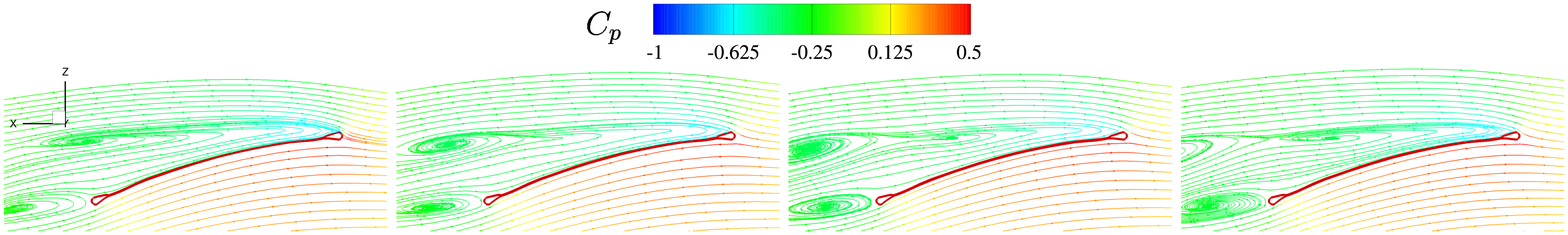}
		\label{membrane_state_streama}
	}
	\\
	\subfloat[][]{
		\includegraphics[width=1.0\textwidth]{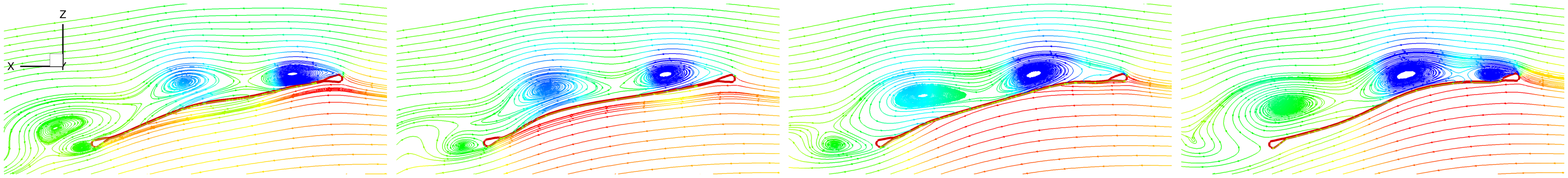}
		\label{membrane_state_streamb}
	}
	\\
	\subfloat[][]{
		\includegraphics[width=1.0\textwidth]{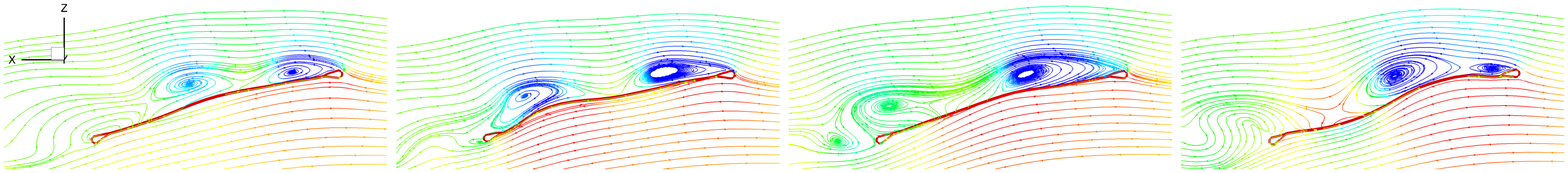}
		\label{membrane_state_streamc}
	}
	\\
	\subfloat[][]{
		\includegraphics[width=1.0\textwidth]{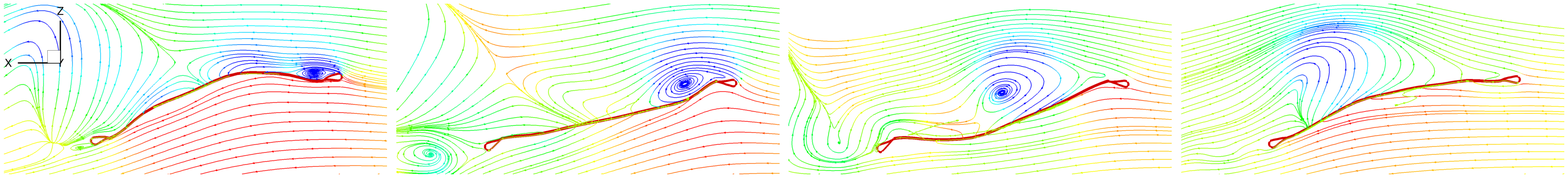}
		\label{membrane_state_streamd}
	}
	\caption{Instantaneous streamlines of 3D flexible membrane wing on the mid-span plane coloured by pressure coefficient for $(Re,m^*)=$ (a) $(24300,1.2)$, (b) $(24300,2.4)$, (c) $(24300,3.6)$, (d) $(24300,14.4)$.}
	\label{membrane_state_stream}
\end{figure}

We further examine the instantaneous flow features around the flexible membrane on the mid-span plane for the four selected cases in \reffig{membrane_state_stream}. For $m^*$=1.2 in \reffig{membrane_state_stream} \subref{membrane_state_streama}, a large region of separated flows is observed on almost the whole upper surface of the membrane. The large-scale vortex is assembled with the trailing edge vortex to form a vortex pair behind the membrane, which is similar to a typical von Karman vortex shedding process behind a bluff body. As the inertia effect becomes stronger, the flexible membrane vibrates at one of the specific modes within the DBS regime. The generated vortices get closer to the membrane surface, and shed into the wake by coupling with the membrane vibration at $m^*$=2.4 shown in \reffig{membrane_state_stream} \subref{membrane_state_streamb}. It is observed from \reffig{membrane_state_stream} \subref{membrane_state_streamc} that the vortex shedding pattern changes to the low frequency response and the first mode occasionally occurs in the transitional regime at $m^*$=3.6. The first vibrational mode dominates the membrane vibration at $m^*$=14.4 as plotted in \reffig{membrane_state_stream} \subref{membrane_state_streamd}. In this mode, only one apparent large-scale vortex detaches from the surface and sheds into the wake in a completed oscillation period, which is synchronized with the first membrane vibrational mode.

\begin{figure}
	\centering 
	\subfloat[][]{
		\includegraphics[width=0.25\textwidth]{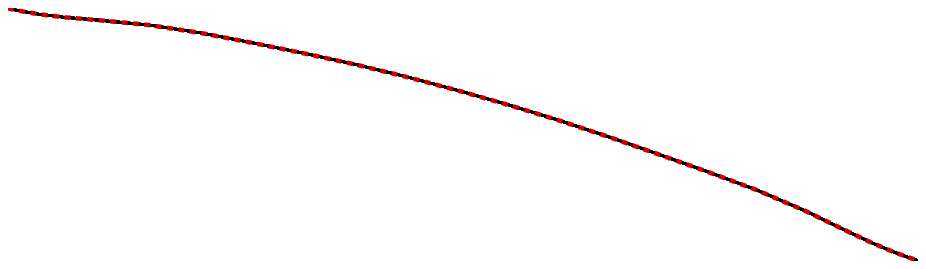}\label{membrane_vibrationa}}
	\subfloat[][]{
		\includegraphics[width=0.25\textwidth]{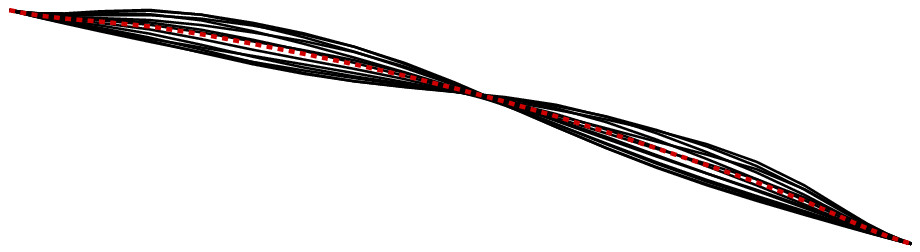}\label{membrane_vibrationb}}
	\subfloat[][]{
		\includegraphics[width=0.25\textwidth]{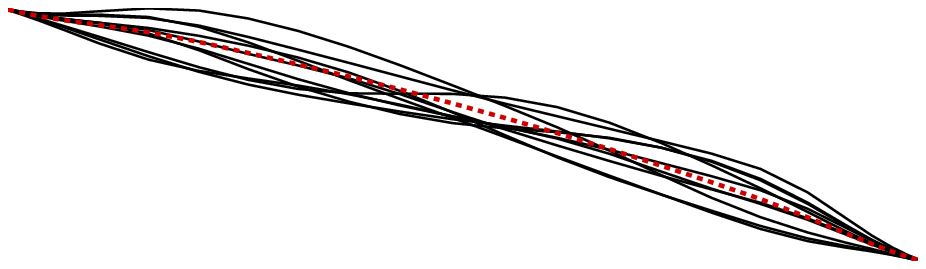}\label{membrane_vibrationc}}
	\subfloat[][]{
		\includegraphics[width=0.25\textwidth]{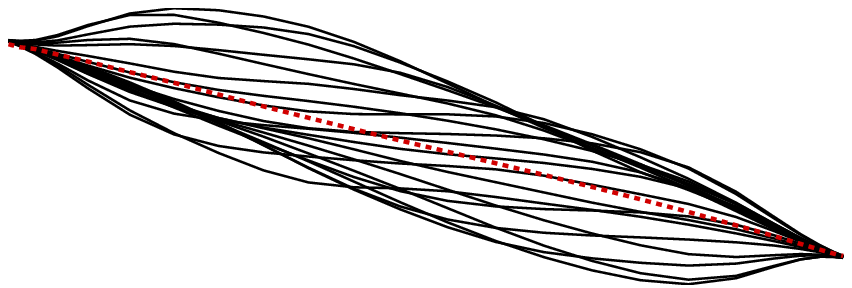}\label{membrane_vibrationd}}
	\\
	\centering 
	\subfloat[][]{
		\includegraphics[height=0.12\textwidth,width=0.27\textwidth]{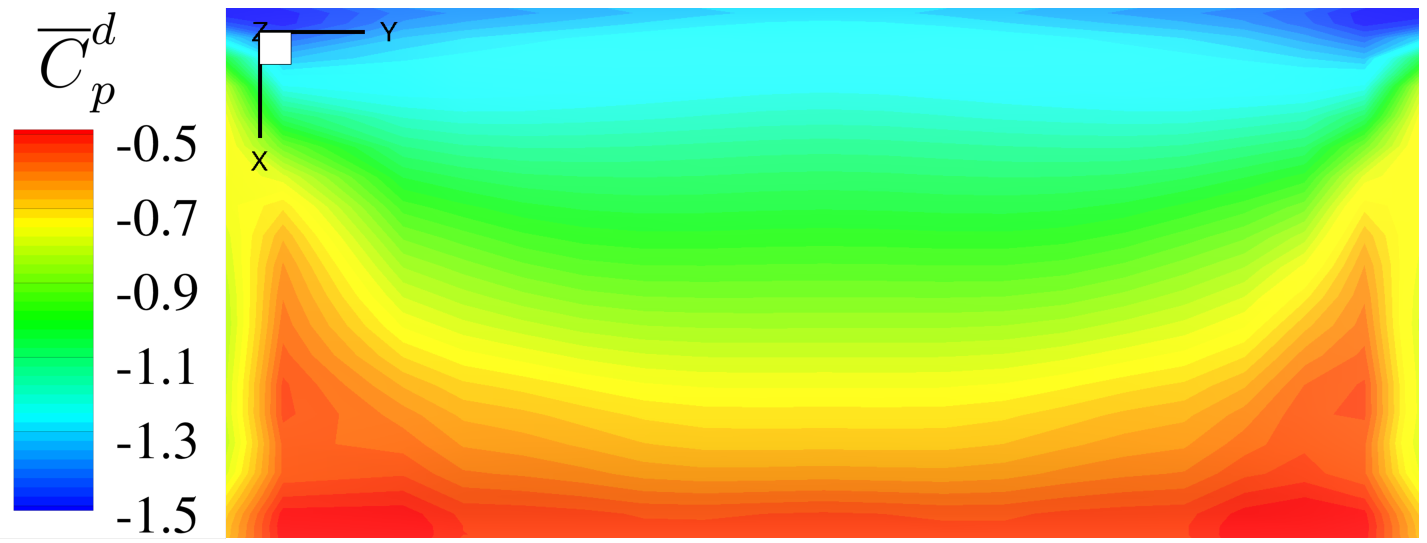}\label{cpda}}
	\
	\subfloat[][]{
		\includegraphics[height=0.12\textwidth,width=0.22\textwidth]{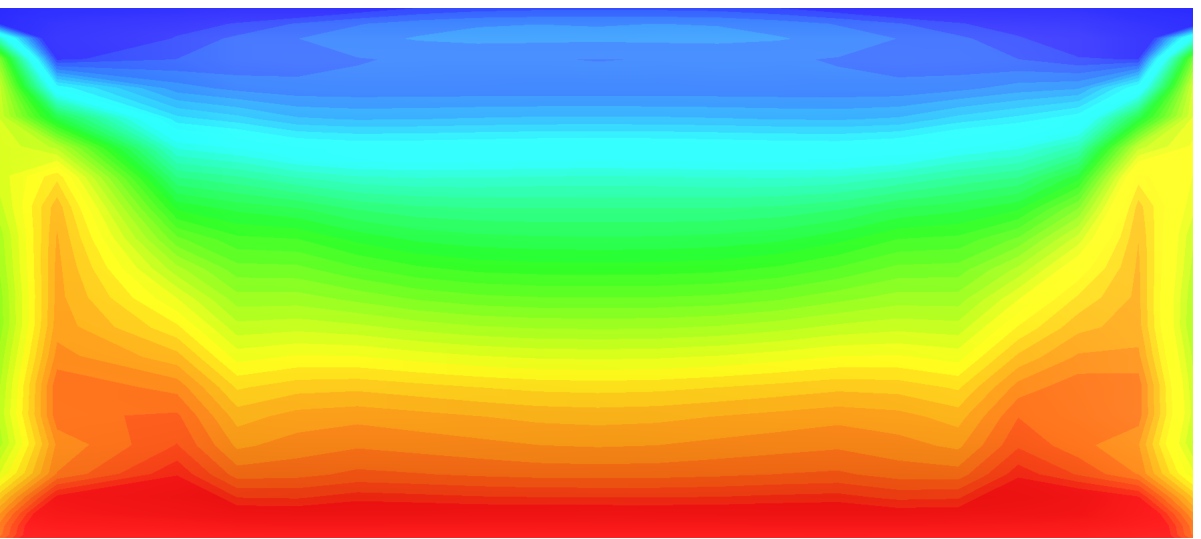}\label{cpdb}}
	\
	\subfloat[][]{
		\includegraphics[height=0.12\textwidth,width=0.22\textwidth]{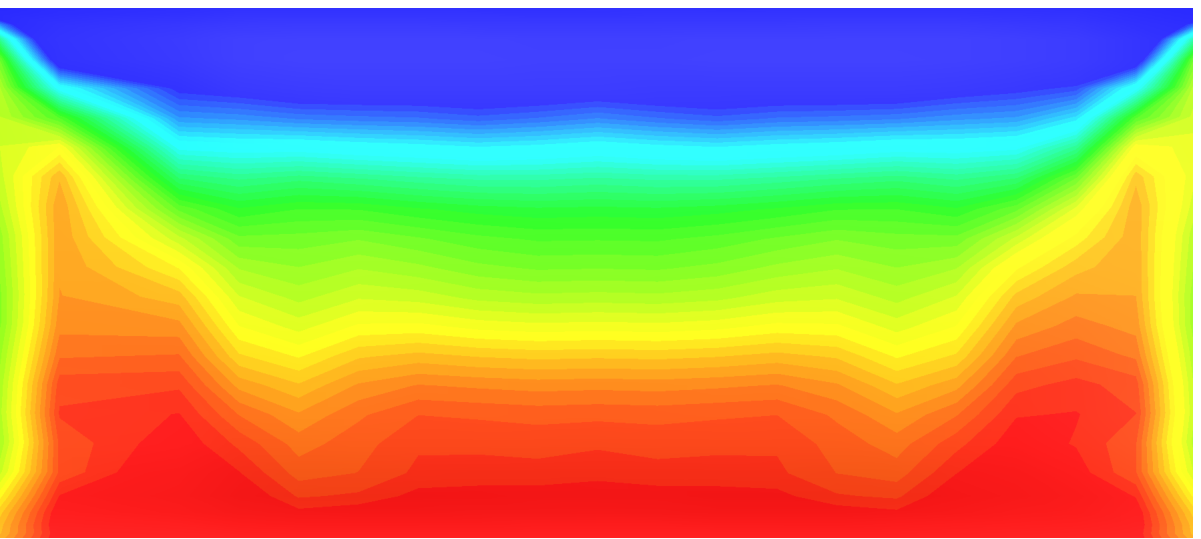}\label{cpdc}}
	\
	\subfloat[][]{
		\includegraphics[height=0.12\textwidth,width=0.22\textwidth]{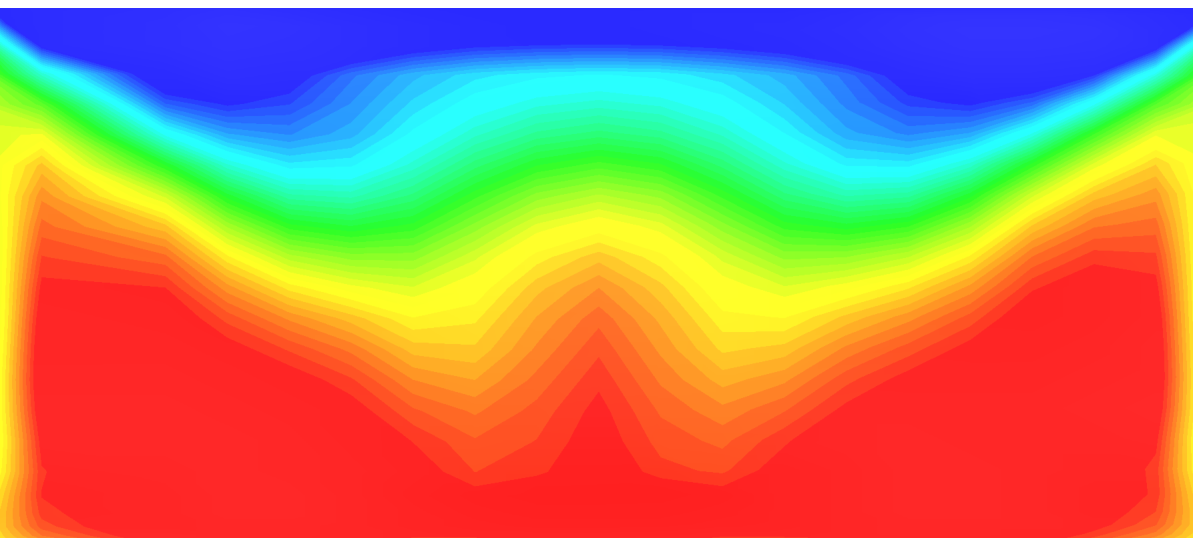}\label{cpdd}}
	\\
	\subfloat[][]{
		\includegraphics[height=0.17\textwidth,width=0.3\textwidth]{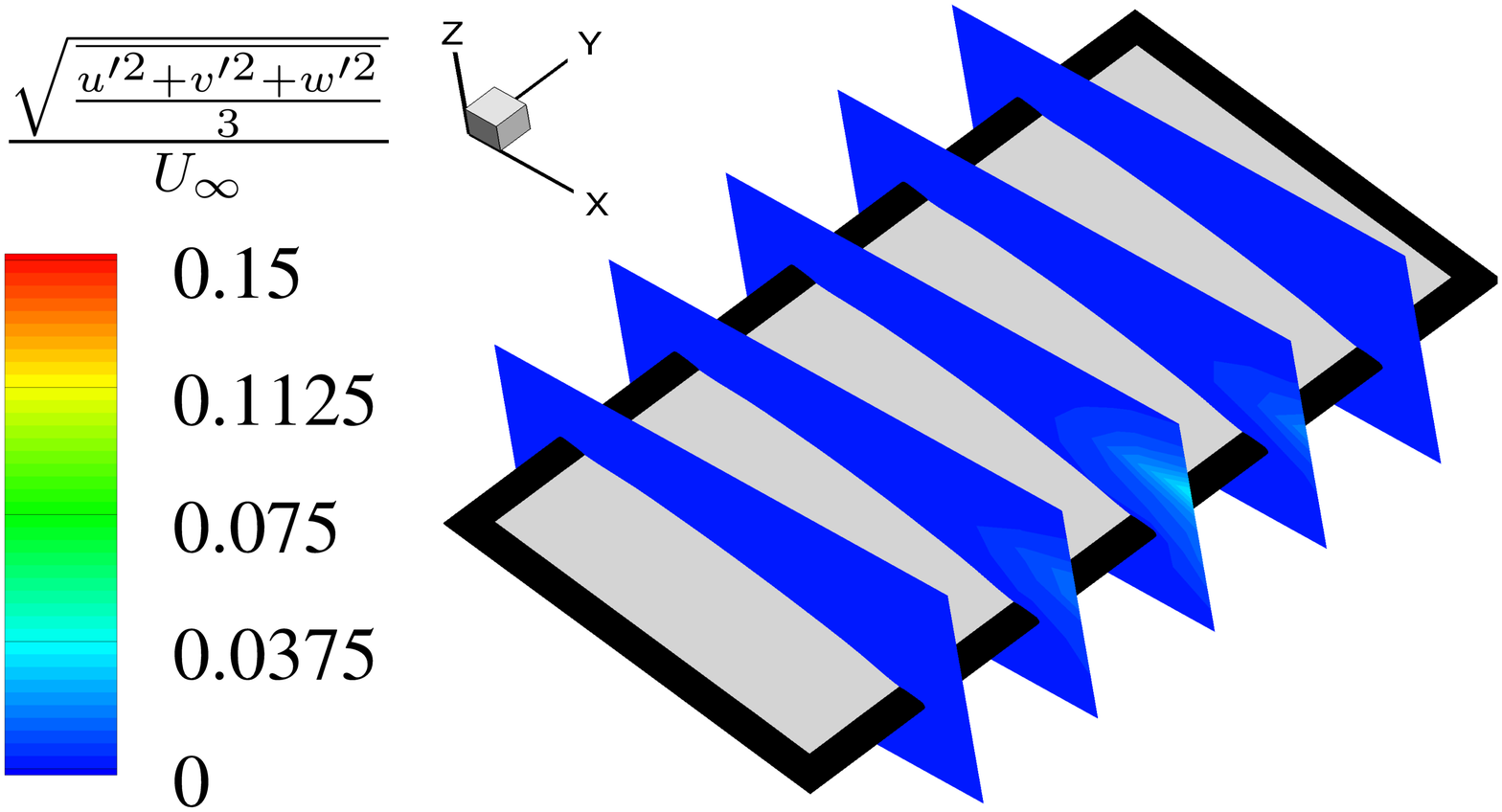}\label{tke_slicea}}
	\subfloat[][]{
		\includegraphics[height=0.17\textwidth,width=0.22\textwidth]{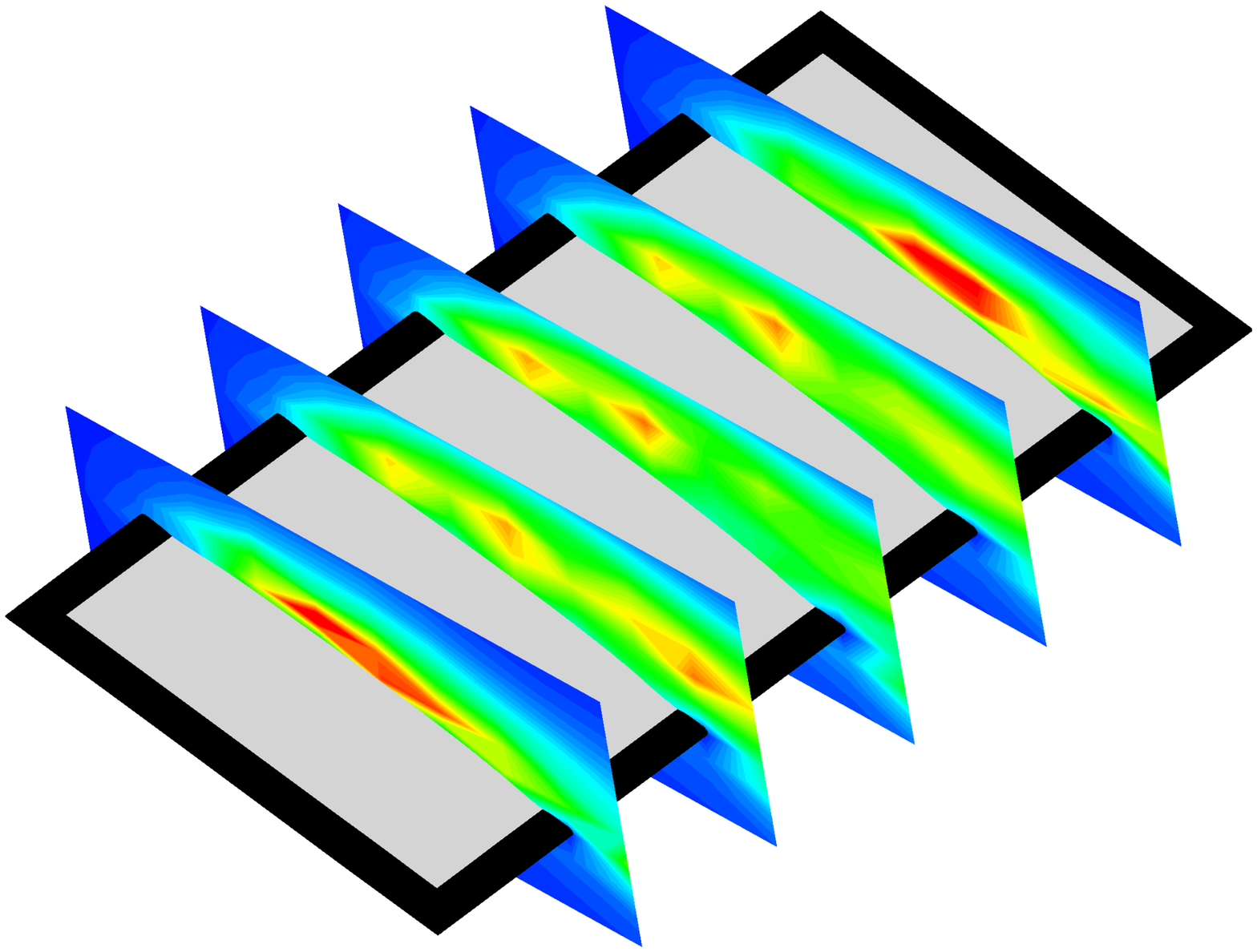}\label{tke_sliceb}}
	\subfloat[][]{
		\includegraphics[height=0.17\textwidth,width=0.22\textwidth]{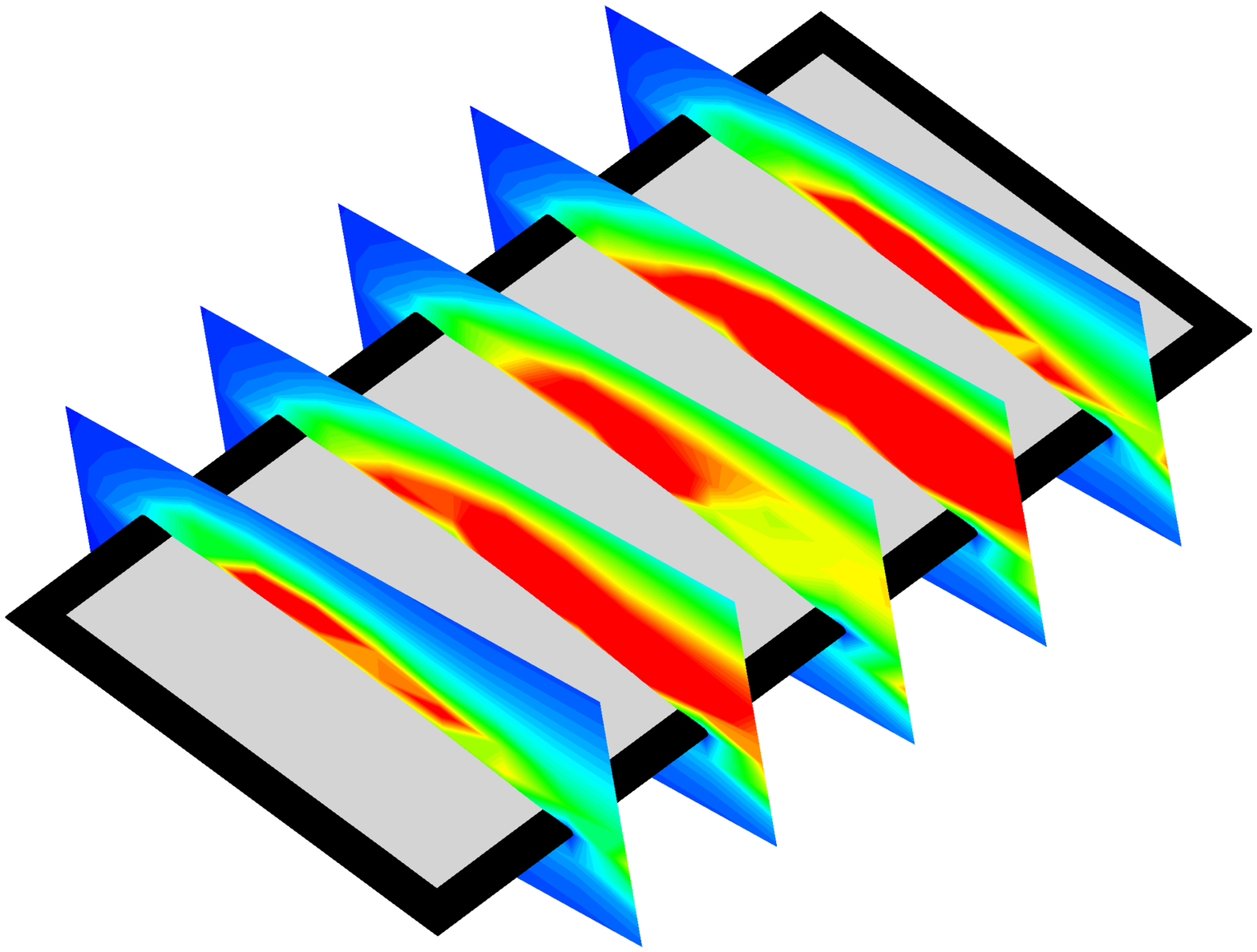}\label{tke_slicec}}
	\subfloat[][]{
		\includegraphics[height=0.17\textwidth,width=0.22\textwidth]{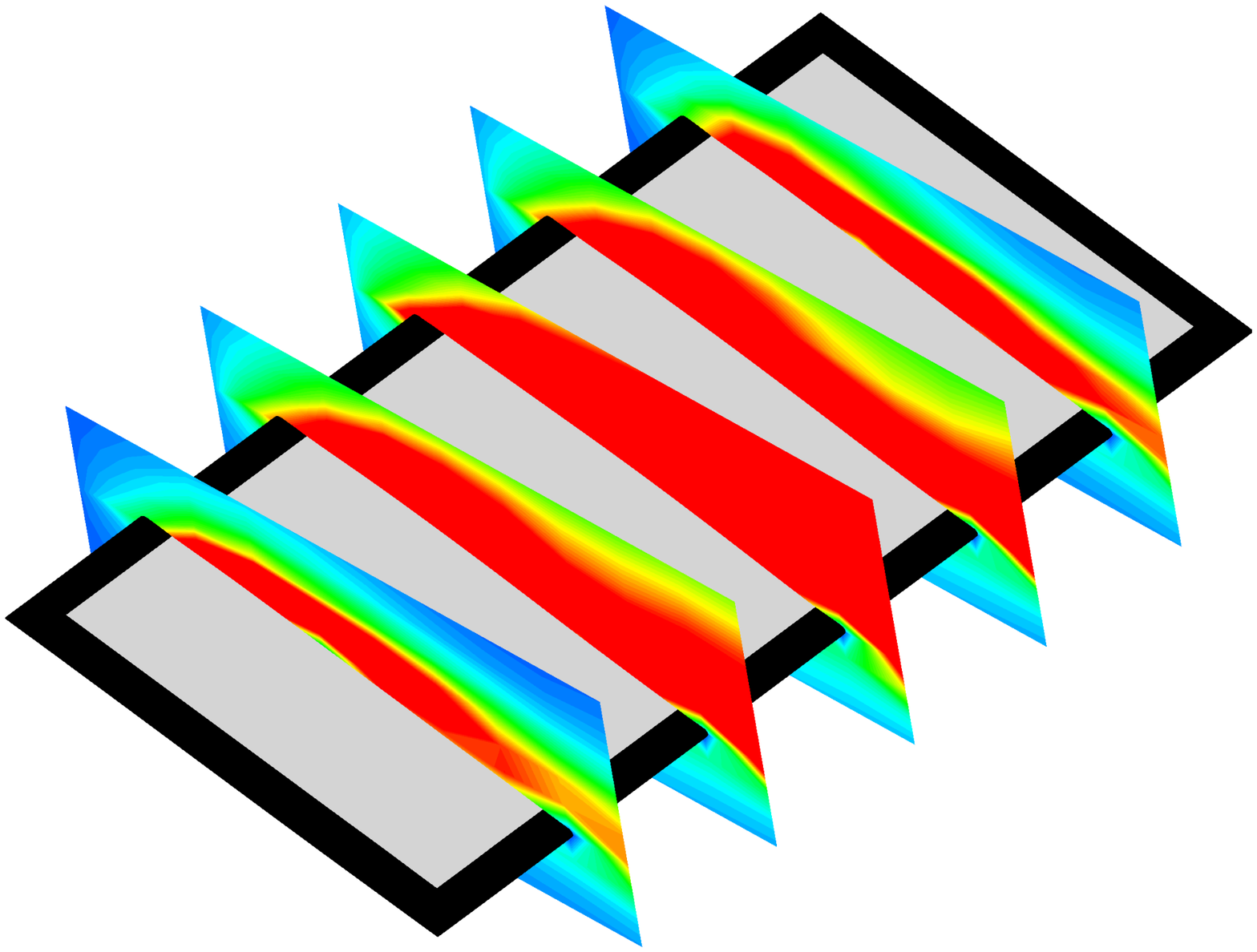}\label{tke_sliced}}
	\\
	\subfloat[][]{
		\includegraphics[height=0.17\textwidth,width=0.3\textwidth]{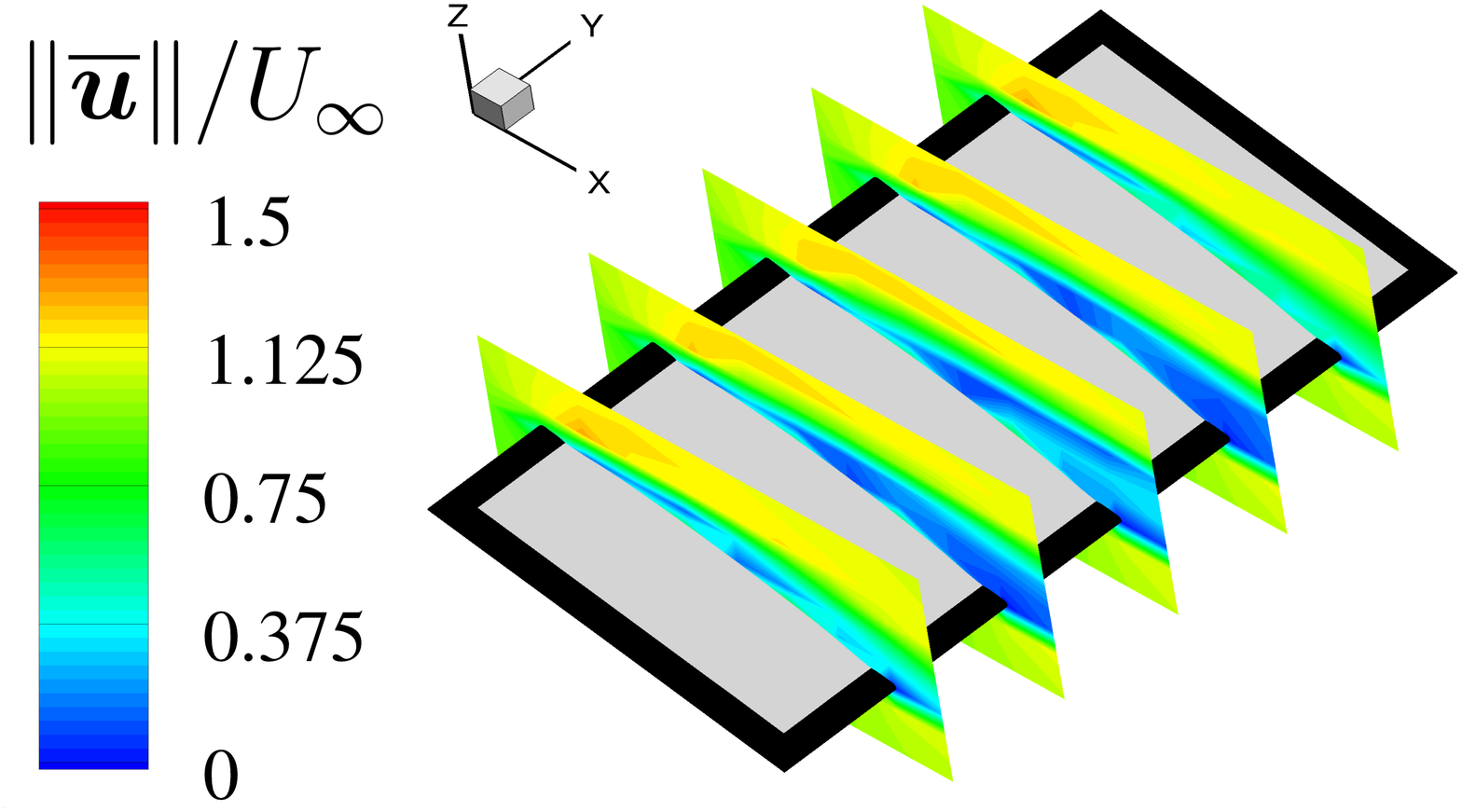}\label{vm_slicea}}
	\subfloat[][]{
		\includegraphics[height=0.17\textwidth,width=0.22\textwidth]{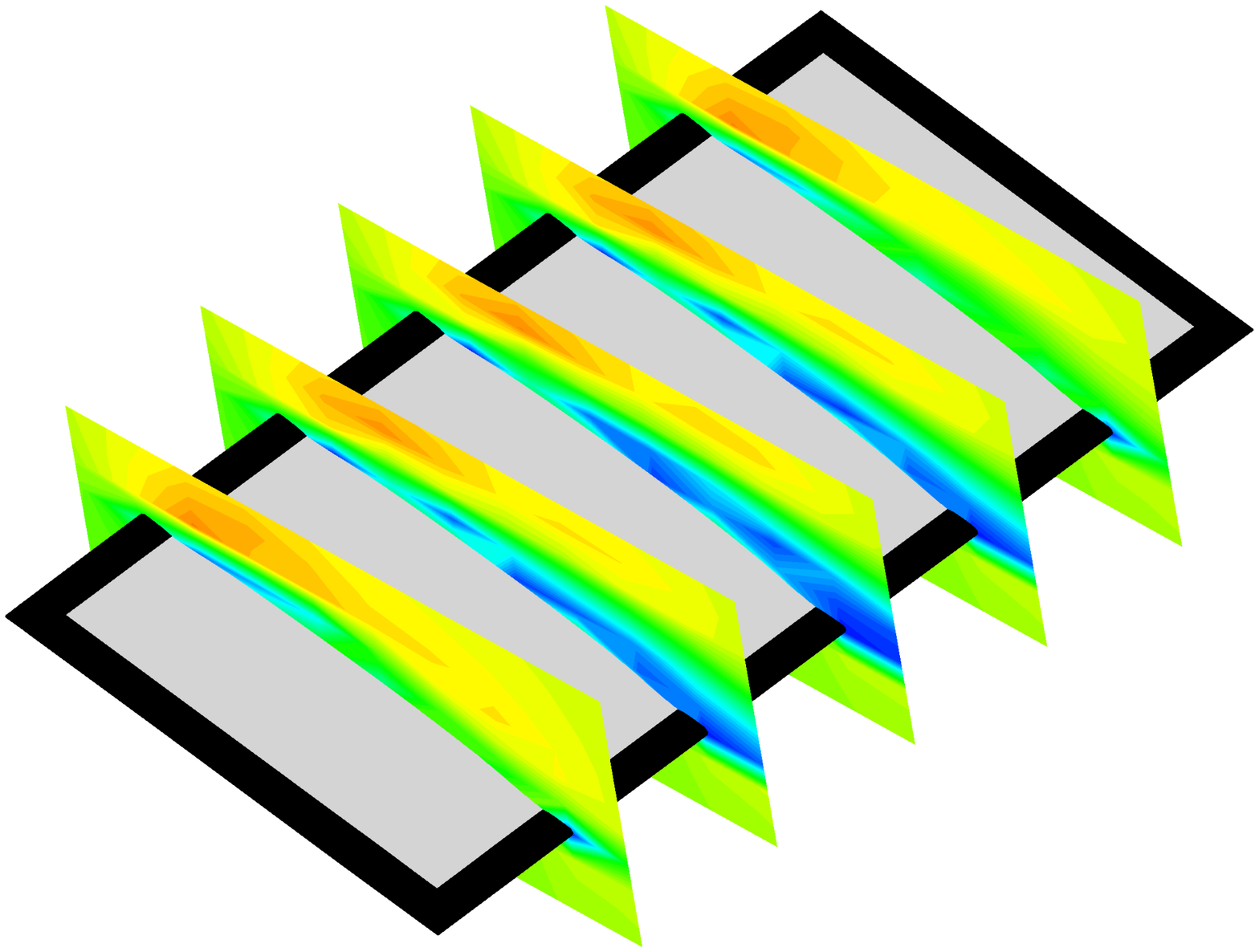}\label{vm_sliceb}}
	\subfloat[][]{
		\includegraphics[height=0.17\textwidth,width=0.22\textwidth]{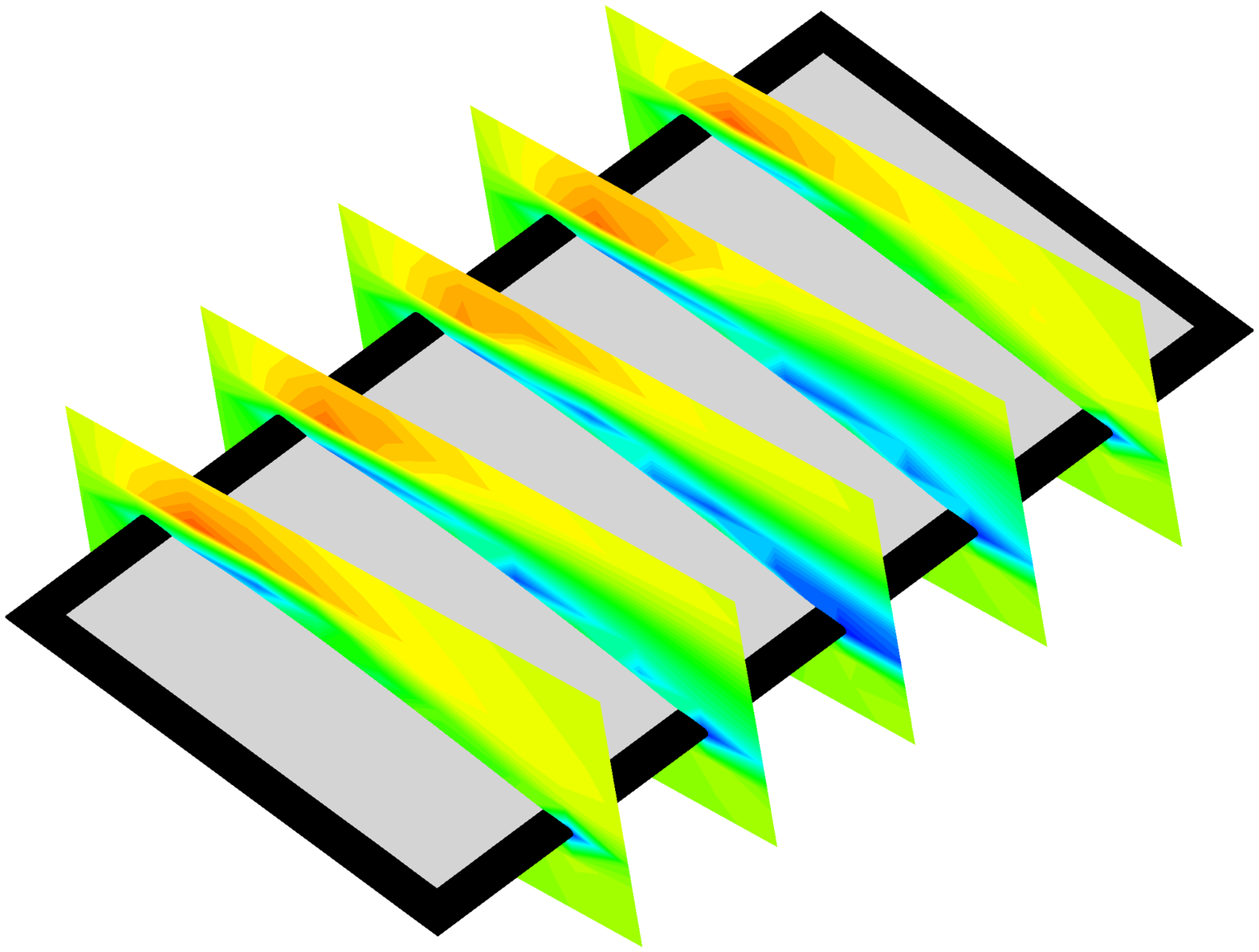}\label{vm_slicec}}
	\subfloat[][]{
		\includegraphics[height=0.17\textwidth,width=0.22\textwidth]{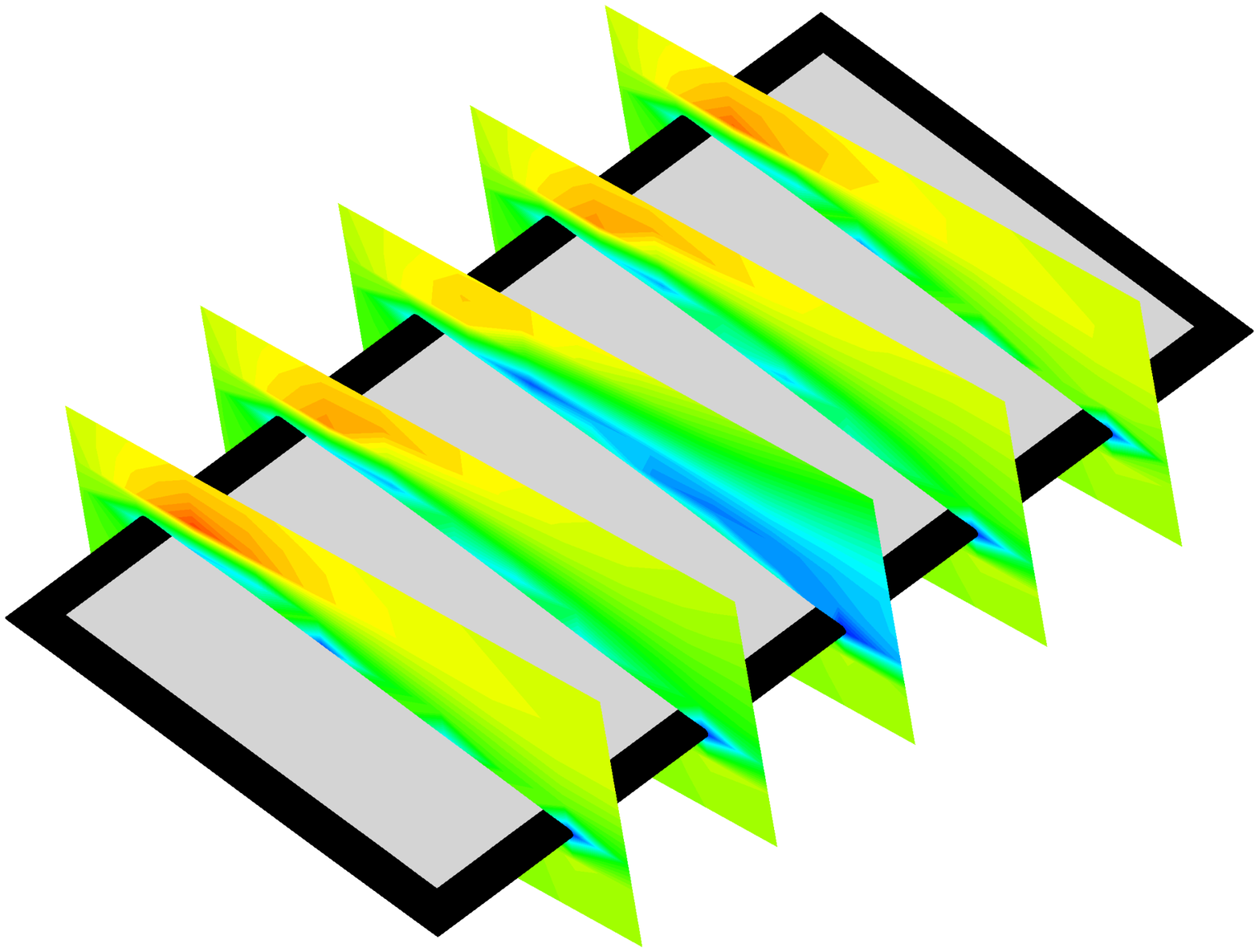}\label{vm_sliced}}
	\caption{Flow past a 3D rectangular membrane wing: (a,b,c,d) full-body profile responses of 3D flexible membrane wing at mid-span location, (e,f,g,h) time-averaged pressure coefficient difference between upper and lower surfaces, (i,j,k,l) turbulent intensity and (m,n,o,p) time-averaged normalized velocity magnitude on five slices along the span-wise direction at $(Re,m^*)=$ (a,e,i,m) $(24300,1.2)$, (b,f,j,n) $(24300,2.4)$, (c,g,k,o) $(24300,3.6)$, (d,h,l,p) $(24300,14.4)$. ($\color{red}{---}$) denotes the time-averaged membrane shape in (a,b,c,d).}
	\label{membrane_m_com}
\end{figure}

We further compare the membrane dynamics and the flow features in \reffig{membrane_m_com}. \refFigs{membrane_m_com} (a-d) show the full-body profile responses at the mid-span location for these four representative cases. The red dashed line represents the mean membrane shape. We observe that the membrane deforms up and remains a static steady wing shape at $m^*$=1.2. The flexible membrane exhibits a dominant chord-wise second mode at $m^*$=2.4. As $m^*$ further increases to 3.6, both the alternatively occurring chord-wise first and second modes are observed in the superimposed membrane responses. Meanwhile, the mean membrane camber becomes smaller compared to that at a lower $m^*$. The chord-wise first mode dominates the membrane vibration at $m^*$=14.4. Different from the membrane vibration responses in the second mode and transitional mode regimes, the flexible membrane vibrates on both sides of its rigid flat counterpart in the first mode regime. Related to the membrane dynamic state, the flow features of the coupled system are also strongly affected by $m^*$. As the flexible membrane leaves its static equilibrium position to vibrate at $m^*$=2.4, the suction force in the proximity of leading edge in the pressure coefficient difference contour becomes larger as shown in \reffig{membrane_m_com} \subref{cpdb}, compared to the suction area plotted in \reffig{membrane_m_com} \subref{cpda} at $m^*$=1.2. The extended suction area is related to the generated leading edge vortices shown in \reffig{membrane_state_stream} \subref{membrane_state_streamb}, which contributes to the lift improvement. As the dominant structural mode transitions to the chord-wise first mode, the region with the low negative pressure near the trailing edge expands to the leading edge, resulting in the reduction of the lift coefficient at a higher $m^*$. It can be observed from \reffigs{membrane_m_com} \subref{tke_slicea} and \subref{tke_sliceb} that the flow fluctuations in the shear layer become significant and the high-intensity regions get closer to the membrane surface when the flexible membrane loses its deformed-steady state. The turbulent kinetic energy (TKE) keeps growing as $m^*$ further increases. This increasing flow instability is highly coupled with the membrane vibrations with larger amplitudes. Furthermore, the low velocity area within the separation flow regions decreases for heavier membranes in \reffigs{membrane_m_com} (m-p).

\begin{figure}
	\centering 
	\subfloat[][]{\includegraphics[width=0.48\textwidth]{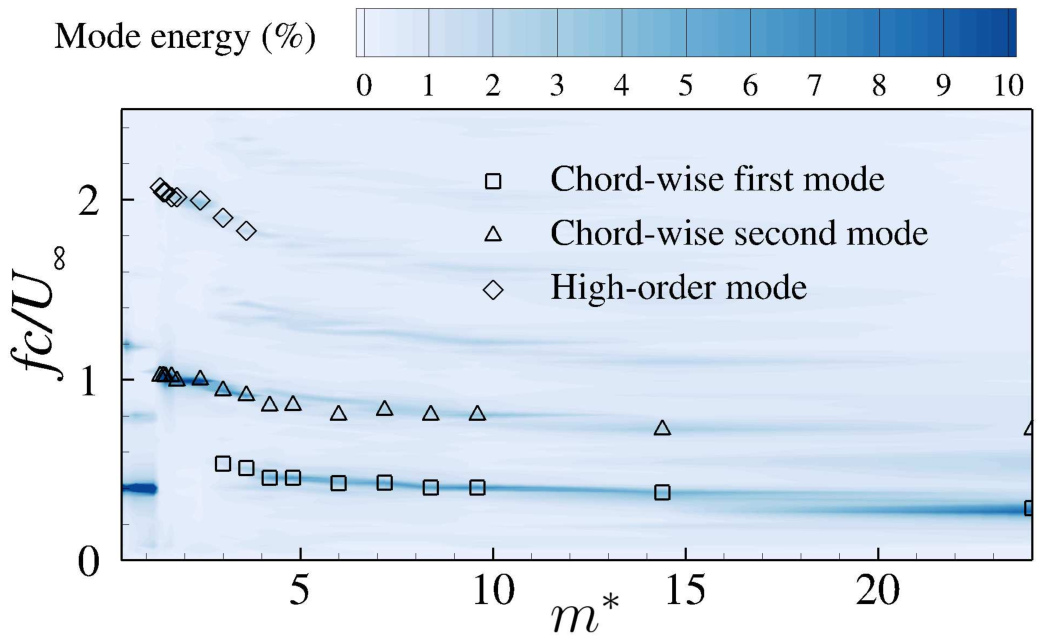}\label{mode_migration2a}}
	\quad
	\subfloat[][]{\includegraphics[width=0.48\textwidth]{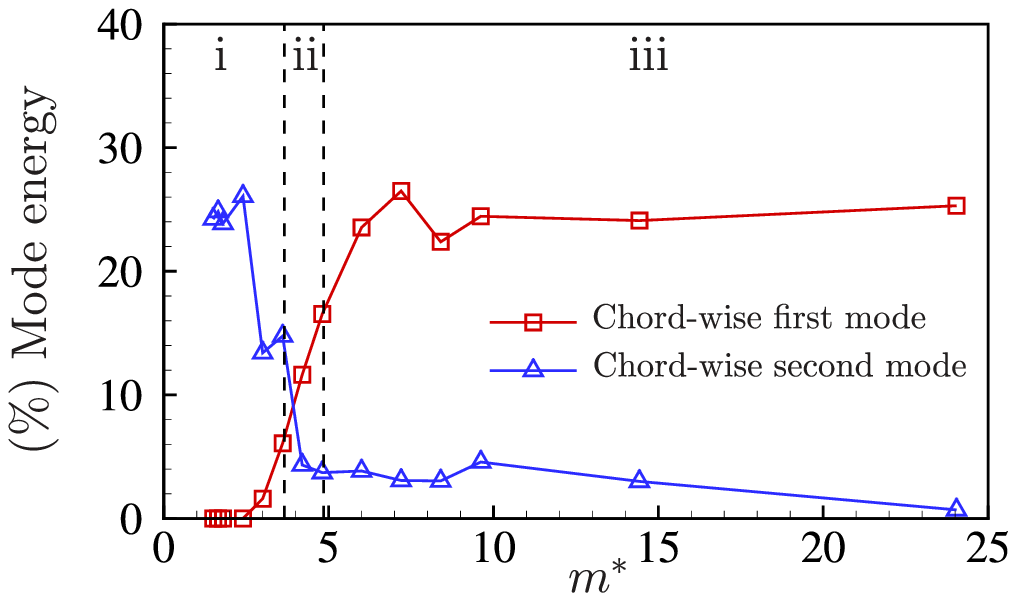}\label{mode_migration2b}}
	\caption{Mode frequency and energy map of coupled fluid-membrane system based on FMD analysis as a function of $m^*$: (a) comparison of aeroelastic mode frequencies between decomposed Fourier modes of the $Y$-vorticity field in the fluid domain and the structural modes with obvious mode energy, and (b) mode energies of the structural first and second modes.}
	\label{mode_migration2}
\end{figure}

The unsteady separated flows with various spatial-temporal scales and the intertwined structural modes motivate us to build a connection between the dynamic characteristics of the fluid and structural domains. The proposed global FMD technique offers an effective approach to map the membrane vibrations and the unsteady flow together from the spatial-temporal space to the spatial-frequency space. The aeroelastic mode energy is quantitatively calculated to identify the dominant mode. Using the FMD method, we analyse the dynamic responses of the 3D coupled system in the studied parameter space. The mode energy spectrum for both full 3D fluid and structural domains is presented in \reffig{mode_migration2}. The contour is coloured by the mode energy of the decomposed Fourier modes of the $Y$-vorticity field in the fluid domain at various mass ratios. Only the chord-wise first, second and some high-order structural modes with relatively large mode energies are shown in the mode energy map. The other modes are neglected due to their weak contributions to the membrane vibrations. Because the membrane achieves a deformed-steady state in the range of $m^* \in [0.36,1.2]$, no vibrational mode is excited in this regime. We find that the flexible membrane maintains a similar cambered-up wing shape in the deformed-steady state, which leads to the same vortex shedding frequency contents. As $m^*$ grows up higher than 1.2, some structural modes are excited in the coupled system. Meanwhile, the non-dimensional dominant vortex shedding frequency jumps to $fc/U_{\infty}=1.02$ near the flow-excited instability boundary. In the DBS regime, the dominant vortex shedding frequency ($f^{vs}$) is found to lock into the membrane vibration frequency ($f^s$). The frequency contents of the coupled system show an overall downward trend as $m^*$ increases. The dominant mode frequency in the fluid domain switches from the higher branch to the lower branch in the transitional mode region when the mode energy of the lower branch exceeds that of the higher branch. From the structural mode energy map shown in \reffig{mode_migration2} \subref{mode_migration2b}, the chord-wise first mode becomes the dominant mode in the membrane vibration responses as its corresponding mode energy is larger than the mode energy of the second mode. Thus, the mode transition phenomenon occurs. Further investigation of the mode transition phenomenon will be discussed in $\S$\ref{mode_transition}.

\subsection{Effect of Reynolds number}
In this section, we further investigate the effect of $Re$ on the flow-induced vibration and characterize the evolution of the dominant mode and the corresponding mode energy. We choose a series of numerical studies in the parameter space of (M5,R1$\to$9) in \reffig{mode_map}, which consists of the concerned flow-induced vibration and the mode transition phenomena. It can be seen from \reffig{phase} that the aerodynamic performance improves and the mean membrane camber as well as the vibration intensity show an overall upward trend as $Re$ increases. To gain further insight into the effect of $Re$, we compare the instantaneous aerodynamic forces, the unsteady flow features and the membrane vibrations at different Reynolds numbers.

\begin{figure}
	\centering 
	\subfloat[][]{\includegraphics[width=0.49\textwidth]{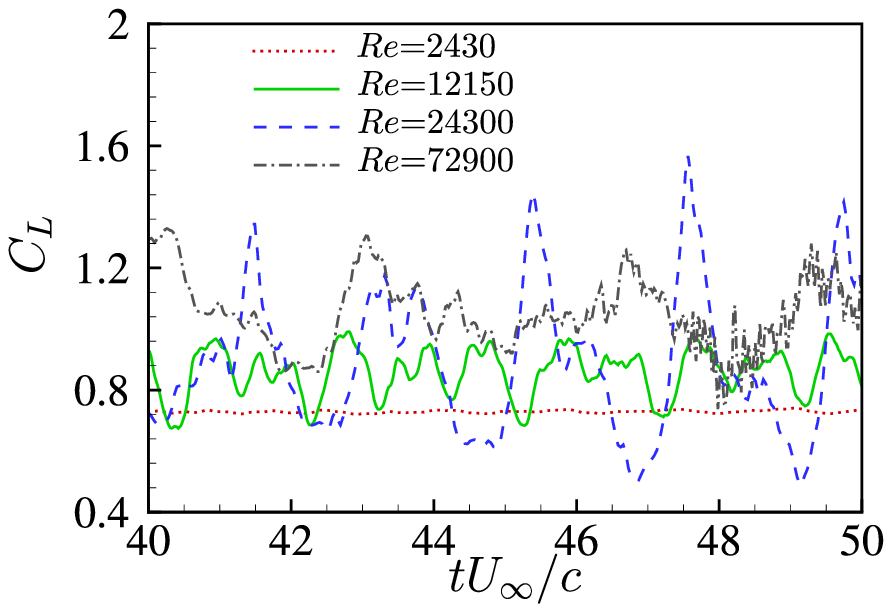}\label{response_rea}}
	\subfloat[][]{\includegraphics[width=0.49\textwidth]{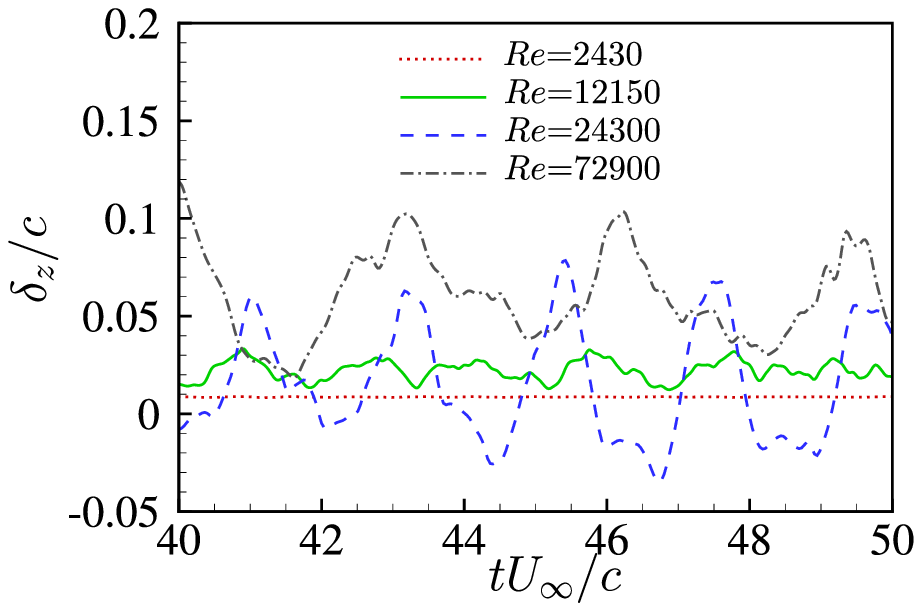}\label{response_reb}}
	\caption{Time histories of instantaneous: (a) lift coefficients and (b) normalized displacements at the membrane centre along the $Z$-direction for four selected cases with $Re$=2430, 12150, 24300 and 72900 at a fixed $m^*$=4.2.}
	\label{response_re}
\end{figure}

\begin{figure}
	\centering 
	\subfloat[][]{
		\includegraphics[width=1.0\textwidth]{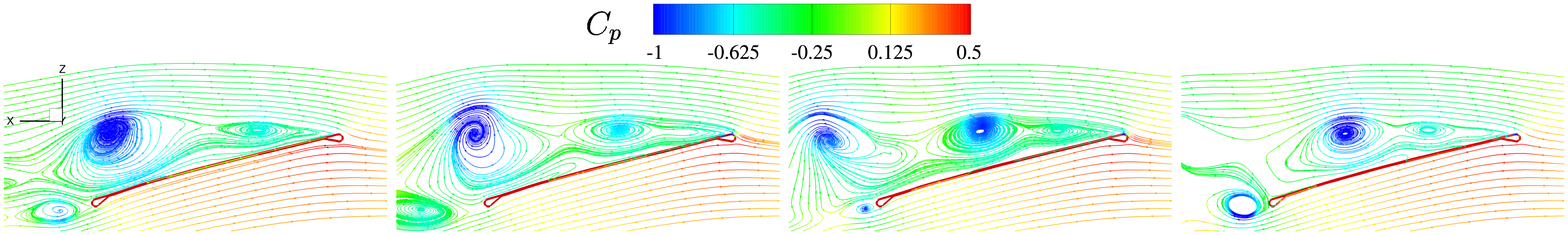}
		\label{stream_rea}
	}
	\\
	\subfloat[][]{
		\includegraphics[width=1.0\textwidth]{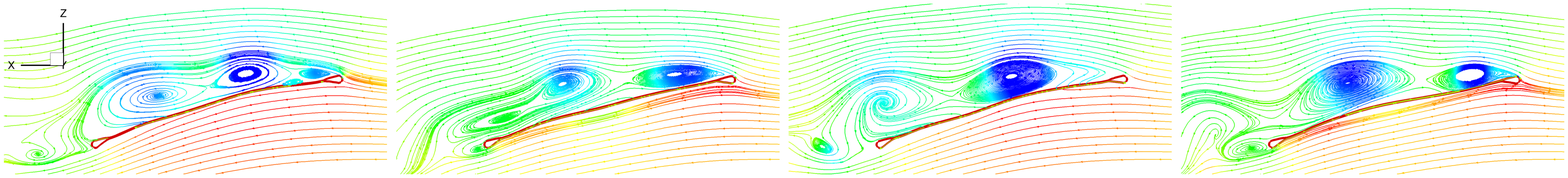}
		\label{stream_reb}
	}
	\\
	\subfloat[][]{
		\includegraphics[width=1.0\textwidth]{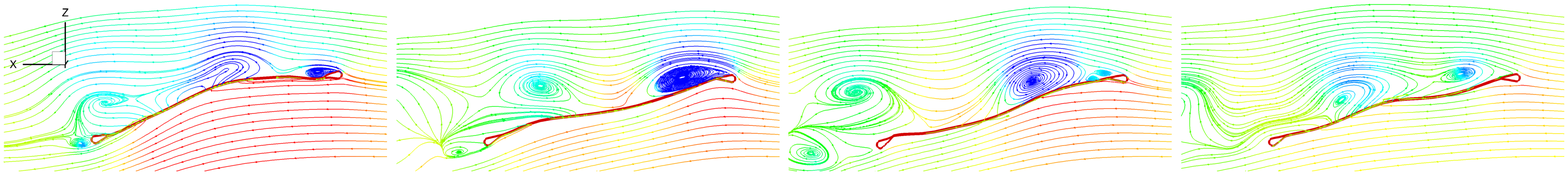}
		\label{stream_rec}
	}
	\\
	\subfloat[][]{
		\includegraphics[width=1.0\textwidth]{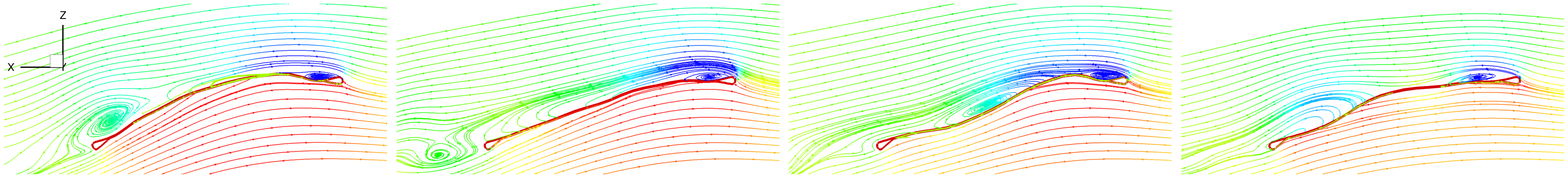}
		\label{stream_red}
	}
	\caption{Instantaneous streamlines of 3D flexible membrane wing on the mid-span plane coloured by pressure coefficient for $(Re,m^*)=$ (a) $(2430,4.2)$, (b) $(12150,4.2)$, (c) $(24300,4.2)$, (d) $(72900,4.2)$.}
	\label{stream_re}
\end{figure}

In \reffig{response_re}, we present the comparison of the time histories of the instantaneous lift coefficients and the instantaneous vertical displacement at the membrane centre at four representative Reynolds numbers (R1=2430, R4=12150, R5=24300, R8=72900). The membrane displacement remains steady at $Re$=2430. The oscillation amplitudes of the lift coefficient and the membrane displacement increase at higher $Re$ values within the DBS regime. Meanwhile, the dominant frequency of the system gradually transitions to the lower frequency branch. The coupled dynamics tends to be non-periodic oscillation. In \reffig{stream_re}, we further display the instantaneous streamlines around the flexible membrane on the mid-span plane. The vortex is generated near the leading edge and convects downstream to form a pair of vortices with the trailing edge vortex at $Re=2430$. Meanwhile, the membrane remains a deformed-steady wing shape. As $Re$ increases to 12150, the flexible membrane interacts with the shed vortices to vibrate in the chord-wise second mode. The wake pattern and the membrane vibrational mode are significantly changed at $Re=24300$. Both the chord-wise first and second structural modes are observed in \reffig{stream_re} \subref{stream_rec}. The chord-wise first mode starts to dominate the membrane vibrations at $Re=72900$ shown in \reffig{stream_re} \subref{stream_red}. The separated flow region is greatly suppressed due to the instantaneous large wing camber. Except for a small vortex near the leading edge, only one main vortex is formed near the mid-chord location of the membrane and sheds into the wake in one completed vibrational cycle.

\begin{figure}
	\centering 
	\subfloat[][]{
		\includegraphics[width=0.25\textwidth]{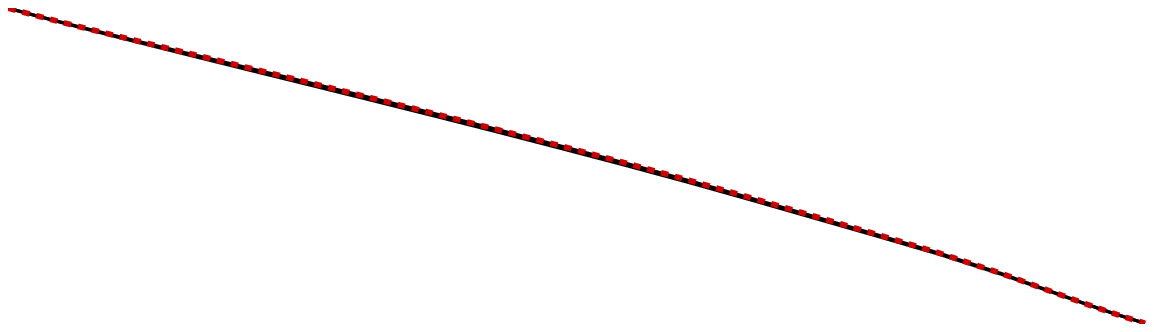}\label{membrane_vibration_rea}}
	\subfloat[][]{
		\includegraphics[width=0.25\textwidth]{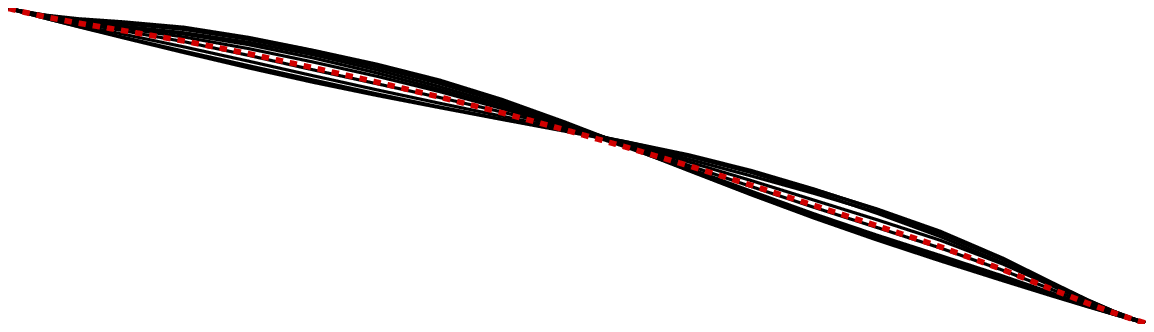}\label{membrane_vibration_reb}}
	\subfloat[][]{
		\includegraphics[width=0.25\textwidth]{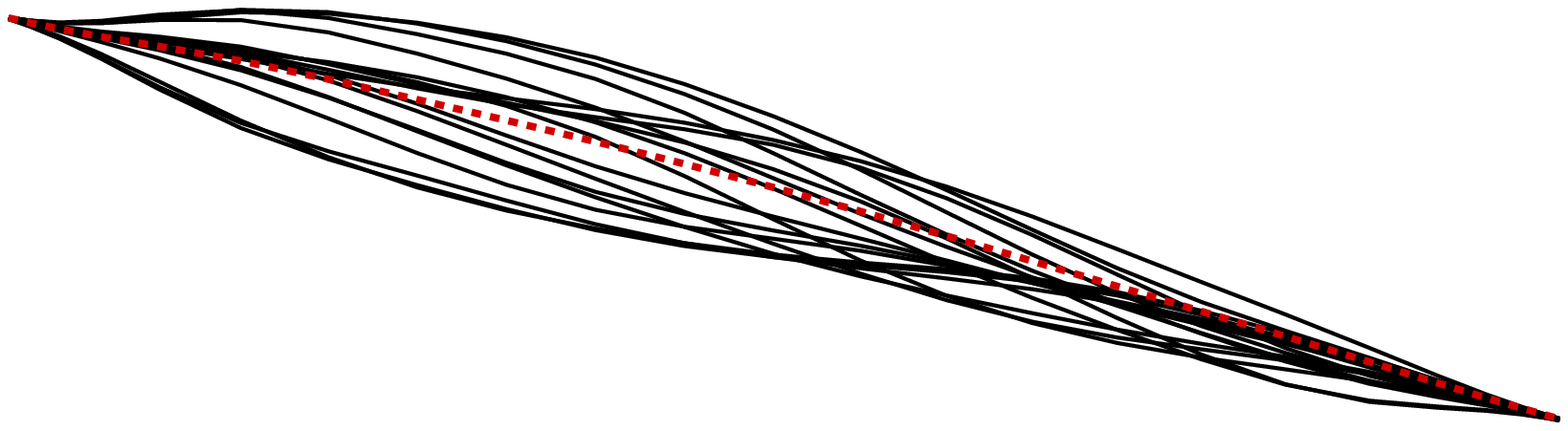}\label{membrane_vibration_rec}}
	\subfloat[][]{
		\includegraphics[width=0.25\textwidth]{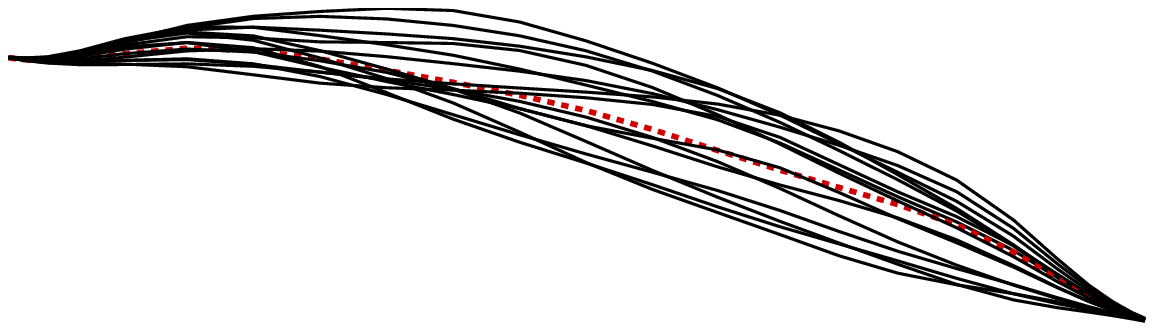}\label{membrane_vibration_red}}
	\\
	\subfloat[][]{
		\includegraphics[height=0.12\textwidth,width=0.27\textwidth]{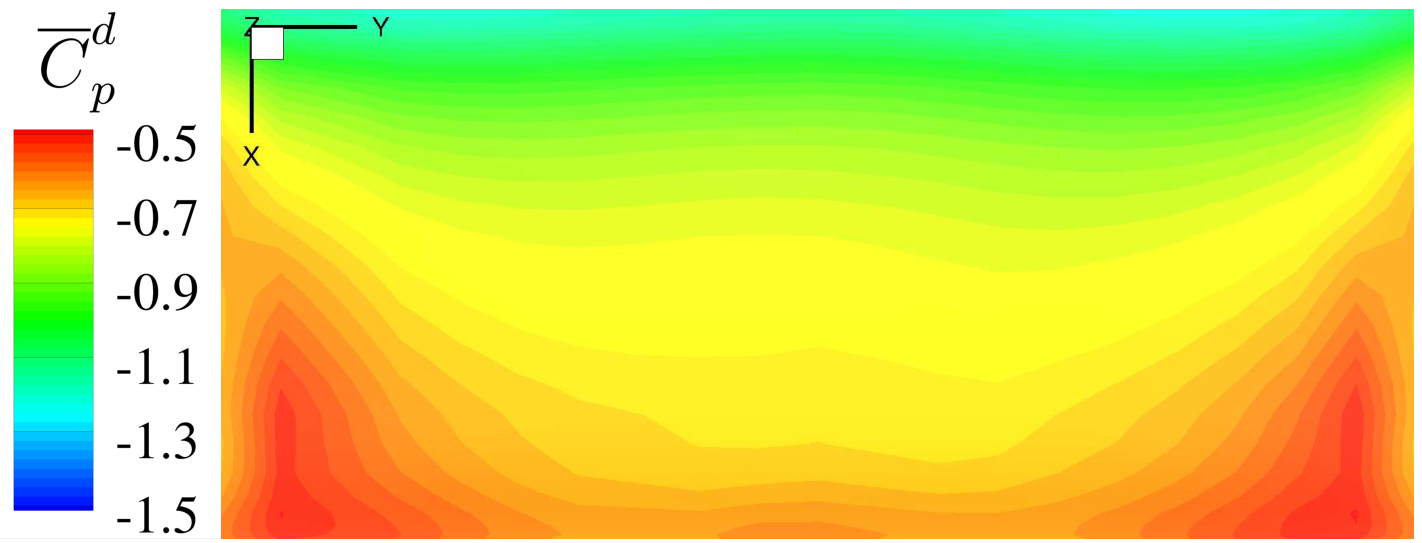}\label{cpd_rea}}
	\
	\subfloat[][]{
		\includegraphics[height=0.12\textwidth,width=0.22\textwidth]{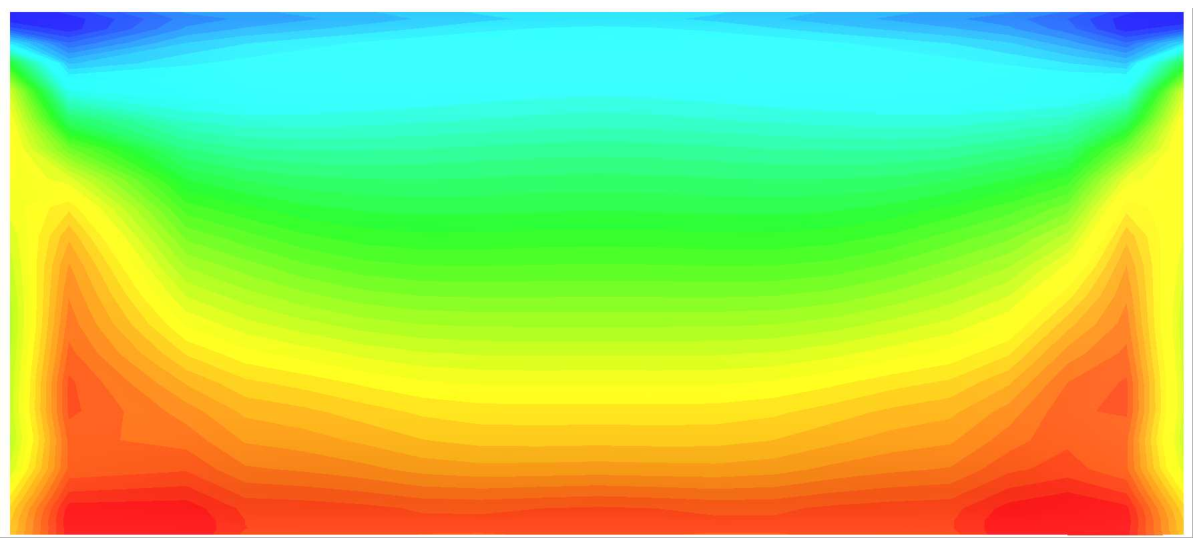}\label{cpd_reb}}
	\
	\subfloat[][]{
		\includegraphics[height=0.12\textwidth,width=0.22\textwidth]{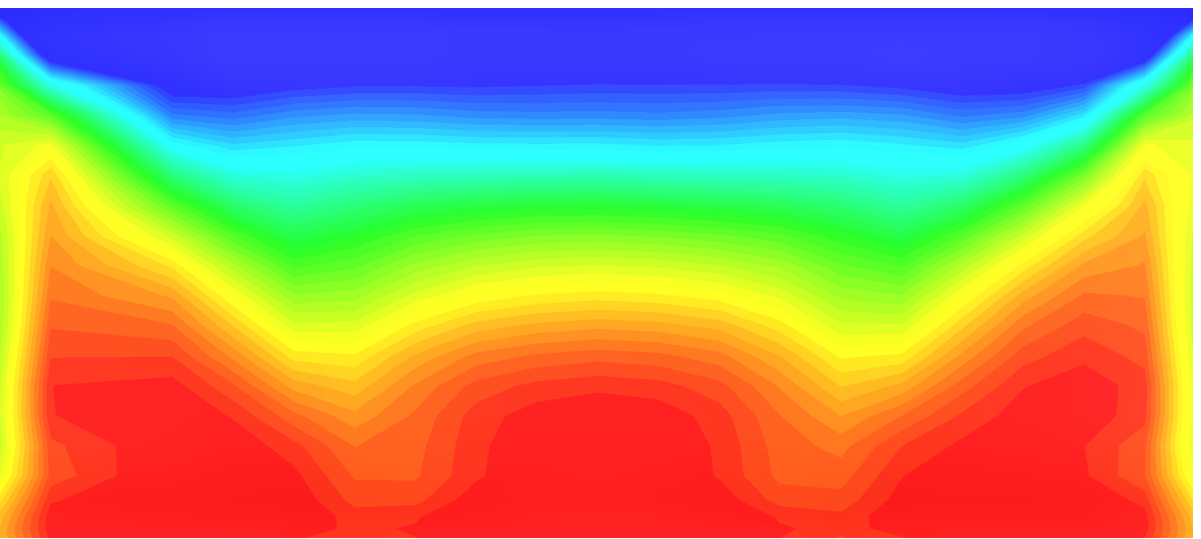}\label{cpd_rec}}
	\
	\subfloat[][]{
		\includegraphics[height=0.12\textwidth,width=0.22\textwidth]{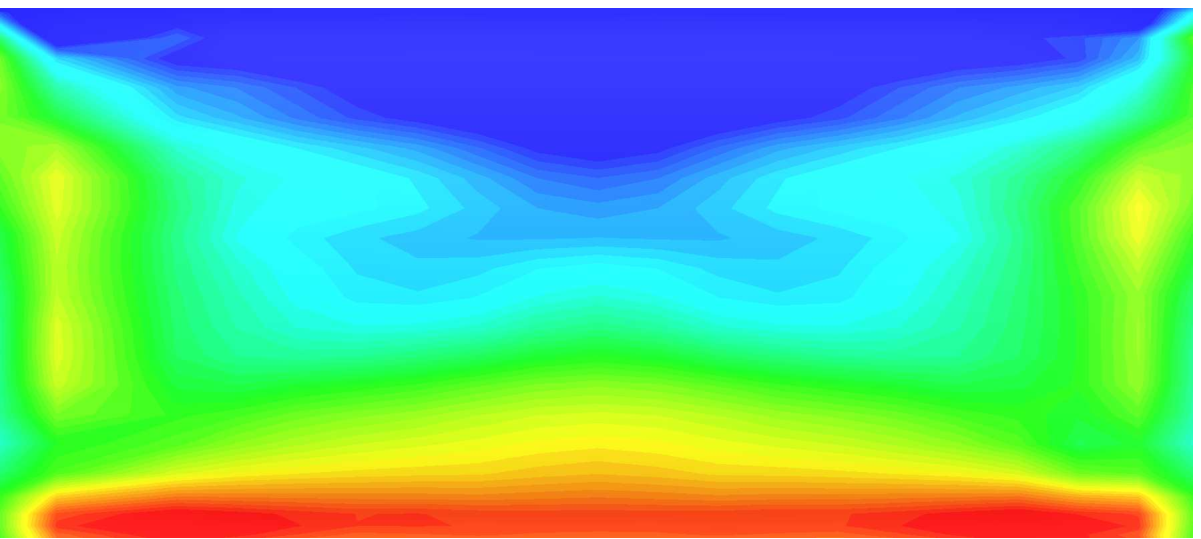}\label{cpd_red}}
	\\
	\subfloat[][]{
		\includegraphics[height=0.17\textwidth,width=0.3\textwidth]{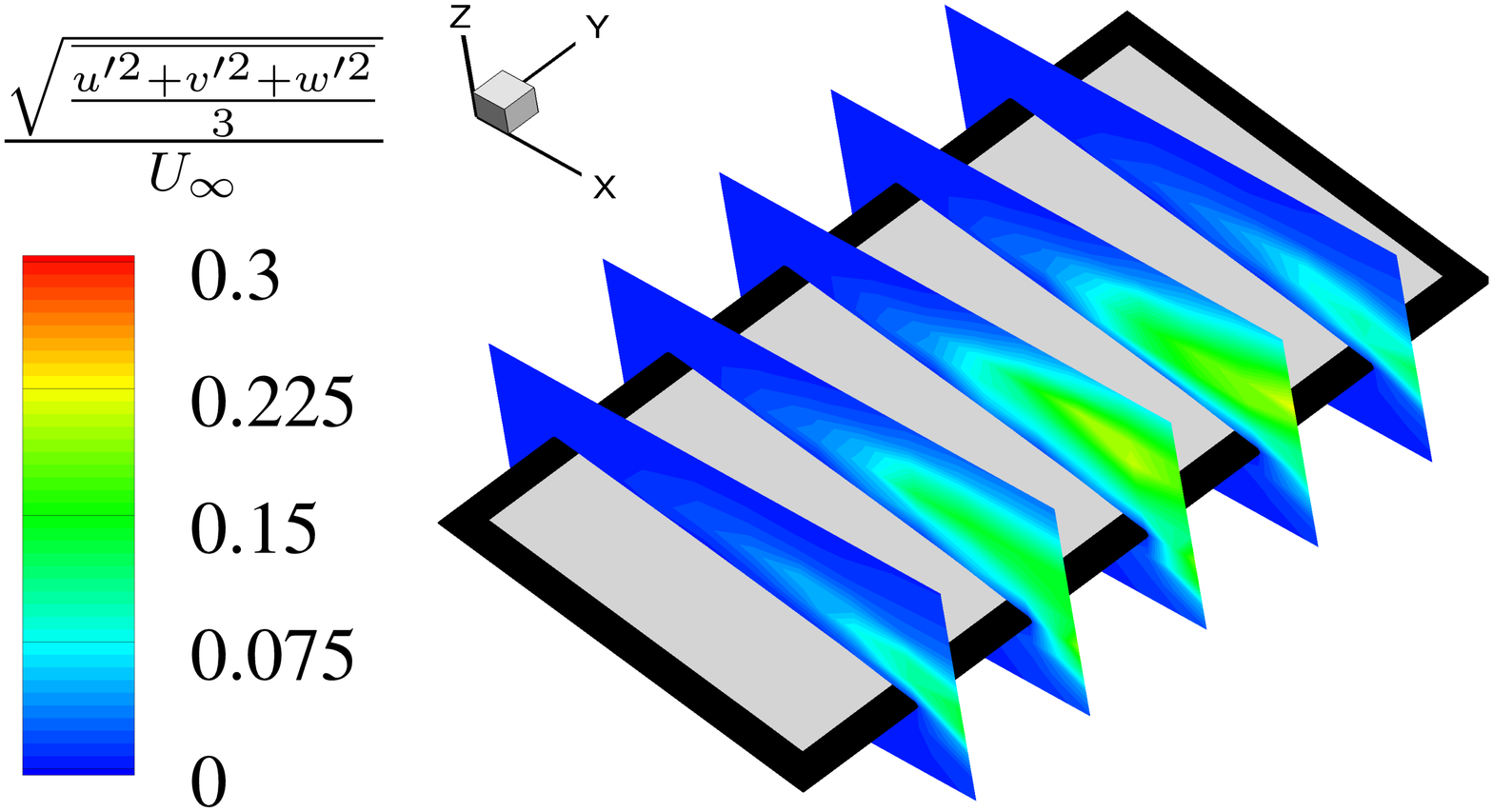}\label{tke_rea}}
	\subfloat[][]{
		\includegraphics[height=0.17\textwidth,width=0.22\textwidth]{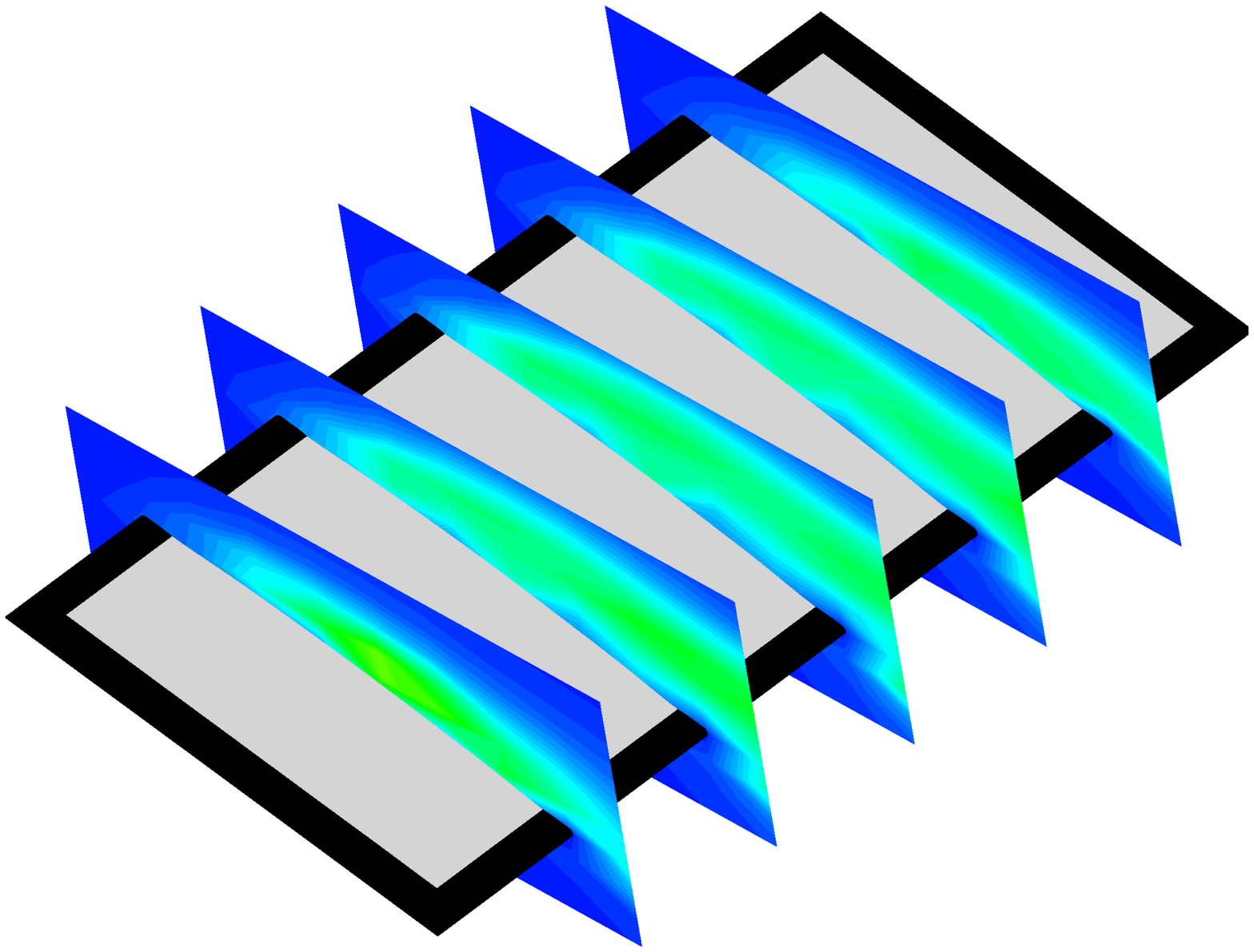}\label{tke_reb}}
	\subfloat[][]{
		\includegraphics[height=0.17\textwidth,width=0.22\textwidth]{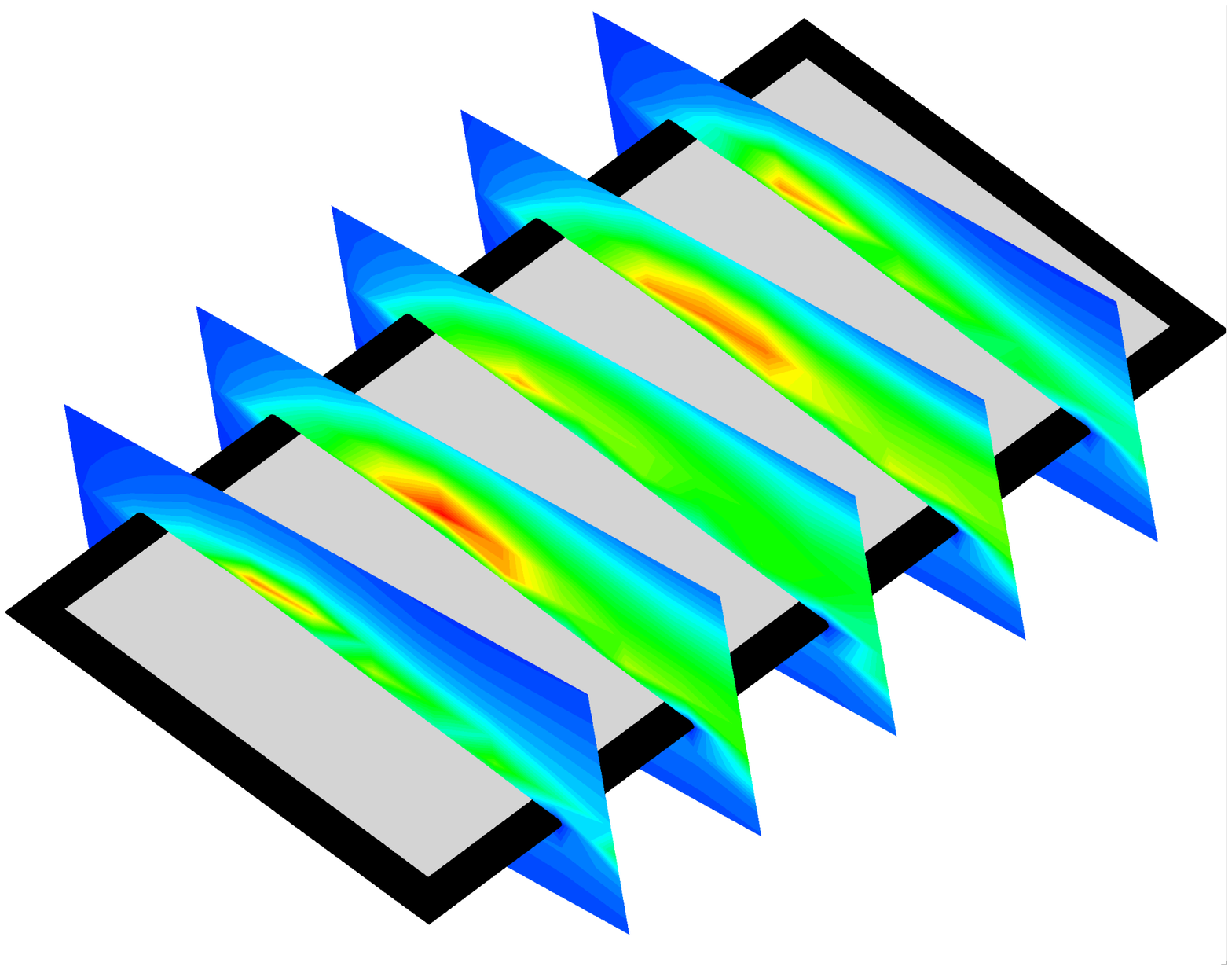}\label{tke_rec}}
	\subfloat[][]{
		\includegraphics[height=0.17\textwidth,width=0.22\textwidth]{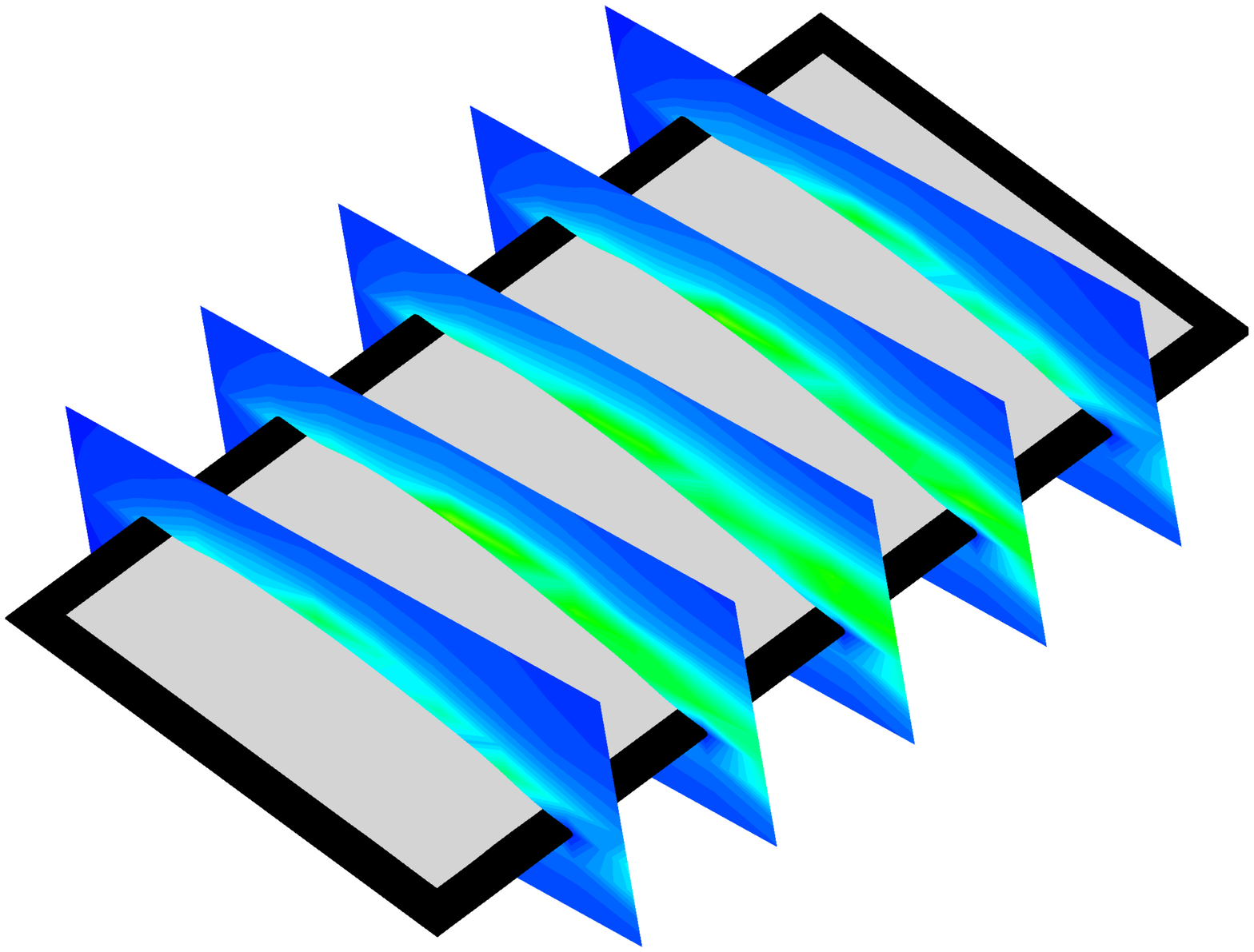}\label{tke_red}}
	\\
	\subfloat[][]{
		\includegraphics[height=0.17\textwidth,width=0.3\textwidth]{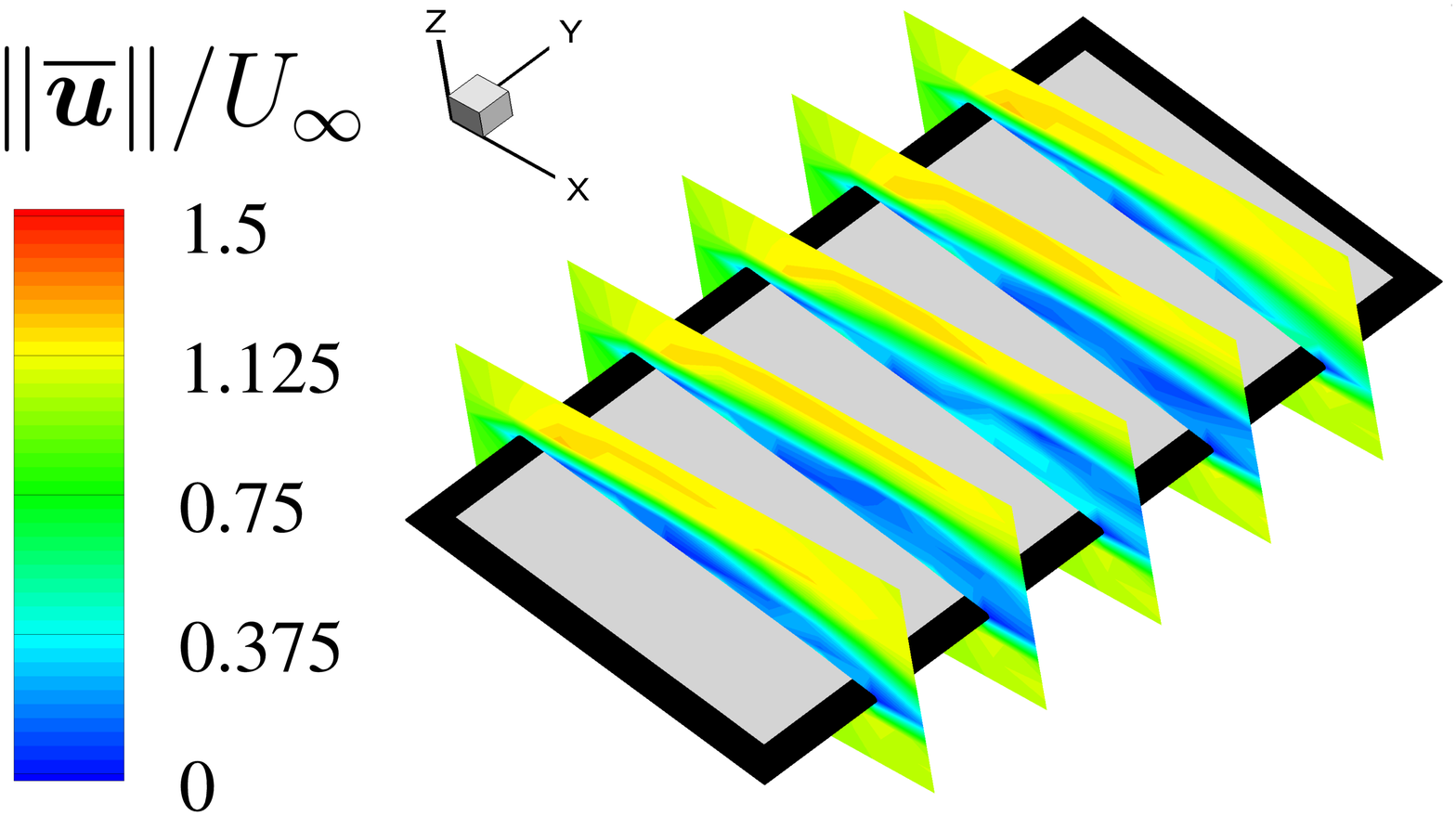}\label{vm_rea}}
	\subfloat[][]{
		\includegraphics[height=0.17\textwidth,width=0.22\textwidth]{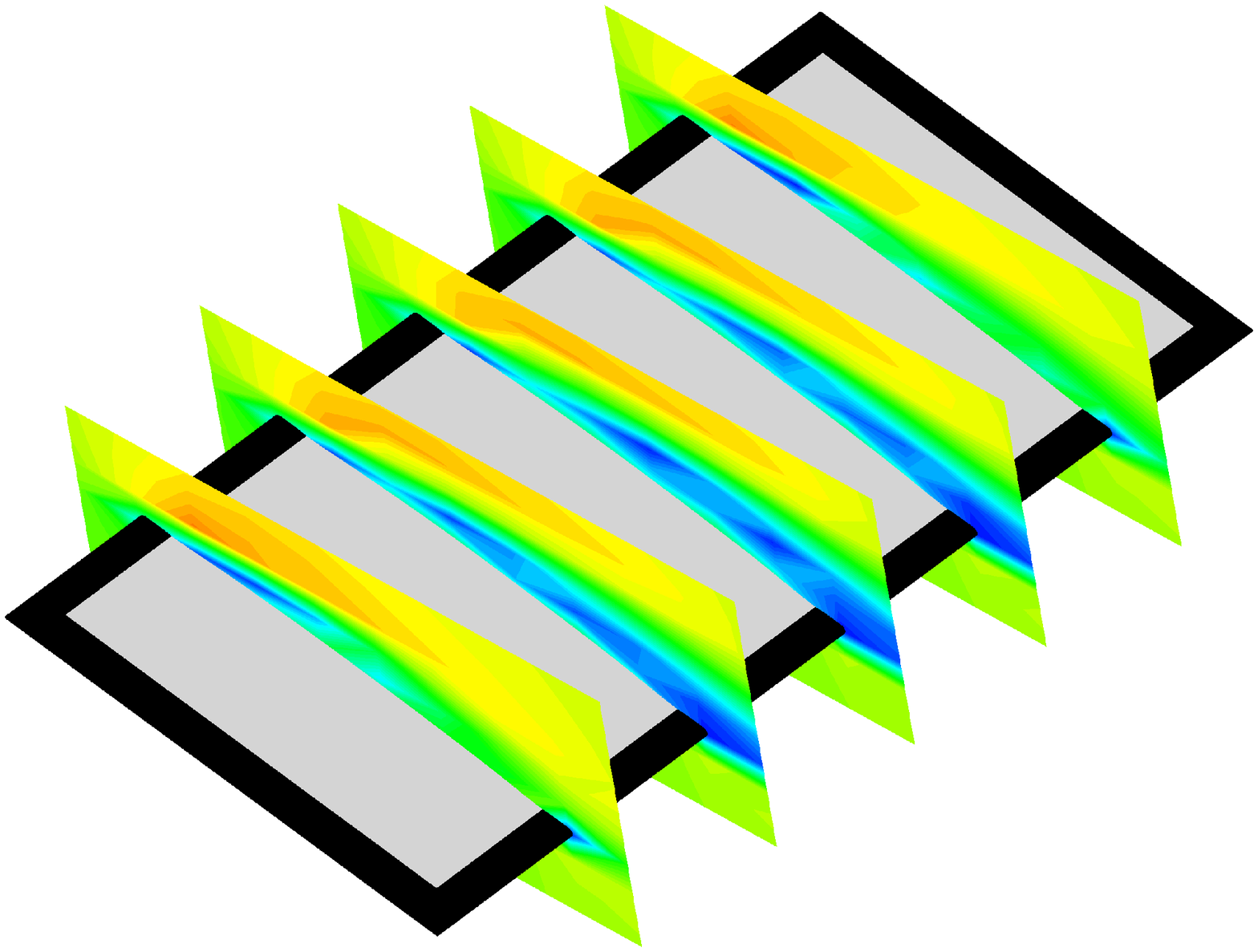}\label{vm_reb}}
	\subfloat[][]{
		\includegraphics[height=0.17\textwidth,width=0.22\textwidth]{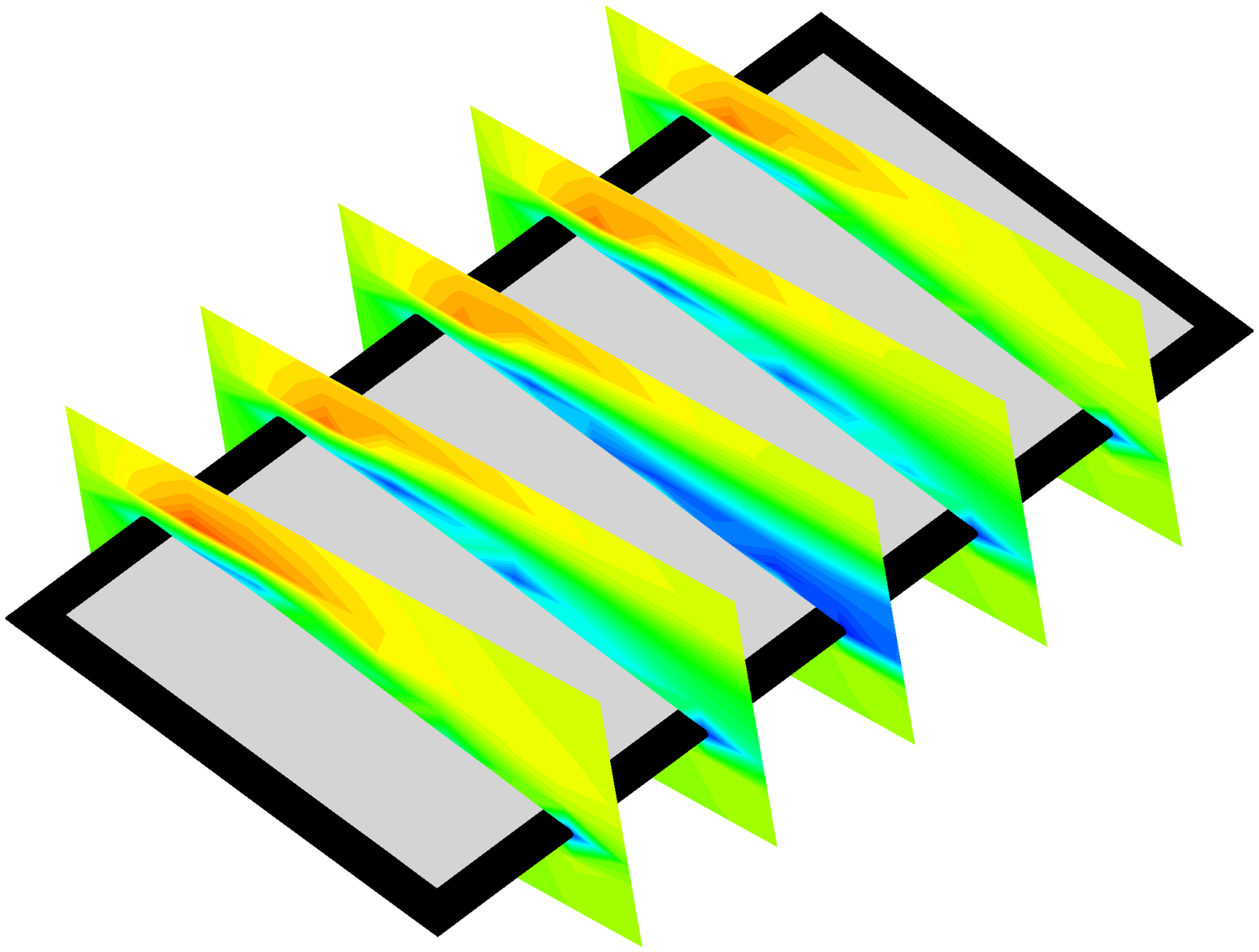}\label{vm_rec}}
	\subfloat[][]{
		\includegraphics[height=0.17\textwidth,width=0.22\textwidth]{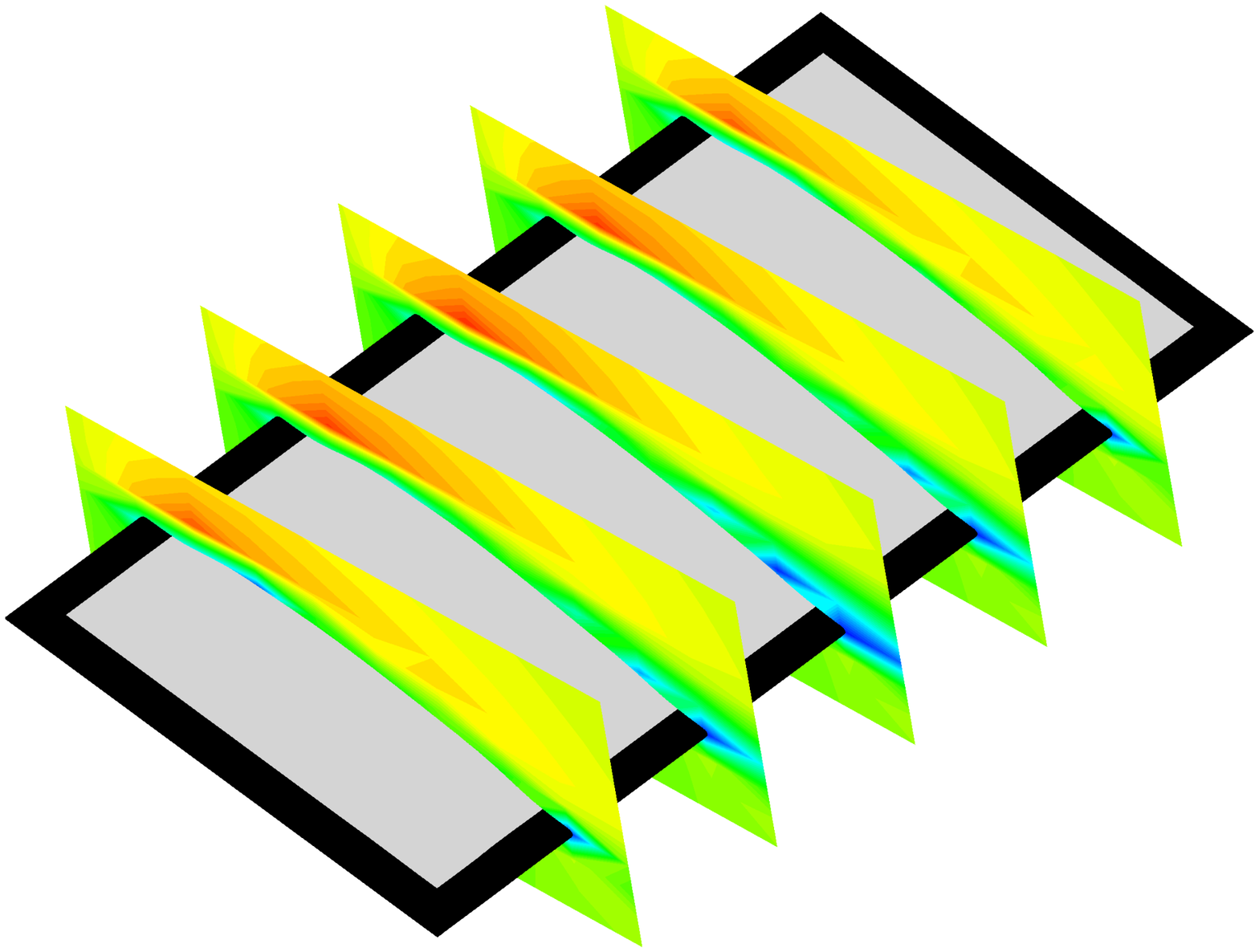}\label{vm_red}}
	\caption{Flow past a 3D rectangular membrane wing: (a,b,c,d) full-body responses of 3D flexible membrane wing at mid-span location, (e,f,g,h) time-averaged pressure coefficient difference between upper and lower surfaces, (i,j,k,l) turbulent intensity and (m,n,o,p) time-averaged normalized velocity magnitude on five slices along the span-wide direction at $(Re,m^*)=$ (a,e,i,m) $(2430,4.2)$, (b,f,j,n) $(12150,4.2)$, (c,g,k,o) $(24300,4.2)$, (d,h,l,p) $(72900,4.2)$. ($\color{red}{---}$) denotes the time-averaged membrane shape in (a,b,c,d).}
	\label{membrane_vibration_re}
\end{figure}

We further compare the membrane profiles, the mean pressure coefficient difference, the turbulent intensity and the mean velocity magnitude for the selected four cases in \reffig{membrane_vibration_re}. Consistent with the observation for the membrane vibration states in \reffig{stream_re}, the flexible membrane achieves a deformed-steady state at $Re=2430$. The membrane exhibits a chord-wise second mode, and then transitions to the chord-wise first mode as $Re$ further increases. Compared to the pressure difference distribution of the membrane in the DSS regime in \reffig{membrane_vibration_re} \subref{cpd_rea}, the suction force area in the proximity of the leading edge expands when the membrane vibration is excited. It can be attributed to the longer attached vortices interacting with the membrane vibration. It is observed from \reffig{membrane_vibration_re} \subref{cpd_red} that the membrane with large camber at $Re$=72900 enhances the suction area and further improves the aerodynamic performance. The flow fluctuations in the shear layer become stronger and get closer to the membrane surface when the membrane vibration occurs at a higher $Re$. The flow fluctuations become weaker at $Re$=72900 as shown in \reffig{membrane_vibration_re} \subref{tke_red} due to the significantly suppressed separated flow for the largely cambered membrane. The low-velocity region becomes smaller at a higher $Re$.

\begin{figure}
	\centering 
	\subfloat[][]{\includegraphics[width=0.48\textwidth]{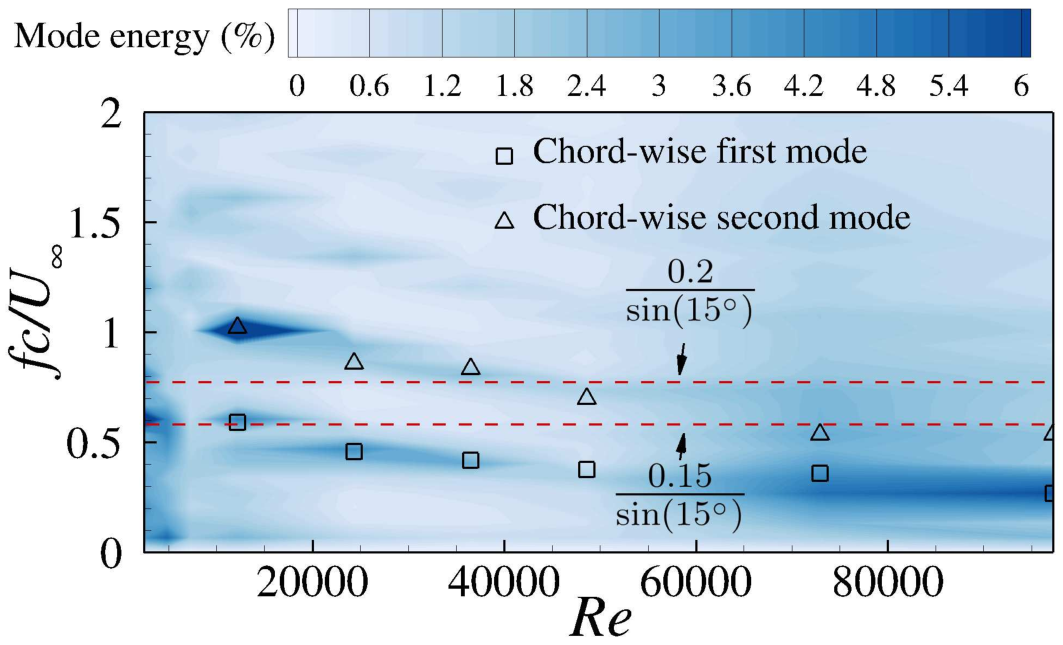}\label{mode_migration_rea}}
	\quad
	\subfloat[][]{\includegraphics[width=0.48\textwidth]{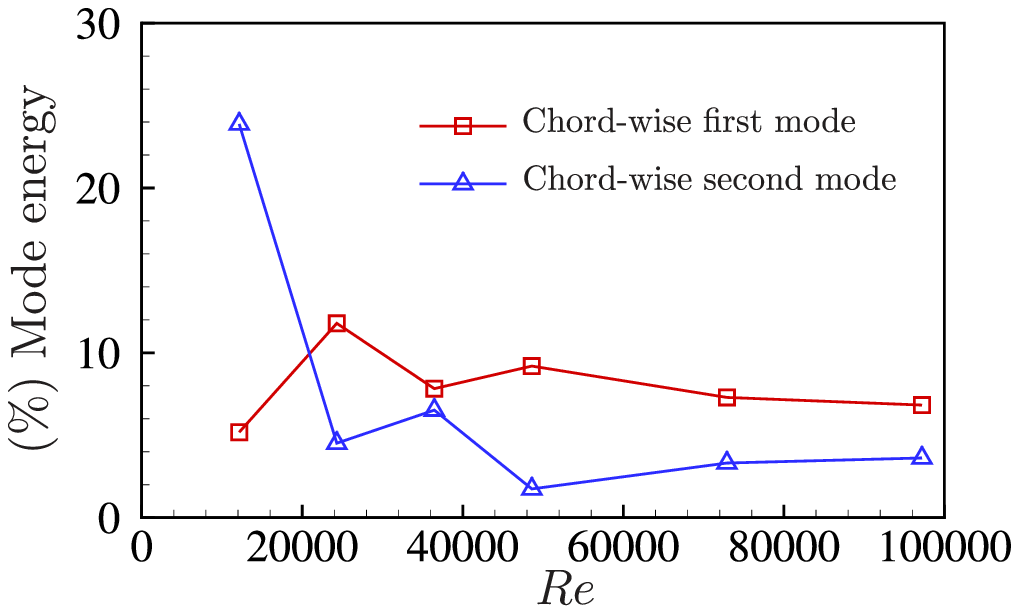}\label{mode_migration_reb}}
	\caption{Mode frequency and energy map of fluid-membrane coupled system based on FMD analysis as a function of $Re$: (a) comparison of aeroelastic mode frequencies between decomposed Fourier modes of the $Y$-vorticity field  in the fluid domain and the structural modes with obvious mode energy, (b) mode energies of the structural first and second modes.}
	\label{mode_migration_re}
\end{figure}

We observe the frequency lock-in phenomenon at various mass ratios in the DBS regime in \reffig{mode_migration2}. A natural question is to ask whether the frequency lock-in phenomenon also occurs at various Reynolds numbers. To examine the frequency lock-in, we extract the mode energy spectra from the coupled system both in the fluid and structural domains based on the FMD method. Similar to \reffig{mode_migration2}, we only plot the most energetic structural modes for simplicity. The comparison of the mode energy spectra of the fluid and structural Fourier modes is presented in \reffig{mode_migration_re} \subref{mode_migration_rea}. The variation of the first and second mode energies as a function of $Re$ is plotted in \reffig{mode_migration_re} \subref{mode_migration_reb}. \cite{rojratsirikul2011flow} summarized the Strouhal number of the vortex shedding process for different finite wings over a wide range of $Re$. They found that the modified non-dimensional frequency falls in the range of $(fc \sin \alpha) / U_{\infty} \in [0.15, 0.2]$. The two red dashed lines shown in \reffig{mode_migration_re} \subref{mode_migration_rea} represent the upper and lower limits of the non-dimensional frequency. We observe the non-dimensional vortex shedding frequency shows few changes within $Re \in [2430,12150]$. The dominant non-dimensional vortex shedding frequency in this parameter space for a deformed-steady membrane falls into the summarized non-dimensional frequency range. This observation is consistent with the conclusion of a flat plate with an almost constant vortex shedding frequency in the range of $Re\in[10^3, 10^5]$ \citep{katopodes2018free}. When the membrane vibration occurs, the dominant non-dimensional vortex shedding frequency jumps from 0.6 to 1.02. The frequency lock-in is noticed in \reffig{mode_migration_re} \subref{mode_migration_rea} in the range of $Re \in [12150,97200]$. In \reffig{mode_migration_re} \subref{mode_migration_reb}, the mode energy of the chord-wise first mode exceeds the mode energy of the chord-wise second mode at $Re=24300$. The mode transition from the second mode to the first mode is triggered. Thus, the first mode dominates the membrane vibration at a higher $Re$.

Through the studies of the effect of $m^*$ and $Re$ on the coupled fluid-membrane dynamics, we find similar phenomena and common characteristics for the onset of the membrane vibration and the mode transition at various mass ratios and Reynolds numbers. The results show that the role of the flow-excited instability in the coupled dynamics at various physical parameters shares some commonalities. In addition to $Re$ and $m^*$, the aeroelastic number $Ae$ related to the membrane flexibility is another important parameter to govern the coupled dynamics. To make the findings more general, we also examine the evolution of the coupled fluid-membrane dynamics as a function of $Ae$. The purpose is to explore whether the membrane flexibility influences the flow-induced vibration in a similar way to mass ratio and Reynolds number.

\subsection{Effect of aeroelastic number}
\begin{figure}
	\centering 
	\includegraphics[width=0.9\textwidth]{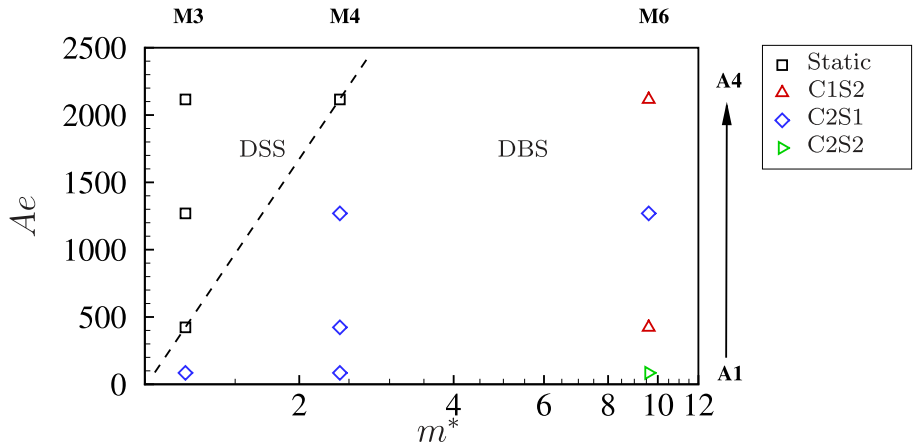}
	\caption{Stability phase diagram: non-dimensional aeroelastic number $Ae$ versus mass ratio $m^*$ for the 3D flexible membrane at $Re=24300$ for $\alpha$=$15^\circ$. Here, the dashed line ($\textcolor[RGB]{1,1,1}{---}$) is plotted to distinguish the flow-excited instability boundary. $\square$ denotes the simulation results corresponding to the deformed-steady state. $\triangle$, $\Diamond$ and $\triangleright$ represent the C1S2 mode, the C2S1 mode and the C2S2 mode in the dynamic balance state, respectively.}
	\label{mode_map_ae_mass}
\end{figure}

We perform a series of numerical simulations to investigate the role of flexibility in the flow-excited instability. Four sets of aeroelastic numbers (A1$\to$4) are selected in the simulations and the values are 84.6, 423.14, 1269.42 and 2115.7, respectively. To make the conclusion more comprehensive, we choose three types of mass ratios (M3, M4 and M6) to expand the parameter space, which represents a light wing, a medium weight wing and a heavy wing. The Reynolds number is fixed at $Re=24300$ and the AOA is set to $\alpha$=$15^\circ$, which makes the membrane immersed in unsteady separated flows. \refFig{mode_map_ae_mass} presents the stability phase diagram for the flexible membrane in the parameter space of $m^*$-$Ae$. The dashed line plotted in the figure denotes the flow-excited instability boundary. A new empirical solution of the boundary for the critical aeroelastic number ($Ae_{cr}$) is given as
\begin{equation}
Ae_{cr}=c_0+c_1(m^*)^n,
\label{Aecr}
\end{equation}
where $c_0$, $c_1$ and $n$ are the coefficients determined by numerical simulations or experiments. Two stability regimes, namely DSS and DBS, are also identified from the membrane responses at different parameter combinations, which is the same as the stability phase diagram presented in \reffig{mode_map}. The membrane has a tendency to maintain a deformed-steady state when the wing is lighter and more rigid. The flexible membrane leaves its static equilibrium position to vibrate as the effect of the membrane inertia and the flexibility becomes significant in the coupled system. The mode transition between different modes is also observed in the DBS regime in the studied parameter space of $m^*$-$Ae$.

\begin{figure}
	\centering 
	\subfloat[][]{\includegraphics[width=0.48\textwidth]{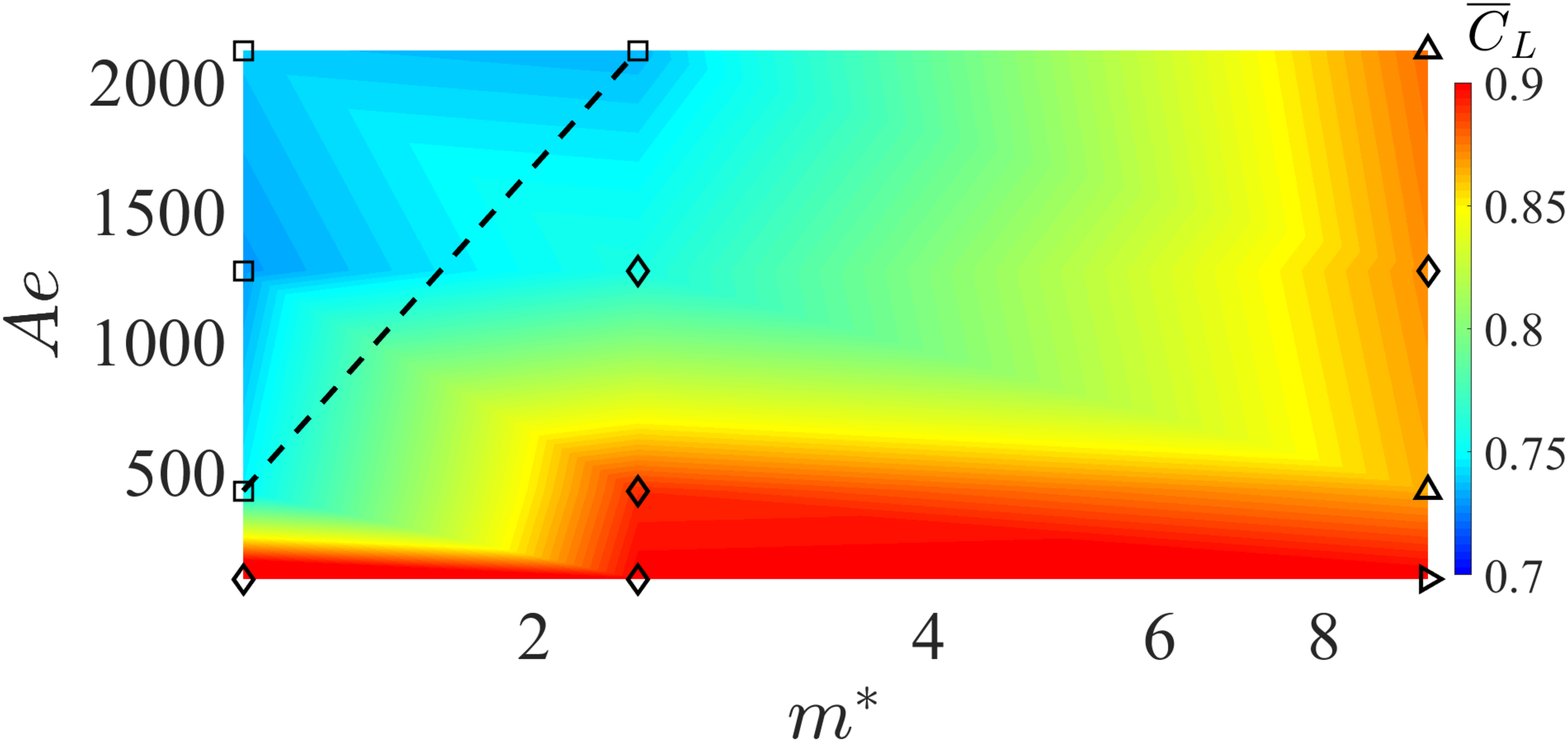}\label{phase_aea}}
	\quad
	\subfloat[][]{\includegraphics[width=0.48\textwidth]{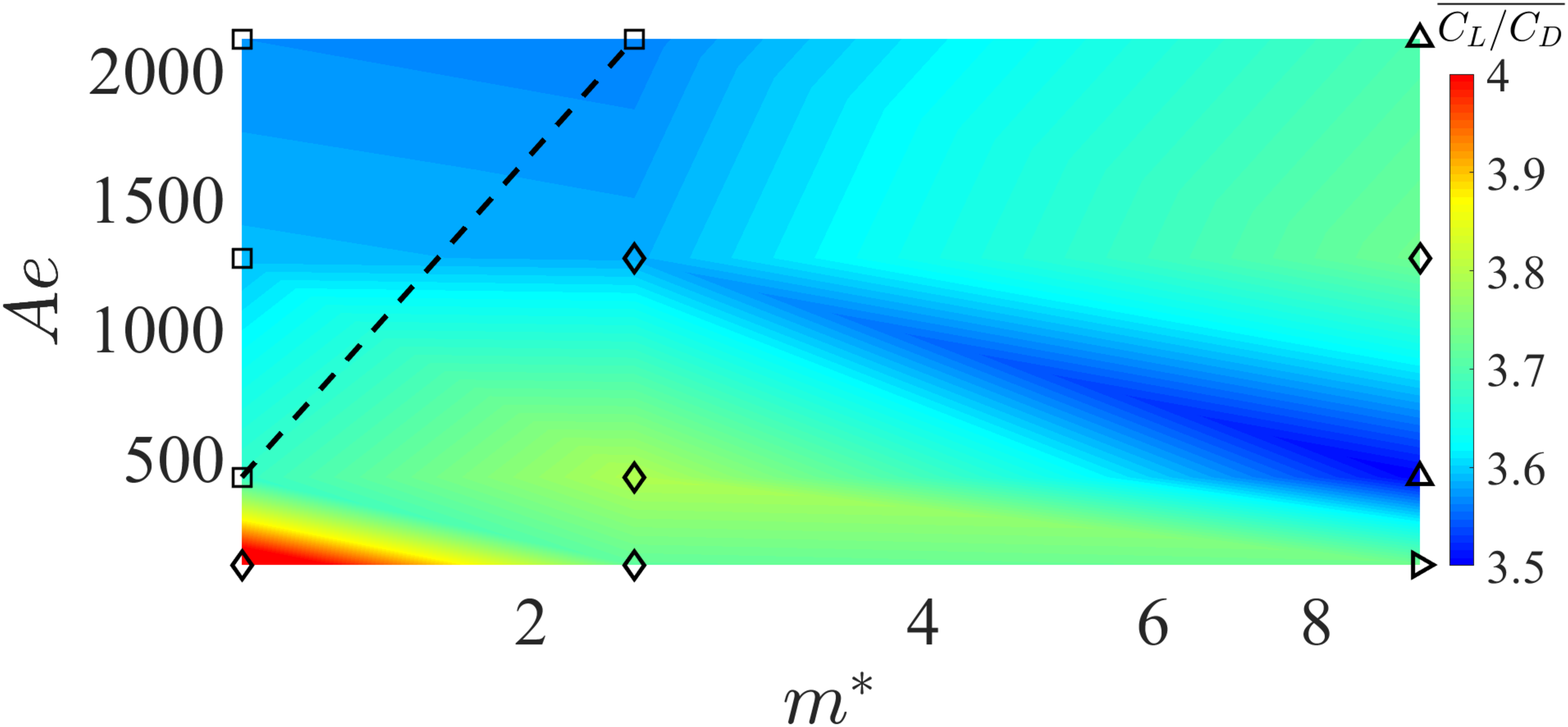}\label{phase_aeb}}
	\\
	\subfloat[][]{\includegraphics[width=0.48\textwidth]{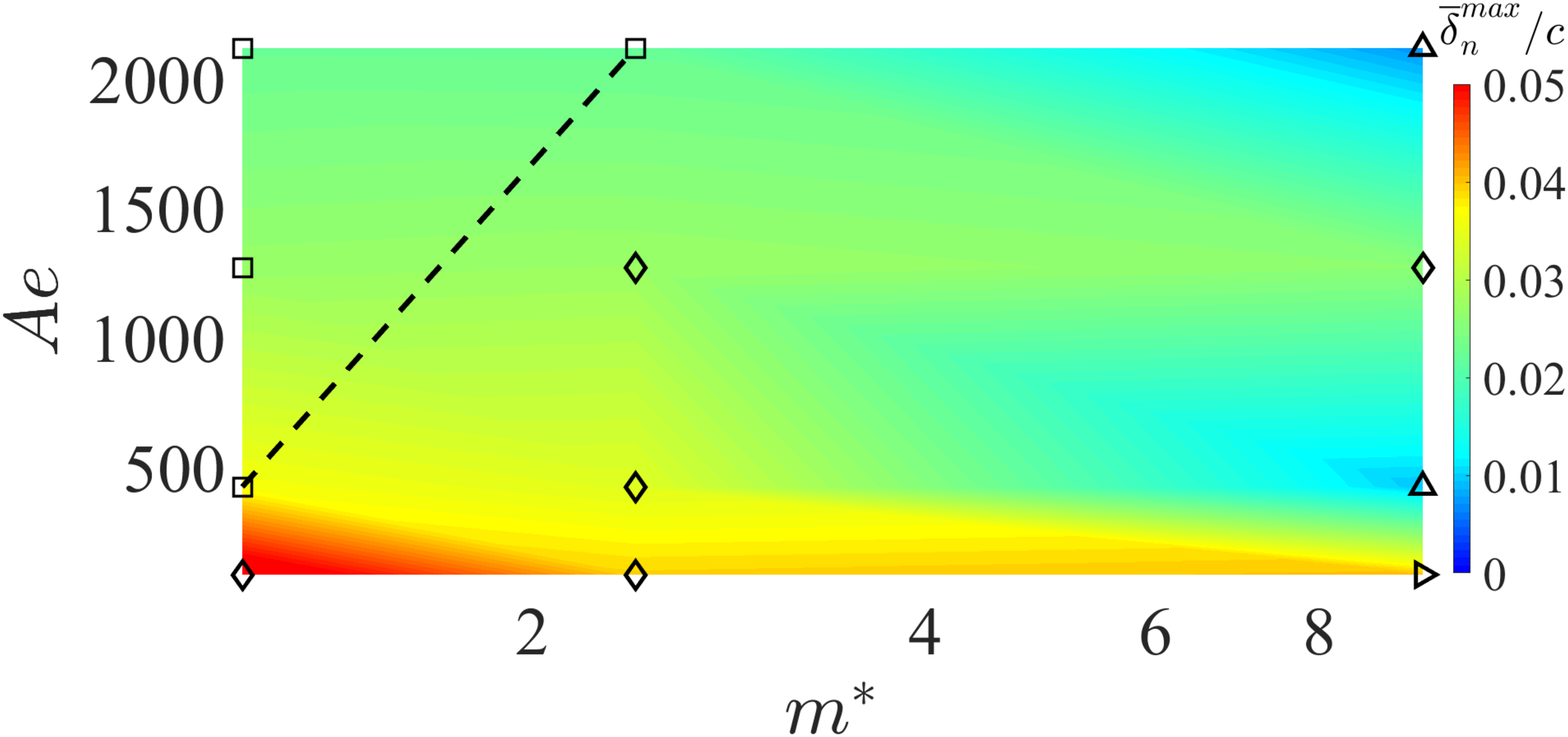}\label{phase_aec}}
	\quad
	\subfloat[][]{\includegraphics[width=0.48\textwidth]{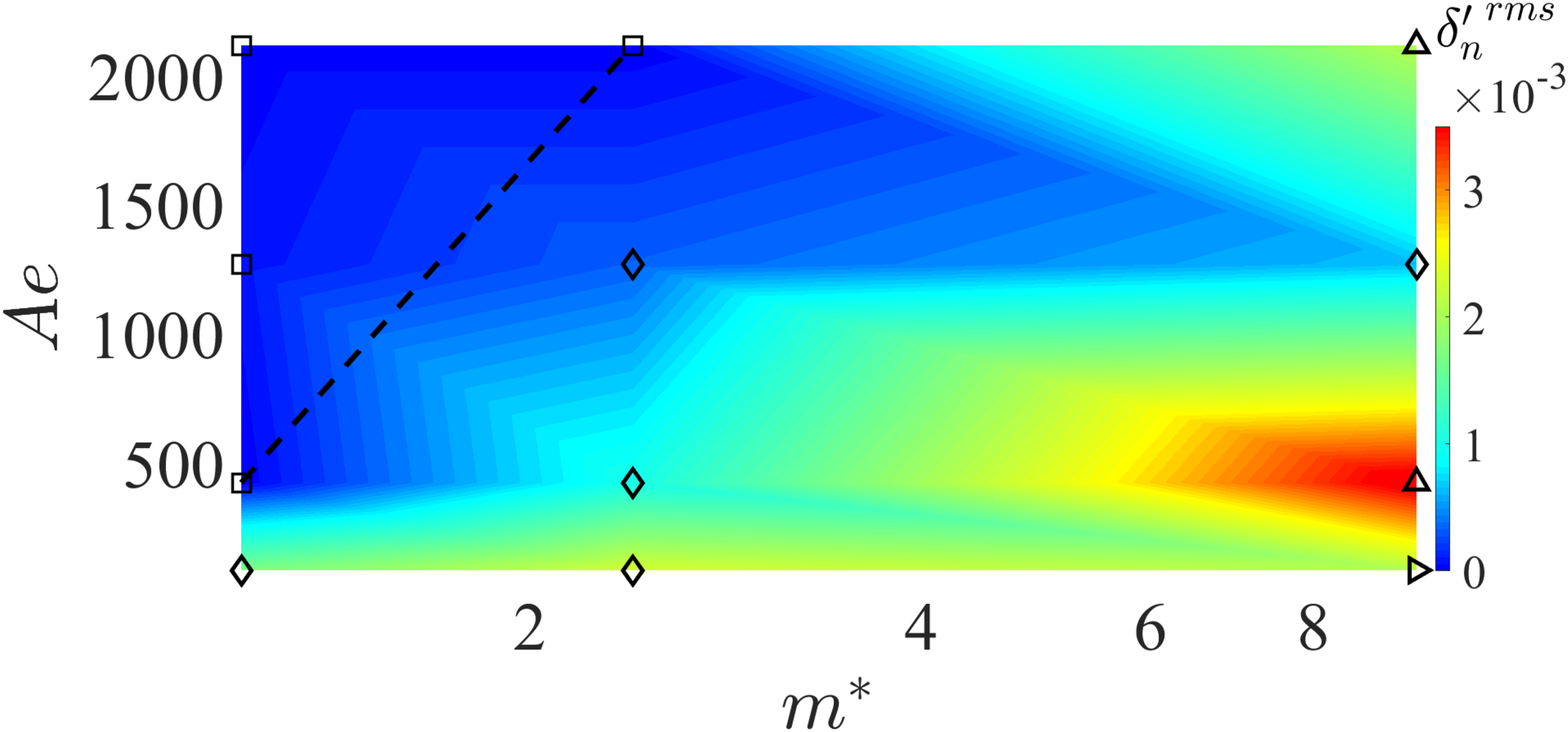}\label{phase_aed}}
	\\
	\subfloat[][]{\includegraphics[width=0.48\textwidth]{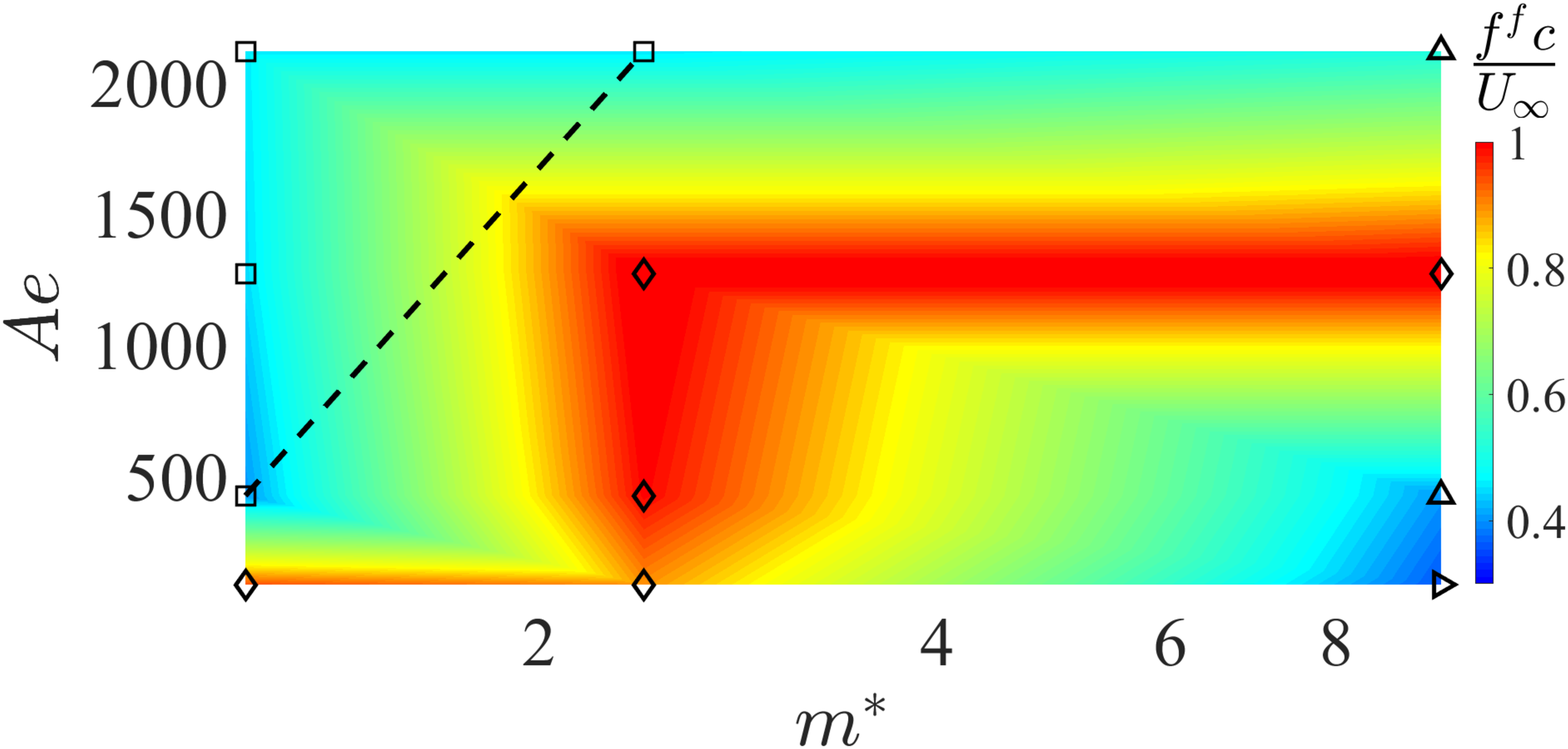}\label{phase_aee}}
	\quad
	\subfloat[][]{\includegraphics[width=0.48\textwidth]{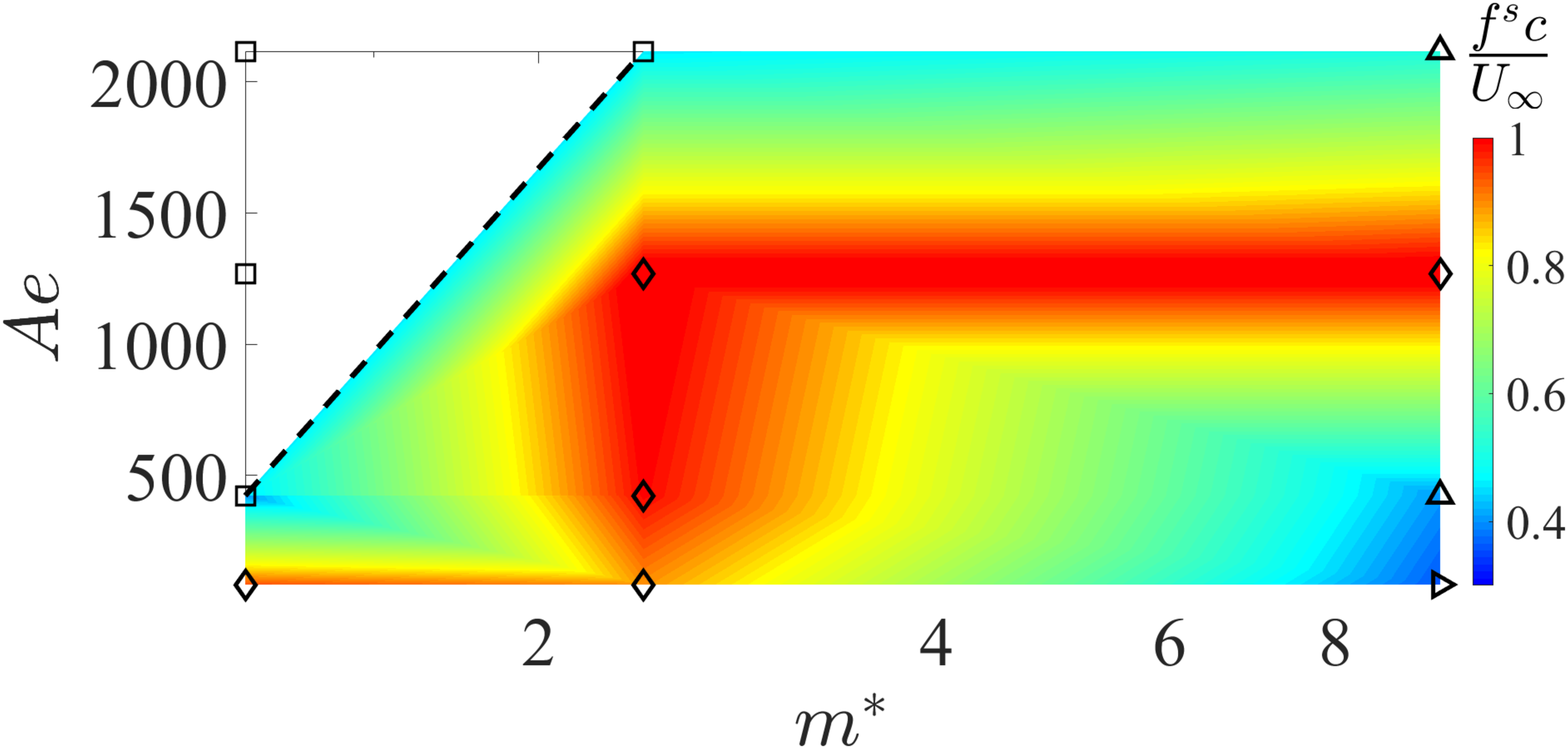}\label{phase_aef}}
	\caption{Coupled fluid-membrane dynamic characteristics of the 3D flexible membrane at $Re=24300$ for $\alpha$=$15^\circ$ over the $m^*$-$Ae$ full parameter space: (a) time-averaged lift coefficient, (b) time-averaged lift-to-drag ratio, (c) maximum time-averaged non-dimensional membrane deflection, (d) maximum r.m.s. membrane deflection fluctuation, (e) non-dimensional dominant frequency in fluid domain and (f) non-dimensional dominant frequency in structural domain via FMD approach. The blank in (f) represents the deformed-steady membrane without vibration.}
	\label{phase_ae}
\end{figure}

\refFig{phase_ae} presents the aerodynamic force characteristics and the membrane deflections as well as their fluctuation intensities and the non-dimensional dominant frequencies of the fluid and structural domains via the FMD method in the studied parameter space. We observe from \reffigs{phase_ae} \subref{phase_aea} and \subref{phase_aeb} that the vibrating membrane exhibits overall superior aerodynamic performance than the deformed-steady membrane. The optimal aerodynamic performance and the maximum membrane deflection are noticed for the lightest and most flexible membrane (M3,A1). The dominant frequency distributions summarized in \reffigs{phase_ae} \subref{phase_aee} and \subref{phase_aef} demonstrate that the frequency of the unsteady flow is synchronized with the frequency of the membrane vibration within the DBS regime. From the above observations, we find that the membrane dynamic responses exhibit rich characteristics as functions of $m^*$ and $Ae$. The effect of mass ratio on the coupled fluid-membrane dynamics has been discussed in $\S$\ref{mass}. In this section, we focus on studying the role of the membrane flexibility in the coupled fluid-membrane dynamics and the flow-excited instability for the membranes with three groups of representative mass ratios respectively.

For the light membrane with $m^*$=1.2 (M3), there is no mode transition observed in the studied aeroelastic number range. As the aeroelastic number decreases (A4$\to$1), the mean lift coefficient, the mean lift-to-drag ratio and the mean membrane displacement grow up continuously. The membrane oscillation intensity increases sharply to 0.017 at A1 when the membrane is coupled with the unsteady flow to vibrate. For the medium weight membrane with $m^*$=2.4 (M4), the membrane loses its aeroelastic static stability at higher $Ae$, compared to the light membrane. It can be inferred that a stronger inertia effect can reduce the static stability range. The membrane remains a dominant chord-wise second mode in the DBS regime as $Ae$ decreases. When the membrane becomes more flexible, the mean lift coefficient, the mean lift-to-drag ratio and the mean membrane displacement are improved. Meanwhile, the membrane vibrates more violently.

\begin{figure}
	\centering 
	\subfloat[][]{\includegraphics[width=0.49\textwidth]{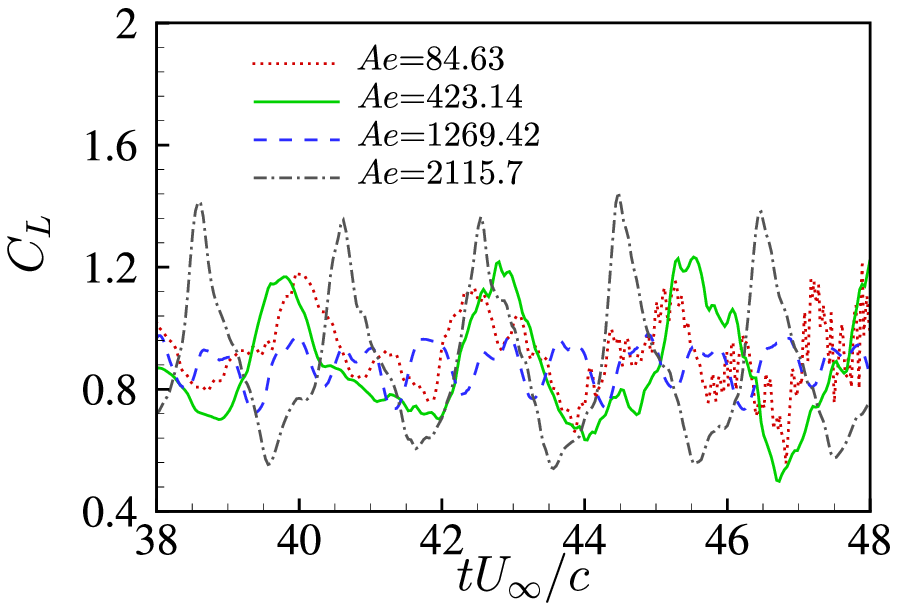}\label{response_aea}}
	\subfloat[][]{\includegraphics[width=0.49\textwidth]{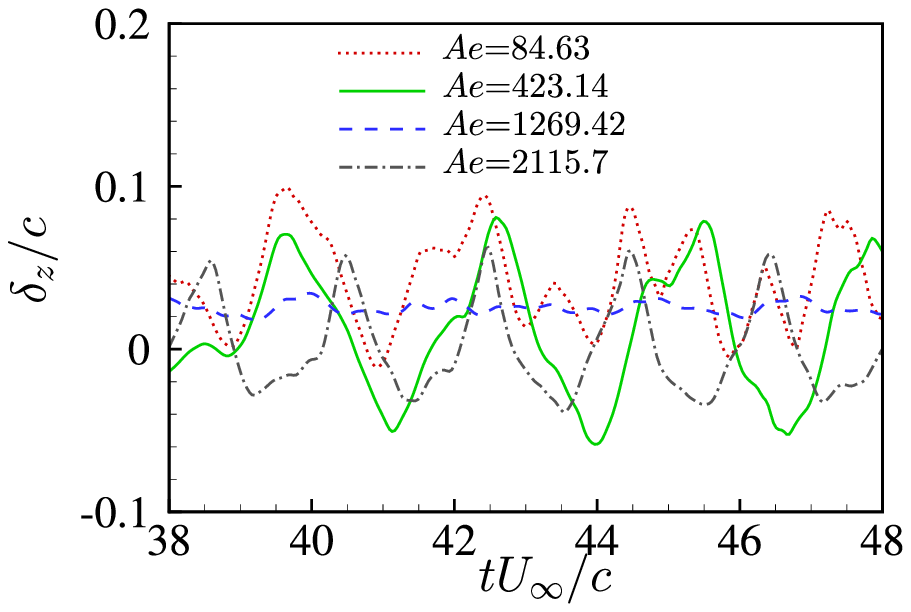}\label{response_aeb}}
	\caption{Time histories of instantaneous: (a) lift coefficients and (b) normalized displacements at the membrane centre along the $Z$-direction for four selected cases with $Ae$=84.63, 423.14, 1269.42 and 2115.7 at a fixed $m^*$=9.6.}
	\label{response_ae}
\end{figure}

\begin{figure}
	\centering 
	\subfloat[][]{\includegraphics[width=0.25\textwidth]{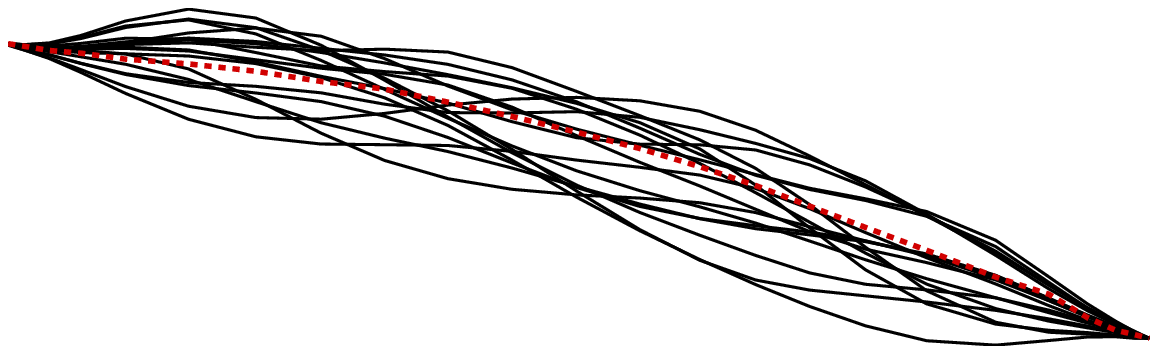}\label{membrane_vibration_aea}}
	\subfloat[][]{\includegraphics[width=0.25\textwidth]{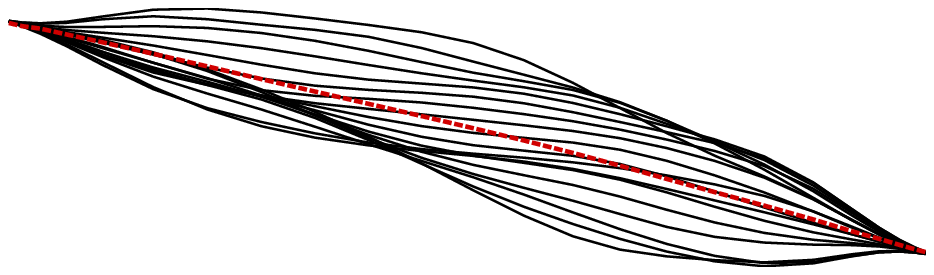}\label{membrane_vibration_aeb}}
	\subfloat[][]{\includegraphics[width=0.25\textwidth]{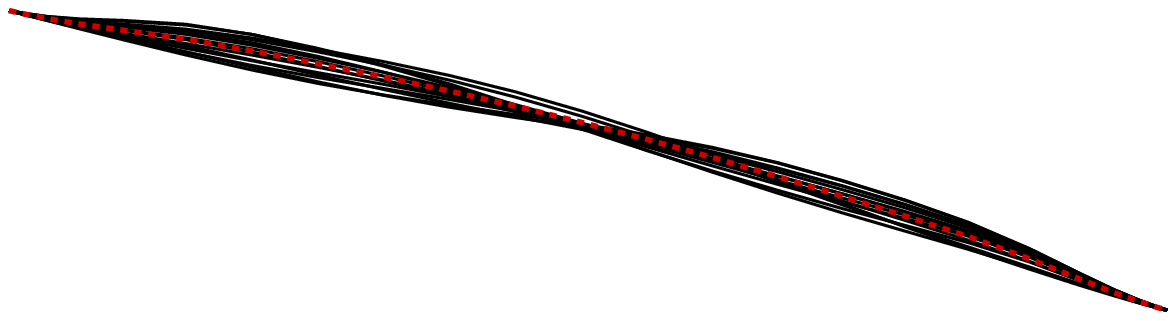}\label{membrane_vibration_aec}}
	\subfloat[][]{\includegraphics[width=0.25\textwidth]{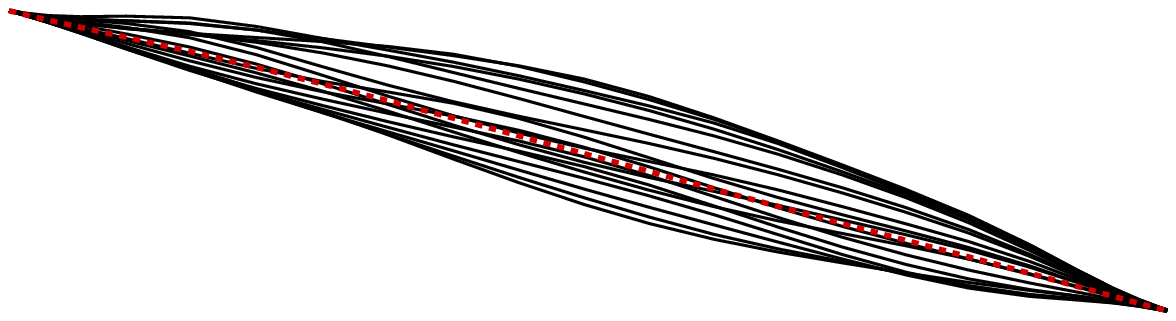}\label{membrane_vibration_aed}}
	\\
	\subfloat[][]{\includegraphics[height=0.12\textwidth,width=0.3\textwidth]{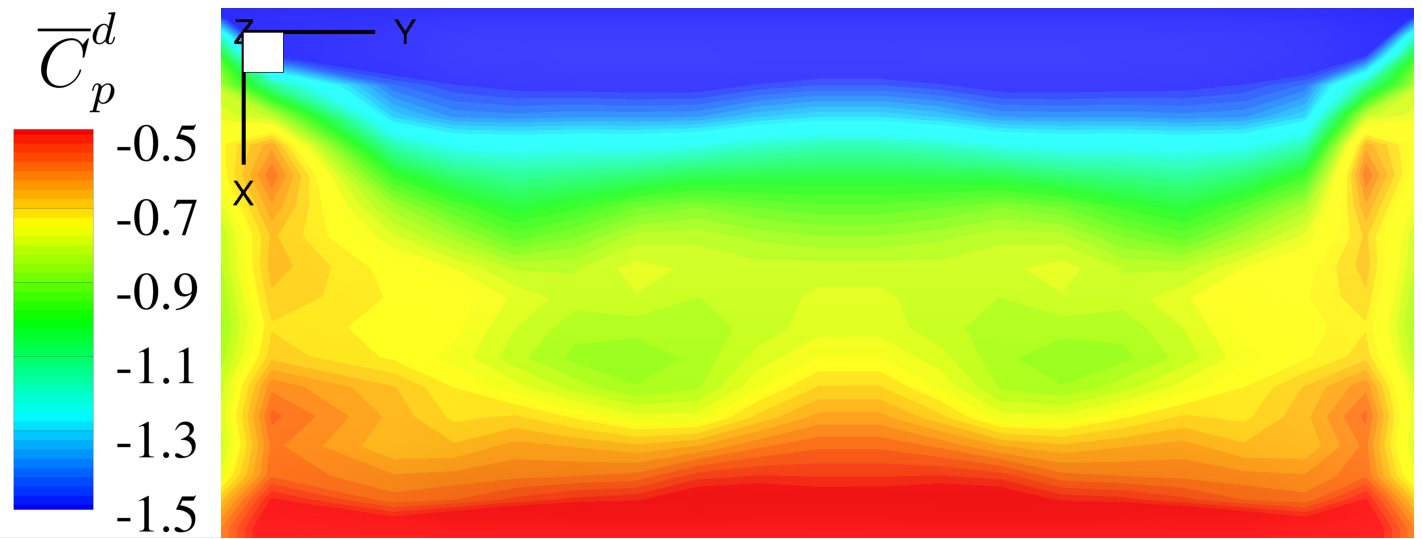}\label{cpd_aea}}
	\
	\subfloat[][]{\includegraphics[height=0.12\textwidth,width=0.22\textwidth]{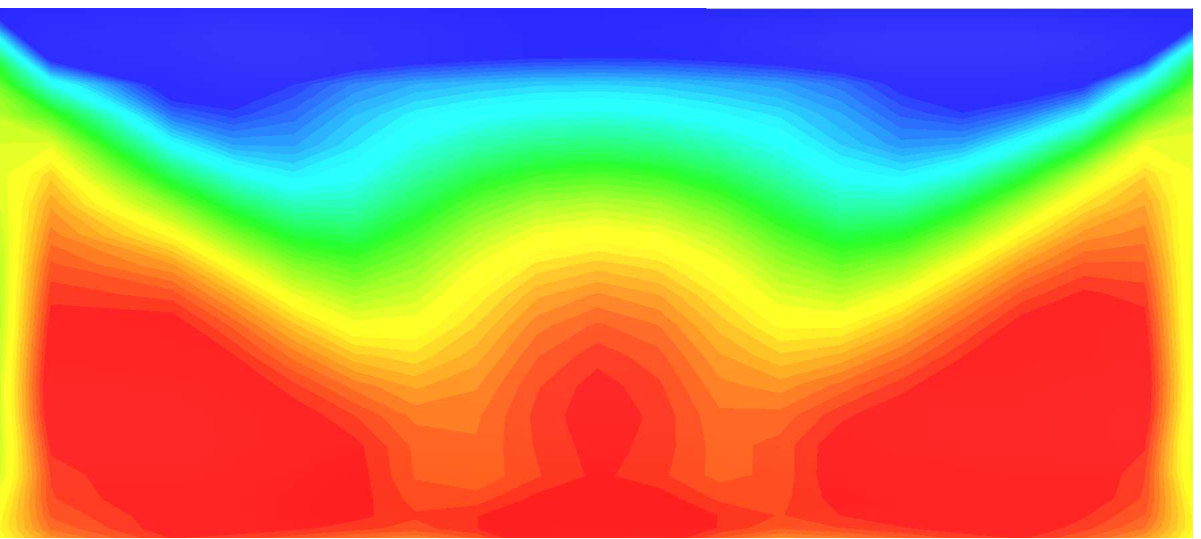}\label{cpd_aeb}}
	\
	\subfloat[][]{\includegraphics[height=0.12\textwidth,width=0.22\textwidth]{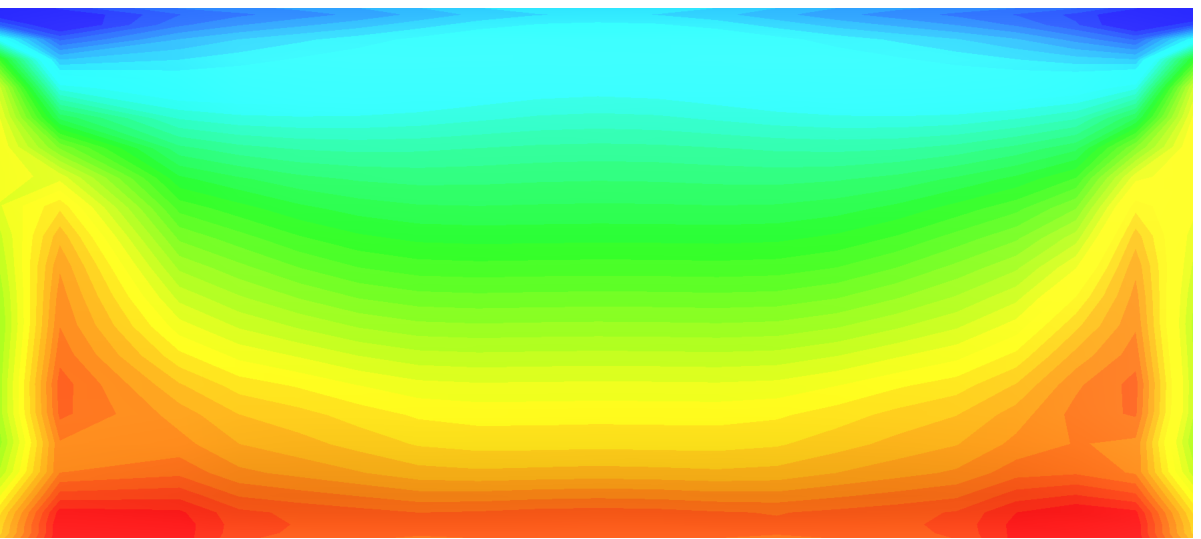}\label{cpd_aec}}
	\
	\subfloat[][]{\includegraphics[height=0.12\textwidth,width=0.22\textwidth]{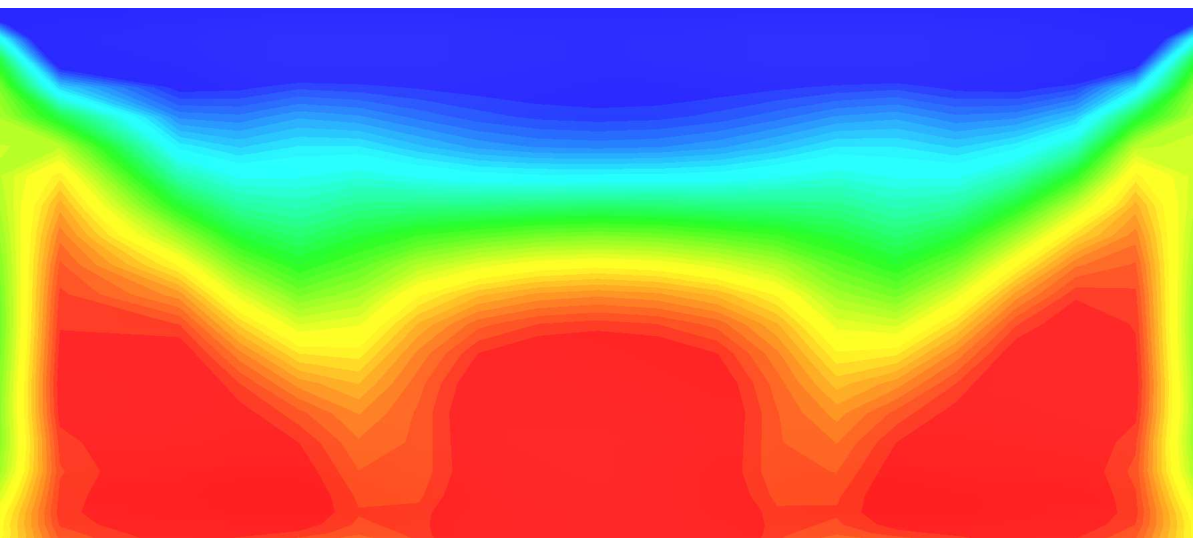}\label{cpd_aed}}
	\\
	\subfloat[][]{\includegraphics[height=0.12\textwidth,width=0.3\textwidth]{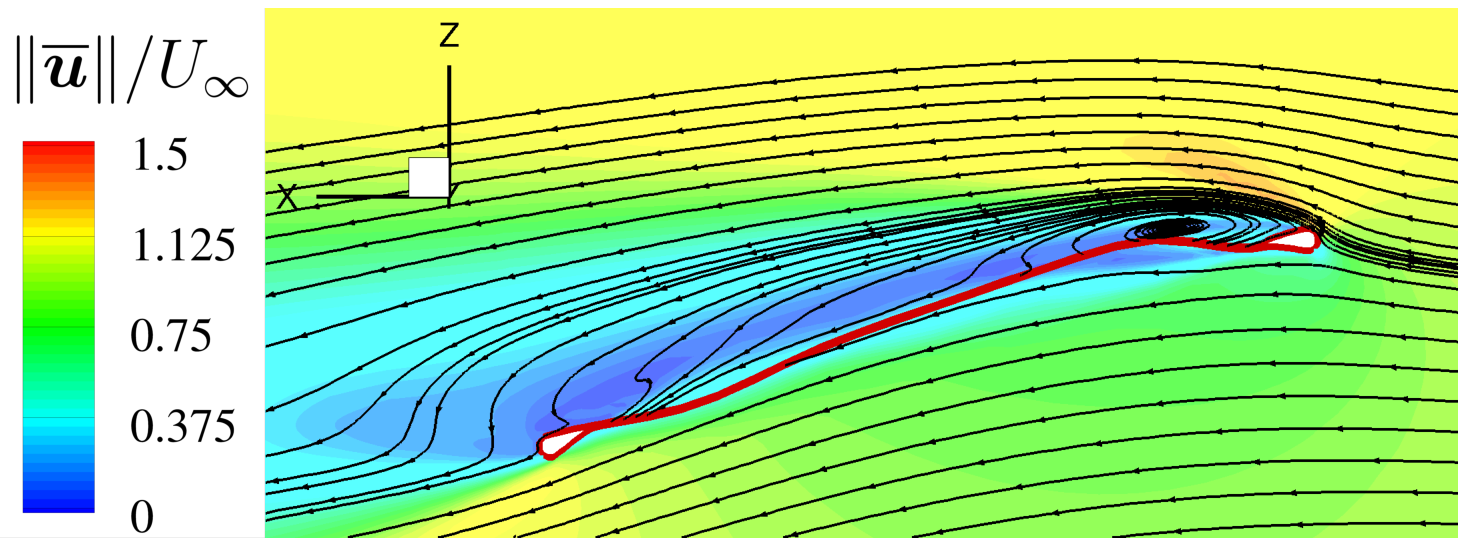}\label{streamline_aea}}
	\
	\subfloat[][]{\includegraphics[height=0.12\textwidth,width=0.22\textwidth]{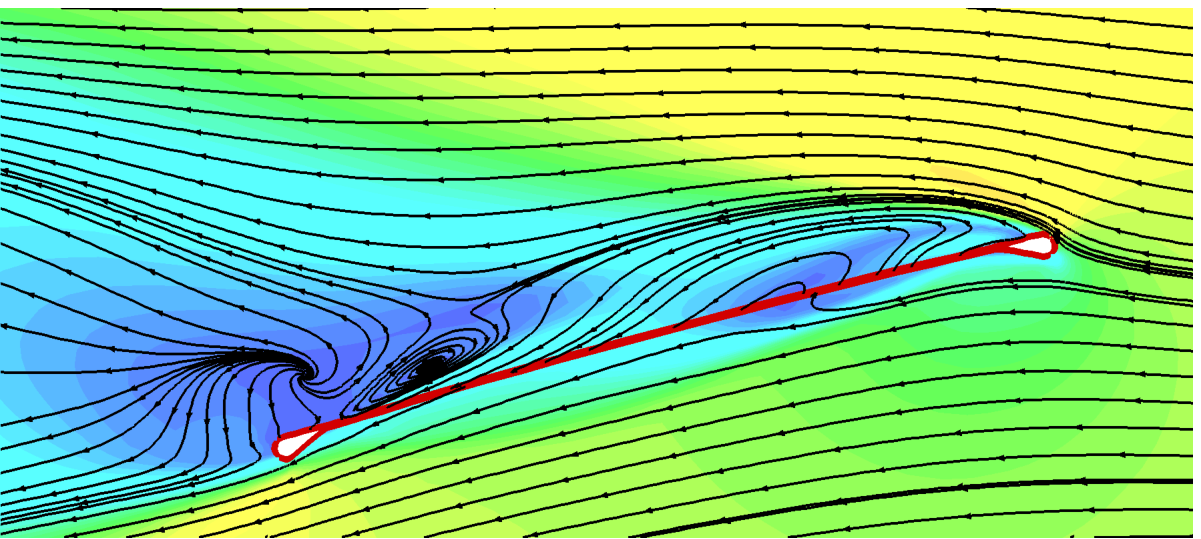}\label{streamline_aeb}}
	\
	\subfloat[][]{\includegraphics[height=0.12\textwidth,width=0.22\textwidth]{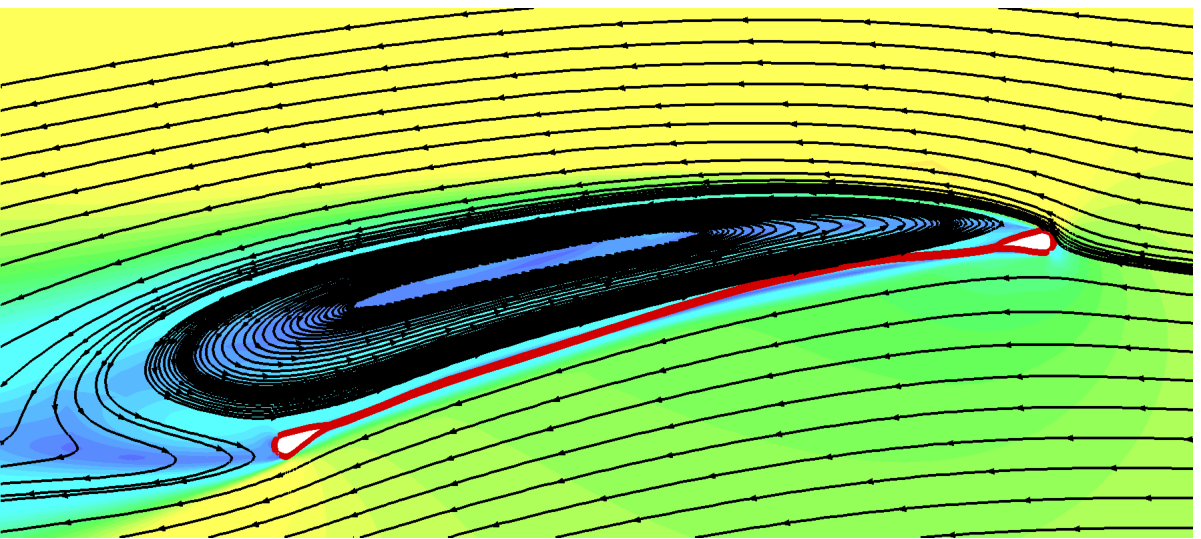}\label{streamline_aec}}
	\
	\subfloat[][]{\includegraphics[height=0.12\textwidth,width=0.22\textwidth]{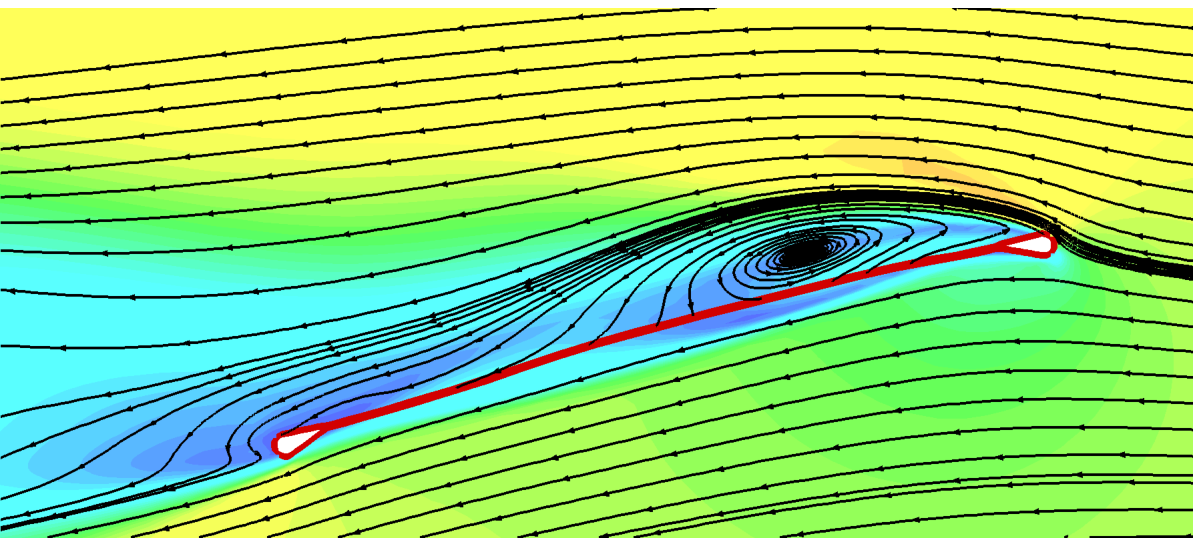}\label{streamline_aed}}
	\caption{Flow past a 3D rectangular membrane wing: (a,b,c,d) full-body responses of 3D flexible membrane wing at mid-span location, (e,f,g,h) time-averaged pressure coefficient difference between the upper and lower surfaces, (i,j,k,l) time-averaged streamlines and normalized velocity magnitude on the mid-span plane at $(Ae,m^*)=$ (a,e,i) $(84.63,9.6)$, (b,f,j) $(423.14,9.6)$, (c,g,k) $(1269.42,9.6)$, (d,h,l) $(2115.7,9.6)$. ($\color{red}{---}$) denotes the time-averaged membrane shape in (a,b,c,d).}
	\label{membrane_response_ae}
\end{figure}

Different from the dynamical response of the light membrane and the medium weight membrane, the dominant mode transitions between different structural modes in the studied $Ae$ space for the heavy membrane with $m^*$=9.6 (M6). It is observed from \reffig{phase_ae} that the aerodynamic forces and the membrane deflection fluctuate with the change of $Ae$. The overall largest mean lift coefficient and mean lift-to-drag ratio are achieved for the most flexible membrane. \refFig{response_ae} plots the time histories of the instantaneous lift coefficient and the instantaneous vertical displacement at the membrane centre at different $Ae$ values. Both the time-varying lift coefficient and the time-varying membrane displacement exhibit variations of the amplitude and the dominant frequency as a function of $Ae$. \refFig{membrane_response_ae} summarizes the membrane vibration responses, the pressure difference distributions and the flow features of the heavy membrane at different aeroelastic numbers. Both instantaneous chord-wise first and second modes are observed from the full-body profile responses of the most flexible membrane (A1) in \reffig{membrane_response_ae} \subref{membrane_vibration_aea}. The dominant mode transitions to the chord-wise first mode in \reffig{membrane_response_ae} \subref{membrane_vibration_aeb} when the aeroelastic number changes to A2. The mean wing camber reduces and the vibration amplitude grows up after the mode transition occurs. In \reffig{membrane_response_ae} \subref{membrane_vibration_aec}, the membrane vibrates in the chord-wise second mode with reduced vibration intensity at $Ae=1269.42$ (A3). For the most rigid membrane (A4), the chord-wise first mode becomes the dominant mode in the coupled system. Corresponding to the variation of the membrane vibrations, the flow features are also affected significantly at different dominant modes. A suction area is observed at the leading edge in \reffig{membrane_response_ae} \subref{cpd_aea}, which is related to the leading edge vortex noticed in \reffig{membrane_response_ae} \subref{streamline_aea}. The high suction area in blue colour at the leading edge reduces slightly and the low suction region in red colour near the trailing edge expands to the leading edge, which is caused by the decreased membrane camber and the wake pattern. We notice that the high suction area in the proximity of the leading edge reduces dramatically in \reffig{membrane_response_ae} \subref{cpd_aec}. Meanwhile, a large vortex is formed on the whole membrane surface in \reffig{membrane_response_ae} \subref{streamline_aec}. Similar pressure distributions and flow features are observed between the membranes vibrating in the chord-wise first mode at $Ae=423.14$ and 2115.7, resulting in similar mean lift forces.

By exploring the coupled fluid-membrane dynamics as functions of mass ratio, Reynolds number and aeroelastic number, we can see that the flexible membrane gets coupled with the unsteady separated flow and vibrates when the relevant physical parameters exceed the flow-excited instability boundary. Besides, the transition of the dominant mode from a specific mode shape to another mode shape is dependent on the variation in the corresponding modal frequency. There should be an underlying process that dictates the onset of the membrane vibration and the mode transition through certain combinations of the physical parameters. In the following section, the underlying process is explored based on the investigation of the local bluff-body vortex shedding instability and the nonlinear natural frequency of the tensioned membrane in the studied parameter space.

\subsection{Onset of flow-induced membrane vibration} \label{onset}
From the aeroelastic mode analysis at different physical parameters, one can see that the frequency of the fluid Fourier mode jumps to the membrane frequency once the self-sustained vibration is established. The onset of the membrane vibration is determined by the fluid-structural parameters (e.g., $m^*$, $Re$, $Ae$, and AOA) and the underlying coupled dynamics phenomena such as the vortex shedding, the shear layer fluctuation, the flow-induced tension. It is shown earlier that the membrane can maintain a constant deformed shape in the DSS regime regardless of $m^*$ when other physical parameters remain fixed. The tension force and the flow features of the stretched steady membrane are not affected by the mass ratio. Among the physical parameters related to the fluid-membrane interaction, only the natural frequency of the membrane is varied as a function of $m^*$ when the membrane remains a constant cambered shape. Once the mass ratio exceeds the critical value, the membrane starts to vibrate through the frequency lock-in with the vortex shedding process. A natural question is to ask whether the natural frequency of the tensioned membrane plays an important role in the onset of the membrane vibration and how it affects the flow-excited instability. The variation of $m^*$ in the coupled fluid-membrane system provides a good starting point to explore the relationship between the natural frequency and the flow-induced vibration. To generalize our finding, the variation of the natural frequency and the flow-induced vibration at different Reynolds numbers and aeroelastic numbers is further studied.

\begin{figure}
	\centering 
	\subfloat[][]{\includegraphics[width=0.9\textwidth]{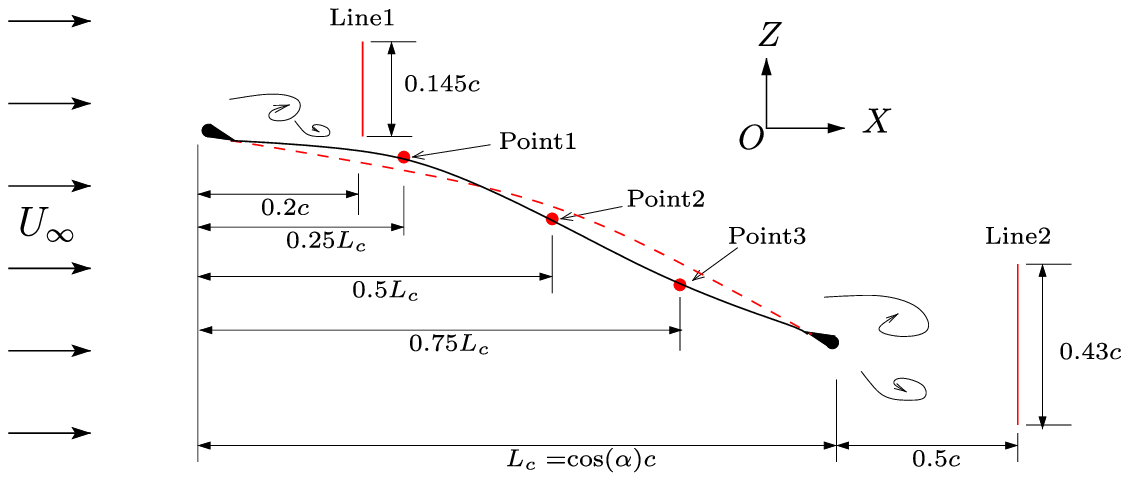}}
	\caption{Schematic of the positions of the monitoring lines and points. The red dashed line ({\textbf{\color{red}- - -}}) indicates the time-averaged membrane shape. The black solid line ({\textbf{\color{black}---}}) represents the instantaneous membrane shape.}
	\label{probe}
\end{figure}

To shed light on the onset of the membrane vibration, the local vortex shedding frequency along a monitoring line near the trailing edge in the wake is calculated to compare with the natural frequency of the membrane in this section. \refFig{probe} presents a schematic of the positions of the monitoring lines and points. The monitoring line for measuring the local frequency of the leading edge vortex is placed $0.2c$ behind the leading edge. The monitoring line corresponding to the trailing edge vortex frequency measurement is located at the half-chord behind the trailing edge. These two vertical lines are chosen at the location which can collect the local flow information significantly. Three monitoring points are placed uniformly on the membrane surface along the membrane chord to collect the local membrane dynamic responses. To understand the mode transition phenomenon, the cross-correlation analysis between the flow fluctuations along the monitoring lines and the membrane vibrations at three points is conducted in the next section.

\begin{figure}
	\centering 
	\includegraphics[width=0.9\textwidth]{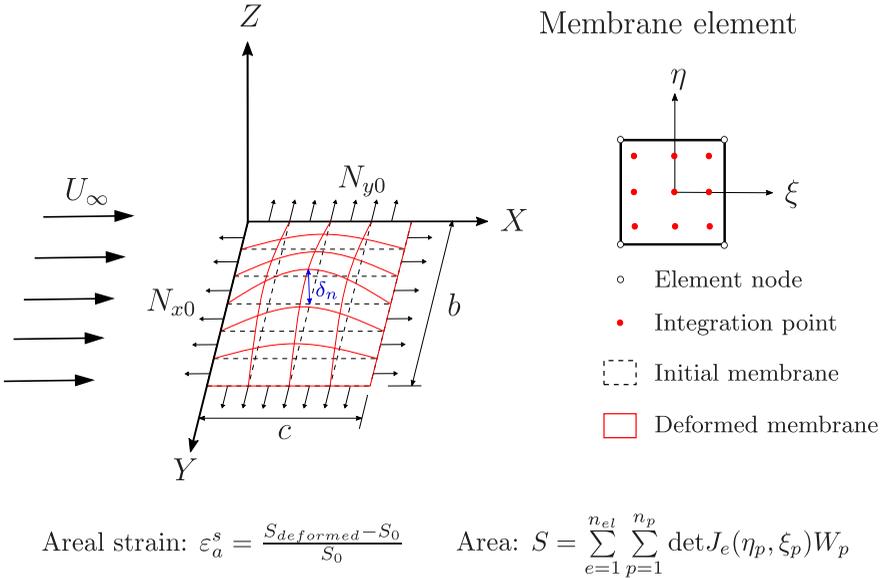}
	\caption{Schematic of a rectangular membrane with all fixed edges undergoing large amplitude vibration in a uniform flow. A deformed shape and the Gaussian quadrature procedure are illustrated.}
	\label{membrane_strain}
\end{figure}

Similar to \cite{kamm2015molecular}, the areal strain is employed to quantize the whole surface distortions under aerodynamic loads. The areal strain is defined as the ratio of the surface area change caused by the flow-induced deformation to the initial surface area, which is given as
\begin{equation}
\varepsilon^s_a=\frac{{S}_{deformed}-S_0}{S_0},
\label{strain}
\end{equation}
where ${S}_{deformed}$ denotes the surface area of the deformed membrane and $S_0$ is the initial surface area of the membrane without deformation. A schematic of a rectangular membrane as well as its deformed shape and the Gaussian quadrature is plotted in \reffig{membrane_strain}. The surface area is calculated by integrating the finite element area using the Gaussian quadrature over the whole surface, which is given as
\begin{equation}
S = \sum\limits_{e=1}^{n_{el}}\sum\limits_{p=1}^{n_p} {\rm{det}}J_e(\eta_p,\xi_p)W_p,
\label{area_cal}
\end{equation}
where $n_{el}$ and $n_p$ represent the number of the membrane elements and the Gauss integration points. $\eta_p$ and $\xi_p$ are the local coordinates of the $p$-th Gauss node. $J_e$ denotes the jacobian of the $e$-th membrane element and $W_p$ is the Gauss weight. The membrane dynamic responses can be divided into two components: (a) mean deformed shape under aerodynamic loads and (b) vibration around its mean position. The areal strain of the mean shape ($\overline{\varepsilon}^s_a$) reflects the stretch ratio of the mean cambered membrane to its initial shape, which can be equivalent to the initial stretching strain applied to the membrane edges. The root-mean-squared value of the areal strain fluctuation (${\varepsilon^s_a}^{\prime rms}$) is regarded as the vibration intensity of the oscillating membrane.

To gain further insight into the onset of the membrane vibration, it is important to calculate the natural frequency of the membrane with large amplitude vibration immersed in an unsteady flow. The natural frequency predicted based on the linear model expressed in \refeq{linear_frequency} neglects the dynamic stress caused by the geometric nonlinearity of the vibrating membrane. Furthermore, the added mass due to the oscillating surrounding air is missed in the equation. We derive a nonlinear dynamic equation of the vibrating rectangular membrane shown in \reffig{membrane_strain} based on a large deflection theory and consider an analytical aerodynamic pressure model as the external loads. Initial tensions along the $X$-axis and $Y$-axis, namely $N_{x0}^s$ and $N_{y0}^s$, are applied along the edges. An approximate analytical formula of the nonlinear natural frequency for the simply supported rectangular membrane can be obtained from the nonlinear partial differential equation by Poincar$\rm{\acute{e}}$-Lindstedt perturbation method. The detailed derivation process of the nonlinear natural frequency formula can be found in Appendix C. The nonlinear natural frequency $f_{ij}^n$ of a vibrating membrane immersed in an unsteady flow with the chord-wise $i$ and span-wise $j$ mode is expressed as
\begin{equation}
f_{ij}^n = \frac{1}{2 \pi} \left( \omega_{ij0} + \frac{3 \beta \kappa \omega_{ij0}}{8}(g_0^2 + h_0^2) \right),
\label{nonlinear_frequency}
\end{equation}
where $\omega_{ij0}$ is the circular frequency corresponding to the initial tensions applied to the membrane. $\beta$ denotes the perturbation parameter in the Poincar$\rm{\acute{e}}$-Lindstedt perturbation method and $\kappa$ represents the coefficient of the vibration equation. $g_0$ and $h_0$ are the initial conditions of the membrane displacement and velocity of the vibrating membrane. The expression of each parameter is given in Appendix C. In \refeq{nonlinear_frequency}, the second term is caused by the geometric nonlinearity and is related to the membrane vibration amplitude. The dynamic stress caused by the membrane transverse displacement can increase the natural frequency. The added mass $m^{am}$ makes the natural frequency smaller than that in a vacuum. \cite{jaiman2014added} presented a theoretical study of the flow-excited instability of an elastic plate oscillating in fluid. The critical $Re$ showed a downward trend with increasing mass ratio, which is consistent with our conclusion for the 3D membrane. Similar to the theoretical study in \cite{jaiman2014added}, the above analytical formulation can be extended to construct the stability boundaries as a function of tension force and mass ratio which can be a scope for future study.

The local non-dimensional frequencies of the dominant vortex and the secondary vortices are calculated based on the Fourier analysis of the collected time-varying vorticity signals along the monitoring Line2. These frequencies reflect the frequency of the aerodynamic load perturbations applied to the flexible membrane. The natural frequency of the tensioned membrane is predicted by \refeq{nonlinear_frequency}. The initial tensions $N_{x0}^s$ and $N_{y0}^s$ along two directions can be equivalent to the tensions of the mean deformed membrane shape. The physical parameters required for the prediction of the added mass can be determined from the numerical simulations. The vibration frequency, the vibration amplitude and the related parameters in the second nonlinear term in \refeq{nonlinear_frequency} are obtained from the numerical simulations. In the current study, the natural frequencies of the first and second modes along the chord-wise and span-wise directions (C1S1, C1S2, C2S1 and C2S2) are calculated because the excited dominant structural modes are close to these four modes in the examined cases. 

\begin{figure}
	\centering
	\subfloat[][]{\includegraphics[width=0.48\textwidth]{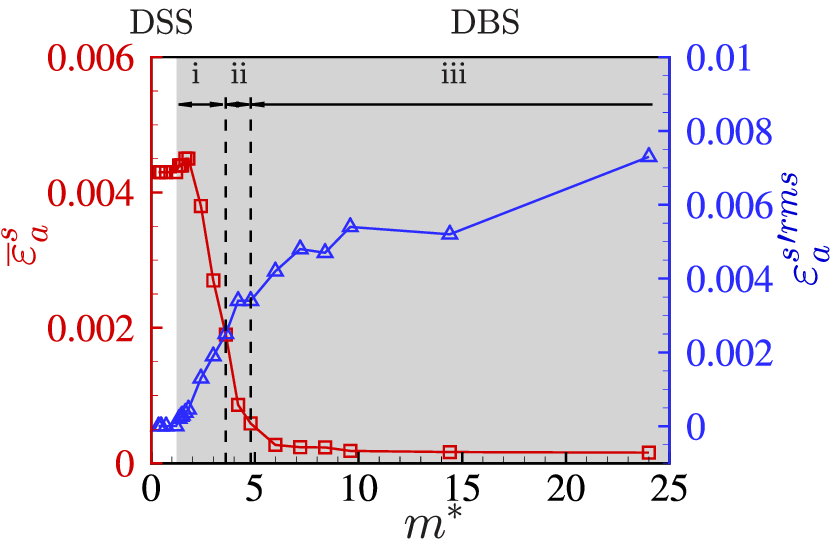}\label{areal_straina}}
	\quad
	\subfloat[][]{\includegraphics[width=0.48\textwidth]{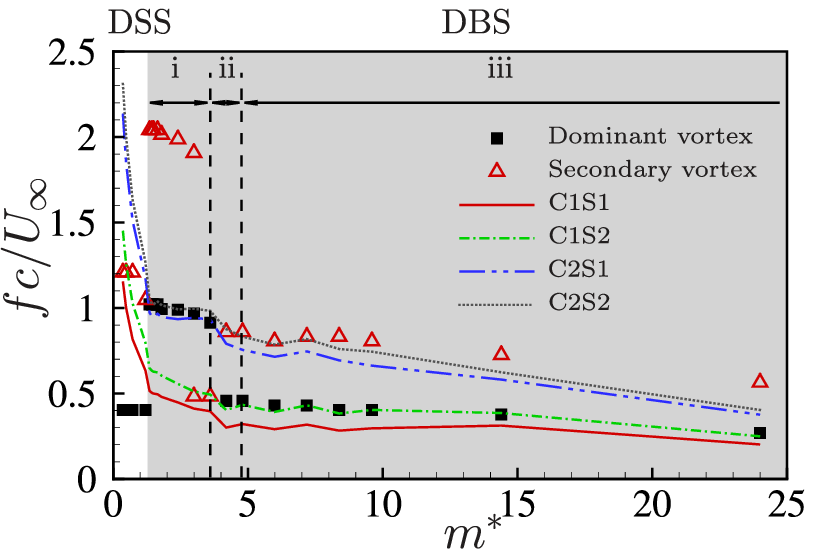}\label{mode_migration3a}}
	\\
	\subfloat[][]{\includegraphics[width=0.48\textwidth]{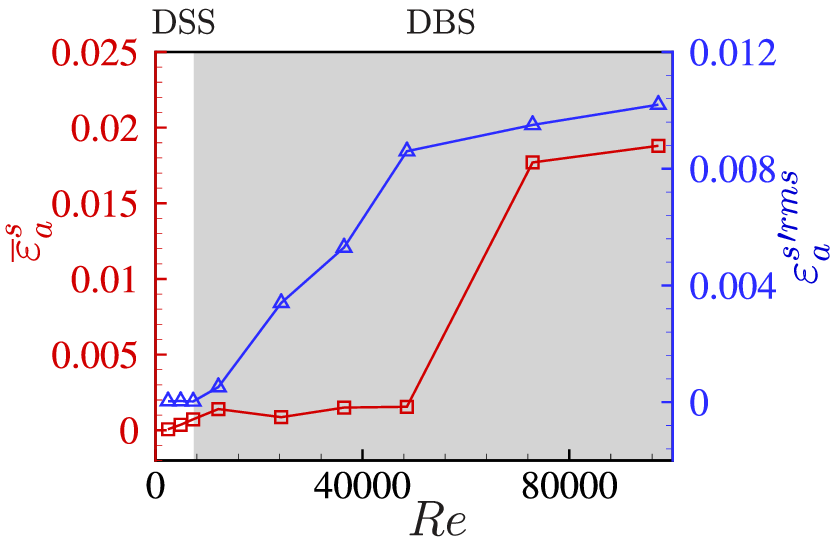}\label{areal_strainb}}
	\quad
	\subfloat[][]{\includegraphics[width=0.48\textwidth]{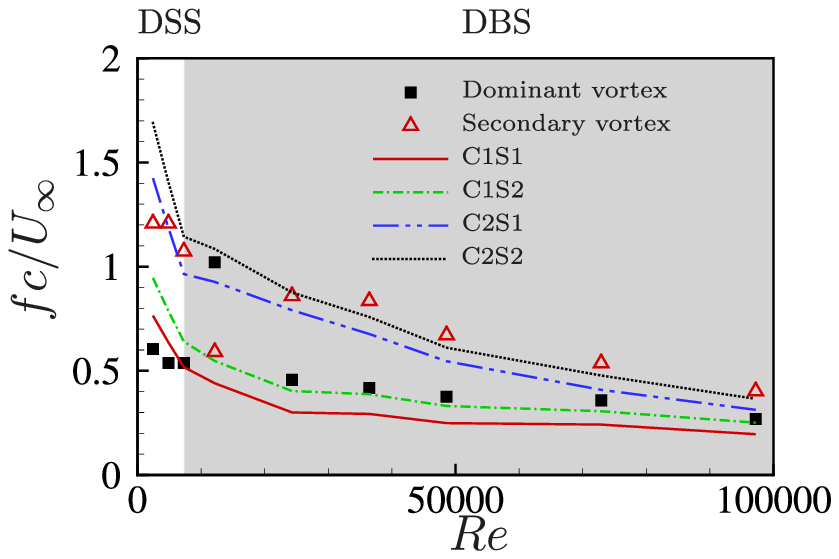}\label{mode_migration3b}}
	\\
	\subfloat[][]{\includegraphics[width=0.48\textwidth]{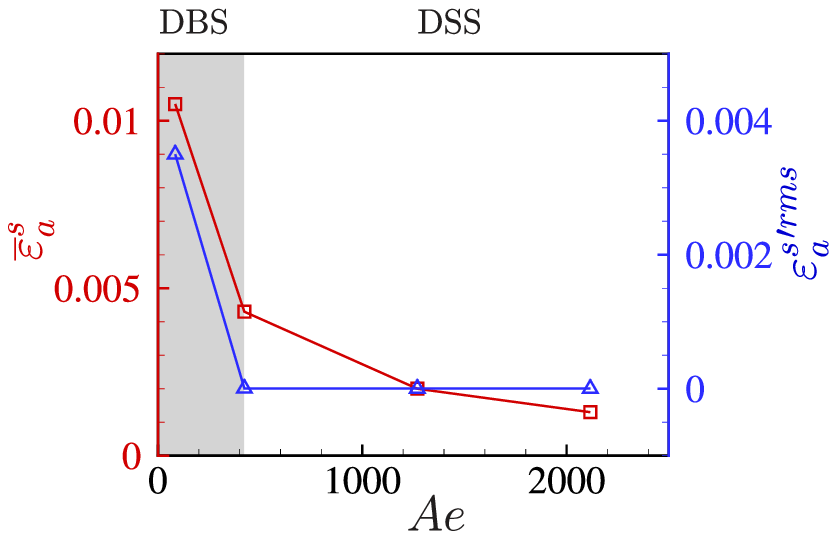}\label{areal_strainc}}
	\quad
	\subfloat[][]{\includegraphics[width=0.48\textwidth]{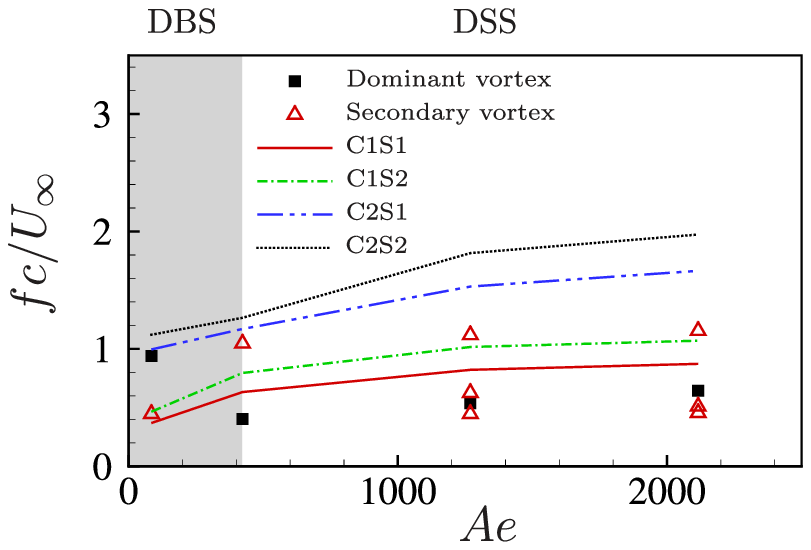}\label{mode_migration3c}}
	\caption{Areal strain of mean membrane shape and r.m.s. of areal strain (a,c,e) and comparison of natural frequencies of tensioned membrane and vortex shedding frequencies measured along Line2 (b,d,f) as a function of (a,b) $m^*$ at fixed $(Re,Ae)$=(24300,423.14), (c,d) $Re$ at fixed $(m^*,Ae)$=(4.2,423.14) and (e,f) $Ae$ at fixed $(m^*,Re)$=(1.2,24300). The areal strain is calculated by integrating the finite element area from the structural responses. The nonlinear natural frequency is predicted by \refeq{nonlinear_frequency} with the aid of numerical simulations.}
	\label{mode_migration3}
\end{figure}

\refFig{mode_migration3} \subref{areal_straina} presents the calculated areal strain of the membrane as a function of $m^*$. The comparison of the vortex shedding frequencies and the nonlinear natural frequencies of the tensioned membrane as a function of $m^*$ is plotted in \reffig{mode_migration3} \subref{mode_migration3a}. It can be observed that $\overline{\varepsilon}^s_a$ remains constant and ${\varepsilon^s_a}^{\prime rms}$ keeps zero in the DSS regime. The second term in \refeq{nonlinear_frequency} related to the membrane vibration and the dynamic stress vanishes. Moreover, there is no added mass for the static steady membrane. Thus, the natural frequency model of the statically deformed membrane can be reduced to \refeq{linear_frequency}. Since the tension force keeps constant in this regime, the non-dimensional natural frequency of each mode is only dependent on $m^*$, which gives a relationship as below
\begin{equation}
\frac{f^n_{ij} c}{U_{\infty}} \sim \sqrt{\frac{1}{m^*}}.
\label{frequency_mass}
\end{equation}

In the DSS regime, the membrane shape maintains a fixed static equilibrium position. Thus, the flow features around the membrane keep similar and exhibit constant vortex shedding frequency regardless of $m^*$. The natural frequency of the membrane decreases when the membrane becomes heavier. It is observed from \reffig{mode_migration3} \subref{mode_migration3a} that the structural natural frequency decreases and gets close to the harmonics of the dominant vortex shedding frequency near the flow-excited instability boundary. The flexible membrane is coupled with the vortex shedding process to vibrate. Meanwhile, the dominant vortex shedding frequency jumps from a lower frequency component associated with the bluff-body vortex shedding frequency to a frequency component locked into the membrane vibration. When the membrane is coupled with the separated flow to vibrate in the DBS regime, $\overline{\varepsilon}^s_a$ exhibits an overall downward trend and ${\varepsilon^s_a}^{\prime rms}$ shows an opposite trend as the inertia effect of the thin structure becomes more significant. It can be observed from \reffig{mode_migration3} \subref{mode_migration3a} that the dominant vortex shedding frequency locks into the membrane vibration with the chord-wise second mode in the range of $m^* \in [1.2,3.6]$. The dominant aeroelastic mode transitions to the chord-wise first mode when the natural frequency of the C1S2 mode becomes close to the dominant frequency of the bluff-body vortex shedding frequency in the transitional mode regime. These approaching natural frequencies in the fluid and structural domains form a new frequency synchronization state through the coupling effect. The common mode transition phenomenon at various physical parameters will be uniformly discussed in $\S$\ref{mode_transition}.

The variations of the areal strain and the nonlinear natural frequency of the coupled system as a function of $Re$ are presented in \reffigs{mode_migration3} \subref{areal_strainb} and \subref{mode_migration3b}, respectively. Similar to \cite{jaiman2014added}, we observe from the numerical simulations that the deformation-induced tension is a function of $Re$, given as $N^s \sim Re^n$, ($0<n<2$). The predicted non-dimensional natural frequency decreases with increasing $Re$ and fixed $m^*$ and $Ae$ within the DSS regime
\begin{equation}
\frac{f^n_{ij} c}{U_{\infty}} \sim \frac{1}{Re} \sqrt{N^s} \sim  Re^{-q}, \quad {\rm{where}} \quad q=1-n/2 \quad (0<q<1).
\label{frequency_re}
\end{equation}

The relationship between the non-dimensional natural frequency and $Re$ given in \refeq{frequency_re} shows a similar conclusion reported in \cite{song2008aeromechanics}. The areal strain of the membrane exhibits an upward trend in the DSS regime when $Re$ increases. Although the membrane slightly deforms up in this regime, the non-dimensional dominant vortex shedding frequency at various $Re$ falls into the range of the modified bluff-body vortex shedding frequency $(f^{vs} c \sin \alpha) / U_{\infty} \in [0.15, 0.2]$ summarized by \cite{rojratsirikul2011flow}. According to \refeq{frequency_re}, the non-dimensional natural frequency of the membrane in the DSS regime becomes smaller at a higher $Re$. Once the natural frequency gets close to the harmonics of the vortex shedding frequency, the membrane starts to vibrate. As $Re$ further increases, the areal strain of the mean shape grows up slightly, and then rapidly increases at $Re$=48600 within the DBS regime. Meanwhile, ${\varepsilon^s_a}^{\prime rms}$ increases continuously. Under the influence of the increasing $Re$, $\overline{\varepsilon}^s_a$ and ${\varepsilon^s_a}^{\prime rms}$, the non-dimensional natural frequency of the vibrating membrane keeps decreasing finally. The dominant vortex shedding frequency transitions to the lower component and locks into the reduced structural natural frequency.

The variations of the areal strain and the nonlinear natural frequency of the coupled system as a function of $Ae$ are presented in \reffigs{mode_migration3} \subref{areal_strainc} and \subref{mode_migration3c}, respectively. Based on the simulation results within the DSS regime, the relationship between the areal strain and the aeroelastic number can be determined as $\overline{\varepsilon}^s_a \sim Ae^{-n}$, ($0<n<1$). As $m^*$ and $Re$ keep fixed, the non-dimensional natural frequency can be expressed as
\begin{equation}
\frac{f^n_{ij} c}{U_{\infty}} \sim \sqrt{Ae \cdot \overline{\varepsilon}^s_a} \sim  Ae^{q}, \quad {\rm{where}} \quad q=(1-n)/2 \quad (0<q<0.5).
\label{frequency_ae}
\end{equation}

In \reffig{mode_migration3} \subref{areal_strainc}, $\overline{\varepsilon}^s_a$ increases as $Ae$ reduces within the DSS regime. According to \refeq{frequency_ae}, the non-dimensional natural frequency of the membrane within this regime exhibits a downward trend at a lower $Ae$. The membrane vibration is excited when the structural natural frequency approaches the harmonics of the vortex shedding frequency and the frequency lock-in is formed. Meanwhile, the areal strain of the mean membrane and its oscillating intensity increase dramatically when the membrane vibration occurs at the lowest $Ae$.

From the aforementioned analysis, we can infer that the frequency lock-in between the structural natural frequency and the vortex shedding frequency governs the self-sustained membrane vibration. Although the vortex originating from the leading edge convects downwards and sheds into the wake as shown in \reffig{stream_re} \subref{stream_rea}, the membrane still maintains a deformed-steady state at ($Re,m^*$)=(2430,4.2). At this condition, the structural natural frequency is far away from the vortex shedding frequency and its harmonics. Meanwhile, the flow fluctuations in the shear layer are far away from the membrane surface. Through a proper combination of physical parameters, the natural frequency of the membrane can get closer to the dominant vortex shedding frequency or its harmonics. The membrane is coupled with the unsteady separated flow and builds up the self-sustained vibration via frequency lock-in. It can be easily inferred from \reffig{mode_migration3} that the tension force due to the flow-induced deformation affects the structural natural frequency. Specifically, increasing tension force can result in a higher structural natural frequency and vice versa. Therefore, the tension force is a key to connect the coupled dynamic characteristics of the fluid flow and the flexible membrane. For an active control of the frequency lock-in, stretching or relaxing the membrane can suppress or excite the membrane vibration by tuning the natural frequency of the membrane.

\subsection{Mode transition in flow-induced vibration} \label{mode_transition}
Through the investigation of the flow-induced vibration characteristics, the mode transition from one specific aeroelastic mode to another mode is observed in \reffig{mode_map} and \reffig{mode_map_ae_mass} as the physical parameters change. The mode transition can impact the aerodynamic performance, the flow features around the membrane and the membrane vibration response. It is vital to understand the mode transition phenomenon in the coupled system. The results shown in $\S$\ref{onset} demonstrate that the onset of the membrane vibration is dependent on the frequency lock-in phenomenon. In this section, we explore the mode transition from the perspective of the frequency lock-in phenomenon. Specifically, the variation of the structural natural frequency relative to the vortex shedding frequency is investigated and the correlation with the dominant aeroelastic modes is established.

In \reffig{mode_migration3} \subref{mode_migration3a}, the natural frequency of the C1S2 mode becomes closer to the dominant frequency of the bluff-body vortex shedding process at $m^*=2.99$. As $m^*$ further increases, the dominant mode gradually transitions from the C2S1 mode to the C1S2 mode. As shown in \reffig{mode_migration2}, the mode energy corresponding to the C1S2 mode exceeds that of the C2S1 mode. It can be observed from \reffig{membrane_state_stream} that the size of the dominant vortex enlarges as $m^*$ reaches the transitional region accompanied by the reduced mean camber and the growing vibration amplitude. The variation of these membrane dynamics enhances the bluff-body vortex shedding process. The flow fluctuations related to the bluff-body vortex shedding instability become more energetic in the coupled system, which is coupled with the structural C1S2 mode to dominant the system. In \reffig{mode_migration3} \subref{mode_migration3b}, the non-dimensional natural frequency of the C1S2 mode is reduced and approaches the dominant bluff-body vortex shedding frequency within the range of $(f^{vs} c \sin \alpha) / U_{\infty} \in [0.15, 0.2]$ as $Re$ increases. The frequency lock-in between the frequency of the C1S2 mode and the bluff-body vortex shedding frequency is established. The dominant structural mode transitions to the C1S2 mode via the coupling effect. In \reffig{mode_migration3} \subref{mode_migration3c}, the reduction of the aeroelastic number governs the non-dimensional natural frequency to approach the harmonic of the bluff-body vortex shedding frequency. It can be observed from \reffig{mode_map_ae_mass} and \reffig{membrane_response_ae} that the dominant mode transitions between different aeroelastic modes as $Ae$ changes. It can be concluded that the mode transition process is dependent on the variation of the natural frequency relative to the vortex shedding frequency and its harmonics in the coupled system. The original mode synchronization process is interrupted, and a new mode synchronization state is then formed at the dominant coupled frequency, resulting in the mode transition phenomenon.

To gain deeper insight into the mode transition phenomenon, the cross-correlation analysis is performed based on the high-fidelity numerical data. The cross-correlation function of the discrete time signals $x(t_n)$ and $y(t_n)$ is given as
\begin{equation}
R_{xy} (m) =\frac{1}{N} \sum_{n=0}^{N-m-1} x(t_n) y(t_{n+m}) \quad \quad 0 \le m \le N-1,
\end{equation}
where $R_{xy} (m)$ is the cross-correlation coefficient and $N$ is the data samples. The time-varying velocity fluctuation signals with 1024 samples are collected at the point in the middle of the monitoring Line1 and Line2, respectively. The time-varying membrane deflection fluctuation signals with 1024 samples are obtained at three equispaced points along the membrane chord shown in \reffig{probe}. The cross-correlation analysis is conducted between the membrane deflection and the local velocity information near the leading edge and the trailing edge, respectively. The purpose is to examine how the correlation between the unsteady flow and the membrane vibration changes in the mode transition phenomenon.

\begin{figure}
	\centering 
	\subfloat[][]{\includegraphics[width=0.45\textwidth]{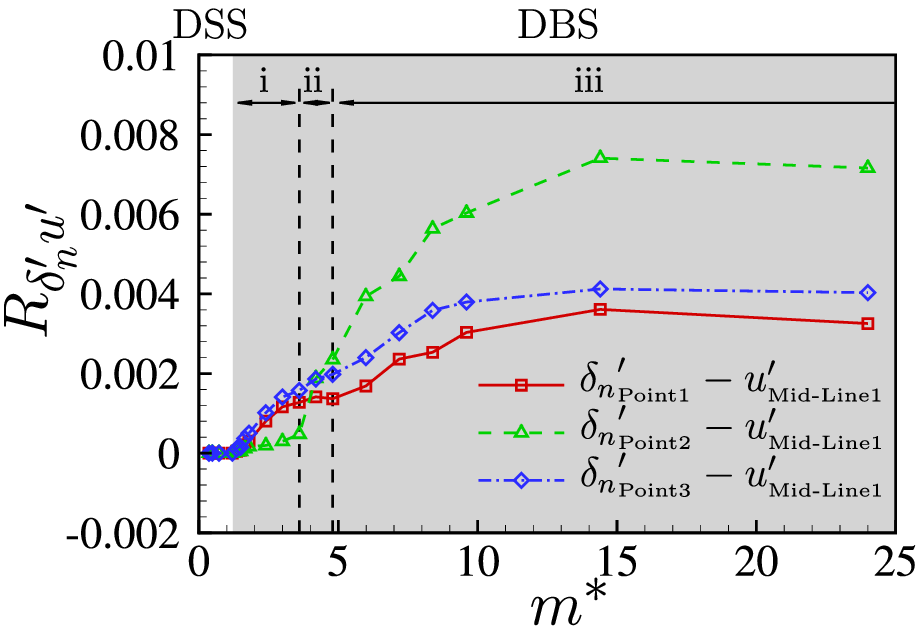}\label{cross_correlationa}}
	\quad
	\subfloat[][]{\includegraphics[width=0.45\textwidth]{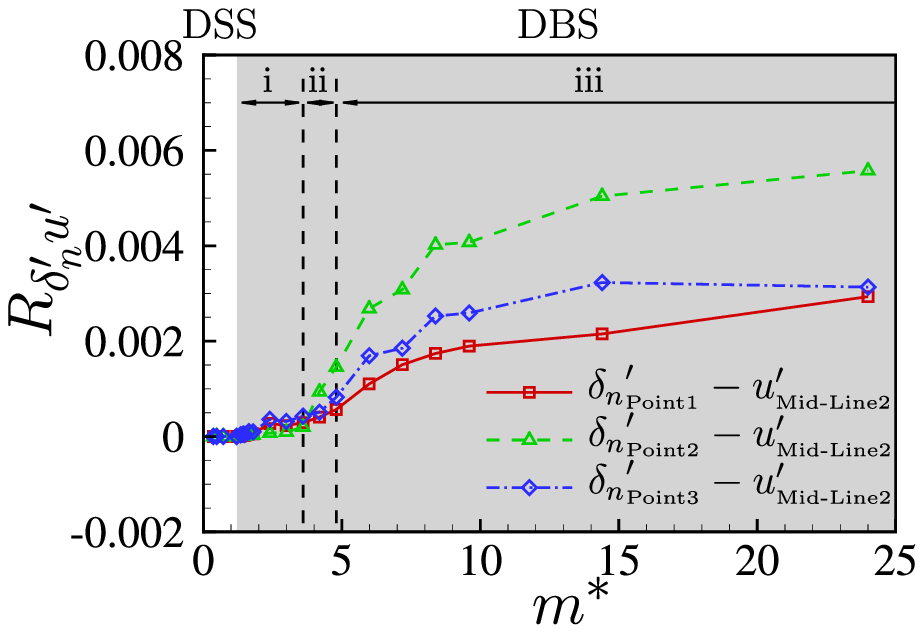}\label{cross_correlationb}}
	\\
	\subfloat[][]{\includegraphics[width=0.45\textwidth]{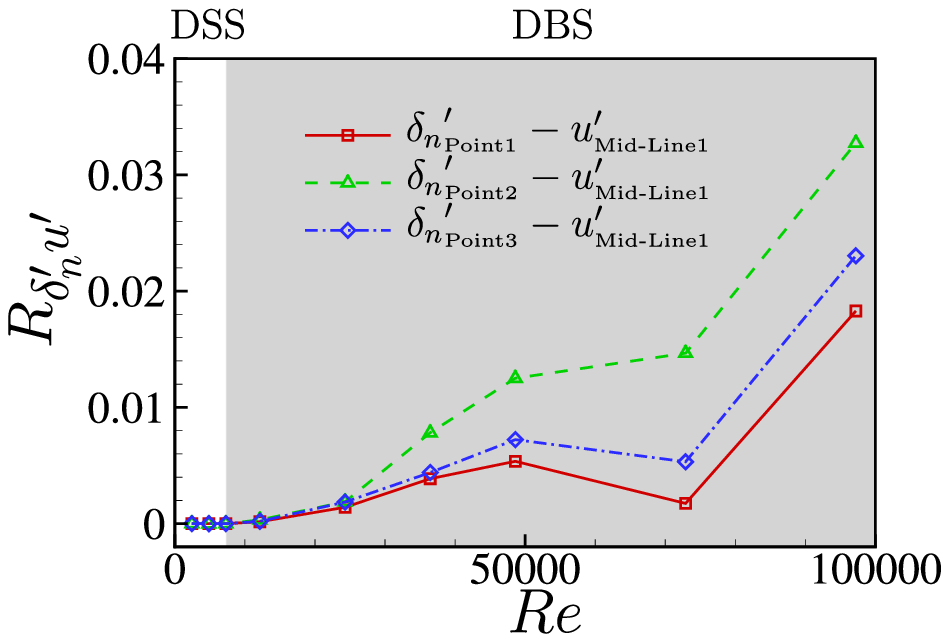}\label{cross_correlationc}}
	\quad
	\subfloat[][]{\includegraphics[width=0.45\textwidth]{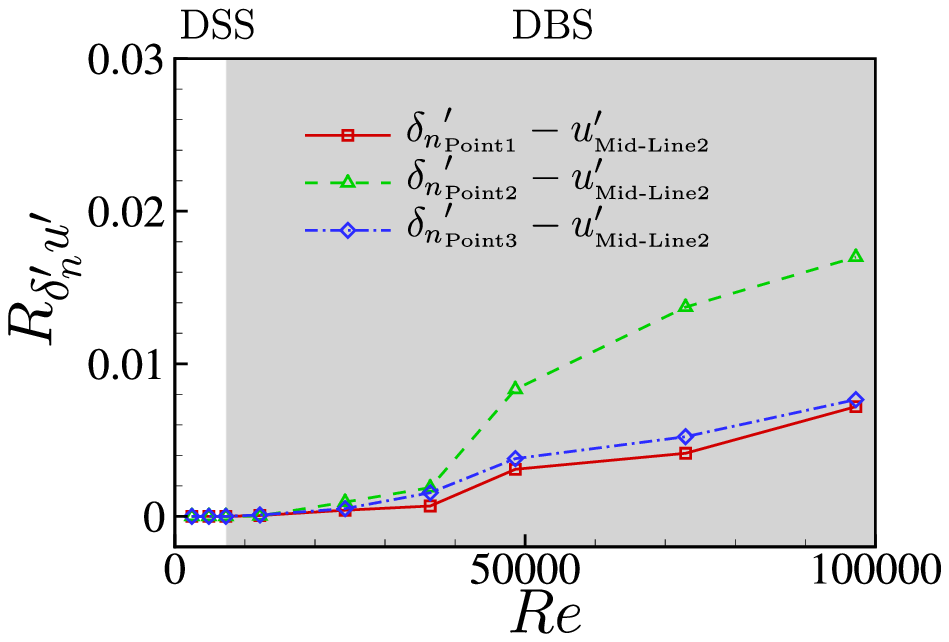}\label{cross_correlationd}}
	\\
	\subfloat[][]{\includegraphics[width=0.45\textwidth]{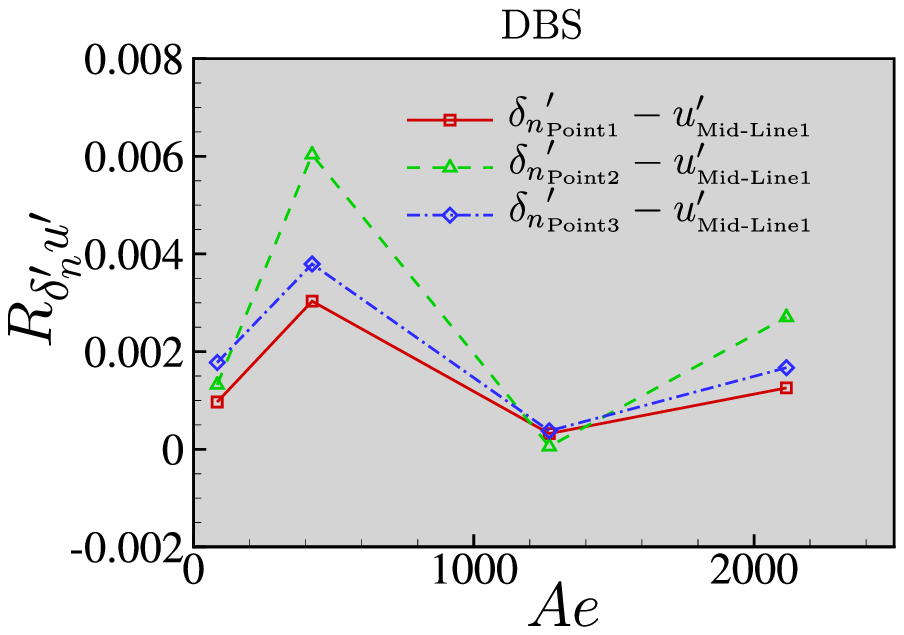}\label{cross_correlatione}}
	\quad
	\subfloat[][]{\includegraphics[width=0.45\textwidth]{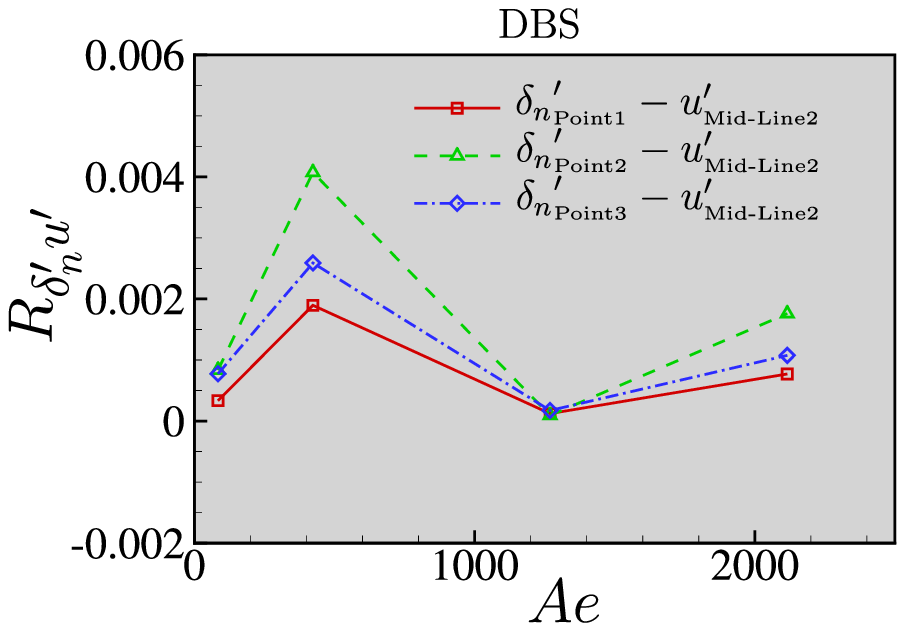}\label{cross_correlationf}}
	\caption{Cross-correlation between the membrane deflection fluctuation $\delta_n^{\prime}$ collected at three equispaced points along the membrane surface and the local flow velocity fluctuation $u^{\prime}$ measured at the point in the middle of: (a,c,e) monitoring Line1 near the leading edge and (b,d,f) monitoring Line2 in the proximity of the trailing edge as a function of (a,b) $m^*$ at fixed $(Re,Ae)$=(24300,423.14), (c,d) $Re$ at fixed $(m^*,Ae)$=(4.2,423.14) and (e,f) $Ae$ at fixed $(m^*,Re)$=(9.6,24300).}
	\label{cross_correlation}
\end{figure}

\refFig{cross_correlation} presents the cross-correlation coefficients as functions of $m^*$, $Re$ and $Ae$. The cross-correlation between the membrane deflection and the velocity is zero in the DSS regime because of the deformed-steady membrane shape. When the membrane vibration occurs, it is found that the flow fluctuations in the wake are strongly correlated with the membrane vibration responses. In \reffigs{cross_correlation} \subref{cross_correlationa} and \subref{cross_correlationb}, the cross-correlation coefficients at monitoring Point1 and Point3 are close to each other and show much higher values than the cross-correlation coefficients at monitoring Point2 in the range of $m^* \in [1.2,3.6]$. The results indicate that the unsteady separated flow has a high correlation with the chord-wise second mode. As $m^*$ further increases to 4.8, the cross-correlation coefficients at monitoring Point2 exceed those values at Point1 and Point3. A stronger correlation between the vortex shedding process and the chord-wise first mode can be observed after the mode transition. It can be seen from \reffigs{cross_correlation} \subref{cross_correlationc} and \subref{cross_correlationd} that the cross-correlation coefficients at monitoring Point2 are smaller than those at monitoring Point1 and Point3 at $Re$=12150. The cross-correlation coefficients at monitoring Point2 become the largest among the three points as $Re$ further increases. The chord-wise first mode dominates the vibration in the range of $Re \in [24300,97200]$. In \reffigs{cross_correlation} \subref{cross_correlatione} and \subref{cross_correlationf}, the cross-correlation coefficients at monitoring Point2 are the largest among the three points at $Ae$=423.14 and 2115.7 when the chord-wise first mode dominates the membrane vibrations. It can be concluded that the cross-correlation between the flow fluctuations and the membrane vibration is dependent on the dominant aeroelastic mode shape. Once the mode transition occurs, we observe the changes of the cross-correlation coefficients at different positions. Combined with the analysis of the flow features and the membrane vibration responses, this strong correlation in the coupled system can help in designing effective active/passive control strategies to switch the aeroelastic modes by governing the mode transitions. Specifically, the preferred membrane performance can be achieved by changing the structural natural frequency to approach the vortex shedding frequency or its harmonics via the frequency lock-in. Due to the mutual dependence between the tension force and the natural frequency, the adjustment of the tension force can offer an effective approach to promote the mode transition.

\section{Conclusions}
In this study, we systematically investigated the flow-excited instability of a simply supported 3D rectangular membrane in separated turbulent flow at moderate Reynolds numbers. To examine the underlying coupled membrane dynamics, three non-dimensional physical parameters were selected namely mass ratio, Reynolds number and aeroelastic number. Two distinctive stability regimes, namely DSS and DBS, were classified from the membrane dynamic responses over a wide range of the selected parameter space. In the parameter space of $m^*$-$Re$ and $m^*$-$Ae$, new empirical relationships for the flow-excited instability boundary were determined from the proposed stability phase diagrams via high-fidelity numerical simulations. To classify the dominant vibrational mode of the coupled system, the global FMD method was employed to extract the mode shapes with the most influential mode energy from the overlapped membrane responses at particular mode frequencies. Based on the comparison of the Fourier mode energy spectra between the fluid and structural domains, we observed the frequency lock-in phenomenon at various physical parameters in the DBS regime. Mode transition phenomenon between different aeroelastic modes was observed in the stability phase diagrams, which were related to the variation of the aerodynamic performance and the membrane vibrations. In the DSS regime, the aerodynamic forces and the membrane deformation were almost independent of $m^*$. The increase of $Re$ and membrane flexibility enhanced the aerodynamic performance due to the growing membrane camber. Compared to the deformed-steady membrane, the membrane vibration caused strong flow fluctuations close to the membrane surface and forced the vortices to attach longer near the leading edge, which enlarged the suction area and improves the aerodynamic performance. The optimal aerodynamic performance was observed for the lighter and more flexible membrane at a higher $Re$.

We further examined the onset of the flow-induced membrane vibration and the mode transition phenomenon in the selected parameter space. The variation of the natural frequency of the tensioned membrane relative to the vortex shedding frequency was monitored as a function of $m^*$, $Re$ and $Ae$. Under the aerodynamic loading, the areal strain was calculated to estimate the membrane stretching and its relationship with the natural frequency of the membrane. For the evaluation of the natural frequency, we derived an approximate analytical formula from the nonlinear vibration equation based on large deflection theory coupled with the added mass effect and a simplified pressure loading. By varying fluid-structure parameters, the comparison between the predicted nonlinear natural frequency of the tensioned membrane and the measured vortex shedding frequency as well as its harmonics was carried out. The increase in the membrane inertia, the fluid inertia and the membrane flexibility gave rise to a reduction of the non-dimensional structural natural frequency, making it closer to the non-dimensional vortex shedding frequency or its harmonics. Consequently, the vortex shedding frequency locks into the natural frequency of the tensioned membrane which provides the self-sustained vibration of a flexible membrane. The newly formed coupling between the varied structural natural frequency and the vortex shedding frequency interrupted the original mode synchronization process, and then generated a new mode synchronization state at the dominant coupled frequency, resulting in the mode transition phenomenon. From both operations and design perspectives, the proposed fluid-structure analysis offers some guidelines for effective active/passive control designs to achieve desirable performance and membrane vibrational states by adjusting the structural natural frequency via a proper combination of the physical parameters and the tension force.

\section*{Acknowledgements}
The first author wishes to acknowledge supports from the National University of Singapore and the Ministry of Education, Singapore. The second author would like to acknowledge the support from the University of British Columbia and the Natural Sciences and Engineering Research Council of Canada (NSERC).

\section*{Declaration of interests}
The authors report no conflict of interest.

\bibliographystyle{jfm}
\bibliography{reference}

\begin{thebibliography}{44}
\expandafter\ifx\csname natexlab\endcsname\relax\def\natexlab#1{#1}\fi
\def\au#1{#1} \def\ed#1{#1} \def\yr#1{#1}\def\at#1{#1}\def\jt#1{\textit{#1}}
  \def\bt#1{#1}\def\bvol#1{\textbf{#1}} \def\vol#1{#1} \def\pg#1{#1}
  \def\publ#1{#1}\def\arxiv#1{#1}\def\org#1{#1}\def\st#1{\textit{#1}}

\bibitem[Amabili(2008)]{amabili2008nonlinear}
{\sc \au{Amabili, M. }} \yr{2008} {\em Nonlinear vibrations and stability of
  shells and plates\/}.  \publ{Cambridge University Press}.

\bibitem[Arb{\'o}s-Torrent {\em et~al.\/}(2013)Arb{\'o}s-Torrent,
  Ganapathisubramani \& Palacios]{arbos2013leading}
{\sc \au{Arb{\'o}s-Torrent, S. }, \au{Ganapathisubramani, B. } \& \au{Palacios,
  R. }} \yr{2013}  \at{Leading-and trailing-edge effects on the aeromechanics
  of membrane aerofoils}.  \jt{J. Fluids Struct.}  \bvol{38},  \pg{107--126}.

\bibitem[Bahlman {\em et~al.\/}(2013)Bahlman, Swartz \&
  Breuer]{bahlman2013design}
{\sc \au{Bahlman, J.~W. }, \au{Swartz, S.~M. } \& \au{Breuer, K.~S. }}
  \yr{2013}  \at{Design and characterization of a multi-articulated robotic bat
  wing}.  \jt{Bioinspir. Biomim}  \bvol{8}~(1),  \pg{016009}.

\bibitem[Bleischwitz {\em et~al.\/}(2016)Bleischwitz, De~Kat \&
  Ganapathisubramani]{bleischwitz2016aeromechanics}
{\sc \au{Bleischwitz, R. }, \au{De~Kat, R. } \& \au{Ganapathisubramani, B. }}
  \yr{2016}  \at{Aeromechanics of membrane and rigid wings in and out of
  ground-effect at moderate reynolds numbers}.  \jt{J. Fluids Struct.}
  \bvol{62},  \pg{318--331}.

\bibitem[Bleischwitz {\em et~al.\/}(2017)Bleischwitz, De~Kat \&
  Ganapathisubramani]{bleischwitz2017fluid}
{\sc \au{Bleischwitz, R. }, \au{De~Kat, R. } \& \au{Ganapathisubramani, B. }}
  \yr{2017}  \at{On the fluid-structure interaction of flexible membrane wings
  for mavs in and out of ground-effect}.  \jt{J. Fluids Struct.}  \bvol{70},
  \pg{214--234}.

\bibitem[Bleischwitz {\em et~al.\/}(2018)Bleischwitz, De~Kat \&
  Ganapathisubramani]{bleischwitz2018near}
{\sc \au{Bleischwitz, R. }, \au{De~Kat, R. } \& \au{Ganapathisubramani, B. }}
  \yr{2018}  \at{Near-wake characteristics of rigid and membrane wings in
  ground effect}.  \jt{J. Fluids Struct.}  \bvol{80},  \pg{199--216}.

\bibitem[Bleischwitz {\em et~al.\/}(2015)Bleischwitz, de~Kat \&
  Ganapathisubramani]{bleischwitz2015aspect}
{\sc \au{Bleischwitz, R. }, \au{de~Kat, R. } \& \au{Ganapathisubramani, B. }}
  \yr{2015}  \at{Aspect-ratio effects on aeromechanics of membrane wings at
  moderate reynolds numbers}.  \jt{AIAA J.}  \bvol{53}~(3),  \pg{780--788}.

\bibitem[Buoso \& Palacios(2017)]{buoso2017demand}
{\sc \au{Buoso, S. } \& \au{Palacios, R. }} \yr{2017}  \at{On-demand
  aerodynamics in integrally actuated membranes with feedback control}.
  \jt{AIAA J.}  \bvol{55}~(2),  \pg{377--388}.

\bibitem[Chen {\em et~al.\/}(2015)Chen, Wu \& Sun]{chen2015research}
{\sc \au{Chen, Z. }, \au{Wu, Y. } \& \au{Sun, X. }} \yr{2015}  \at{Research on
  the added mass of open-type one-way tensioned membrane structure in uniform
  flow}.  \jt{Journal of Wind Engineering and Industrial Aerodynamics}
  \bvol{137},  \pg{69--77}.

\bibitem[Drazin \& Drazin(1992)]{drazin1992nonlinear}
{\sc \au{Drazin, P.~G. } \& \au{Drazin, P.~D. }} \yr{1992} {\em Nonlinear
  systems\/}, ,  \vol{vol.~10}.  \publ{Cambridge University Press}.

\bibitem[Gordnier(2009)]{gordnier2009high}
{\sc \au{Gordnier, R.~E. }} \yr{2009}  \at{High fidelity computational
  simulation of a membrane wing airfoil}.  \jt{J. Fluids Struct.}
  \bvol{25}~(5),  \pg{897--917}.

\bibitem[Gordnier \& Attar(2014)]{gordnier2014impact}
{\sc \au{Gordnier, R.~E. } \& \au{Attar, P.~J. }} \yr{2014}  \at{Impact of
  flexibility on the aerodynamics of an aspect ratio two membrane wing}.
  \jt{J. Fluids Struct.}  \bvol{45},  \pg{138--152}.

\bibitem[Gursul {\em et~al.\/}(2014)Gursul, Cleaver \& Wang]{gursul2014control}
{\sc \au{Gursul, I. }, \au{Cleaver, D. } \& \au{Wang, Z. }} \yr{2014}
  \at{Control of low reynolds number flows by means of fluid--structure
  interactions}.  \jt{Progress in Aerospace Sciences}  \bvol{64},  \pg{17--55}.

\bibitem[Jaiman {\em et~al.\/}(2016)Jaiman, Pillalamarri \&
  Guan]{jaiman2016stable}
{\sc \au{Jaiman, R. }, \au{Pillalamarri, N. } \& \au{Guan, M. }} \yr{2016}
  \at{A stable second-order partitioned iterative scheme for freely vibrating
  low-mass bluff bodies in a uniform flow}.  \jt{Comput. Method. Appl. M.}
  \bvol{301},  \pg{187--215}.

\bibitem[Jaiman {\em et~al.\/}(2014)Jaiman, Parmar \&
  Gurugubelli]{jaiman2014added}
{\sc \au{Jaiman, R.~K. }, \au{Parmar, M.~K. } \& \au{Gurugubelli, P.~S. }}
  \yr{2014}  \at{Added mass and aeroelastic stability of a flexible plate
  interacting with mean flow in a confined channel}.  \jt{Journal of Applied
  Mechanics}  \bvol{81}~(4).

\bibitem[Joshi \& Jaiman(2017)]{joshi2017variationally}
{\sc \au{Joshi, V. } \& \au{Jaiman, R.~K. }} \yr{2017}  \at{A variationally
  bounded scheme for delayed detached eddy simulation: Application to
  vortex-induced vibration of offshore riser}.  \jt{Computers \& fluids}
  \bvol{157},  \pg{84--111}.

\bibitem[Joshi {\em et~al.\/}(2020)Joshi, Jaiman \&
  Ollivier-Gooch]{joshi2020variational}
{\sc \au{Joshi, V. }, \au{Jaiman, R.~K. } \& \au{Ollivier-Gooch, C. }}
  \yr{2020}  \at{A variational flexible multibody formulation for partitioned
  fluid--structure interaction: Application to bat-inspired drones and unmanned
  air-vehicles}.  \jt{Computers \& Mathematics with Applications}
  \bvol{80}~(12),  \pg{2707--2737}.

\bibitem[Kamm \& Grodzinsky(2015)]{kamm2015molecular}
{\sc \au{Kamm, R. } \& \au{Grodzinsky, A. }} \yr{2015}  \at{Molecular,
  cellular, and tissue biomechanics}.  \jt{Cambridge, MA: Massachusetts
  Institute of Technology} .

\bibitem[Katopodes(2018)]{katopodes2018free}
{\sc \au{Katopodes, N.~D. }} \yr{2018} {\em Free-surface Flow: Environmental
  Fluid Mechanics\/}.  \publ{Butterworth-Heinemann}.

\bibitem[Kinsler {\em et~al.\/}(1999)Kinsler, Frey, Coppens \&
  Sanders]{kinsler1999fundamentals}
{\sc \au{Kinsler, L.~E. }, \au{Frey, A.~R. }, \au{Coppens, A.~B. } \&
  \au{Sanders, J.~V. }} \yr{1999}  \at{Fundamentals of acoustics}.
  \jt{Fundamentals of Acoustics, 4th Edition, by Lawrence E. Kinsler, Austin R.
  Frey, Alan B. Coppens, James V. Sanders, pp. 560. ISBN 0-471-84789-5.
  Wiley-VCH, December 1999.}  \pg{p. 560}.

\bibitem[Li {\em et~al.\/}(2020{\natexlab{{\em a\/}}})Li, Jaiman \&
  Khoo]{li2020aeroelastic}
{\sc \au{Li, G. }, \au{Jaiman, R.~K. } \& \au{Khoo, B.~C. }}
  \yr{2020{\natexlab{{\em a\/}}}}  \at{Aeroelastic mode decomposition and mode
  selection mechanism in fluid-membrane interaction}.  \jt{arXiv e-prints}
  \pg{pp. arXiv--2006}.

\bibitem[Li {\em et~al.\/}(2020{\natexlab{{\em b\/}}})Li, Khoo \&
  Jaiman]{li2020computational}
{\sc \au{Li, G. }, \au{Khoo, B.~C. } \& \au{Jaiman, R.~K. }}
  \yr{2020{\natexlab{{\em b\/}}}} Computational aeroelasticity of flexible
  membrane wings at moderate reynolds numbers.  \bt{In {\em AIAA Scitech 2020
  Forum\/}},  \pg{p. 2036}.

\bibitem[Li {\em et~al.\/}(2019)Li, Law \& Jaiman]{li2018novel}
{\sc \au{Li, G. }, \au{Law, Y.~Z. } \& \au{Jaiman, R.~K. }} \yr{2019}  \at{A
  novel 3d variational aeroelastic framework for flexible multibody dynamics:
  Application to bat-like flapping dynamics}.  \jt{Comput. Fluids.}
  \bvol{180},  \pg{96--116}.

\bibitem[Liu {\em et~al.\/}(2018)Liu, Cai, Peng, Zhang \& Lv]{liu2018nonlinear}
{\sc \au{Liu, X. }, \au{Cai, G. }, \au{Peng, F. }, \au{Zhang, H. } \& \au{Lv,
  L. }} \yr{2018}  \at{Nonlinear vibration analysis of a membrane based on
  large deflection theory}.  \jt{Journal of Vibration and Control}
  \bvol{24}~(12),  \pg{2418--2429}.

\bibitem[Newman(1987)]{newman1987aerodynamic}
{\sc \au{Newman, B.~G. }} \yr{1987}  \at{Aerodynamic theory for membranes and
  sails}.  \jt{Prog. Aerosp. Sci.}  \bvol{24}~(1),  \pg{1--27}.

\bibitem[Rojratsirikul {\em et~al.\/}(2011)Rojratsirikul, Genc, Wang \&
  Gursul]{rojratsirikul2011flow}
{\sc \au{Rojratsirikul, P. }, \au{Genc, M. }, \au{Wang, Z. } \& \au{Gursul, I.
  }} \yr{2011}  \at{Flow-induced vibrations of low aspect ratio rectangular
  membrane wings}.  \jt{J. Fluids Struct.}  \bvol{27}~(8),  \pg{1296--1309}.

\bibitem[Rojratsirikul {\em et~al.\/}(2009)Rojratsirikul, Wang \&
  Gursul]{rojratsirikul2009unsteady}
{\sc \au{Rojratsirikul, P. }, \au{Wang, Z. } \& \au{Gursul, I. }} \yr{2009}
  \at{Unsteady fluid--structure interactions of membrane airfoils at low
  reynolds numbers}.  \jt{Exp. Fluids.}  \bvol{46}~(5),  \pg{859}.

\bibitem[Rojratsirikul {\em et~al.\/}(2010{\natexlab{{\em a\/}}})Rojratsirikul,
  Wang \& Gursul]{rojratsirikul2010effect}
{\sc \au{Rojratsirikul, P. }, \au{Wang, Z. } \& \au{Gursul, I. }}
  \yr{2010{\natexlab{{\em a\/}}}}  \at{Effect of pre-strain and excess length
  on unsteady fluid--structure interactions of membrane airfoils}.  \jt{J.
  Fluids Struct.}  \bvol{26}~(3),  \pg{359--376}.

\bibitem[Rojratsirikul {\em et~al.\/}(2010{\natexlab{{\em b\/}}})Rojratsirikul,
  Wang \& Gursul]{rojratsirikul2010unsteady}
{\sc \au{Rojratsirikul, P. }, \au{Wang, Z. } \& \au{Gursul, I. }}
  \yr{2010{\natexlab{{\em b\/}}}} Unsteady aerodynamics of low aspect ratio
  membrane wings.  \bt{In {\em 48th AIAA Aerospace Sciences Meeting Including
  the New Horizons Forum and Aerospace Exposition\/}},  \pg{p. 729}.

\bibitem[Serrano~Galiano \& Sandberg(2016)]{serrano2016effect}
{\sc \au{Serrano~Galiano, S. } \& \au{Sandberg, R.~D. }} \yr{2016} Effect of
  the leading and trailing edge geometry on the fluid-structural coupling of
  membrane aerofoils.  \bt{In {\em 54th AIAA Aerospace Sciences Meeting\/}},
  \pg{p. 0853}.

\bibitem[Serrano-Galiano {\em et~al.\/}(2018)Serrano-Galiano, Sandham \&
  Sandberg]{serrano2018fluid}
{\sc \au{Serrano-Galiano, S. }, \au{Sandham, N.~D. } \& \au{Sandberg, R.~D. }}
  \yr{2018}  \at{Fluid--structure coupling mechanism and its aerodynamic effect
  on membrane aerofoils}.  \jt{J. Fluid Mech.}  \bvol{848},  \pg{1127--1156}.

\bibitem[Shyy {\em et~al.\/}(1999)Shyy, Berg \& Ljungqvist]{shyy1999flapping}
{\sc \au{Shyy, W. }, \au{Berg, M. } \& \au{Ljungqvist, D. }} \yr{1999}
  \at{Flapping and flexible wings for biological and micro air vehicles}.
  \jt{Prog. Aerosp. Sci.}  \bvol{35}~(5),  \pg{455--505}.

\bibitem[Smith \& Shyy(1995)]{smith1995computational}
{\sc \au{Smith, R. } \& \au{Shyy, W. }} \yr{1995}  \at{Computational model of
  flexible membrane wings in steady laminar flow}.  \jt{AIAA J.}
  \bvol{33}~(10),  \pg{1769--1777}.

\bibitem[Song {\em et~al.\/}(2008)Song, Tian, Israeli, Galvao, Bishop, Swartz
  \& Breuer]{song2008aeromechanics}
{\sc \au{Song, A. }, \au{Tian, X. }, \au{Israeli, E. }, \au{Galvao, R. },
  \au{Bishop, K. }, \au{Swartz, S. } \& \au{Breuer, K. }} \yr{2008}
  \at{Aeromechanics of membrane wings with implications for animal flight}.
  \jt{AIAA J.}  \bvol{46}~(8),  \pg{2096--2106}.

\bibitem[Sun {\em et~al.\/}(2017)Sun, Ren \& Zhang]{sun2017nonlinear}
{\sc \au{Sun, X. }, \au{Ren, X. } \& \au{Zhang, J. }} \yr{2017}  \at{Nonlinear
  dynamic responses of a perimeter-reinforced membrane wing in laminar flows}.
  \jt{Nonlinear Dyn.}  \bvol{88}~(1),  \pg{749--776}.

\bibitem[Sun \& Zhang(2016)]{sun2016nonlinear}
{\sc \au{Sun, X. } \& \au{Zhang, J. }} \yr{2016}  \at{Nonlinear vibrations of a
  flexible membrane under periodic load}.  \jt{Nonlinear Dyn.}  \bvol{85}~(4),
  \pg{2467--2486}.

\bibitem[Sun \& Zhang(2017)]{sun2017effect}
{\sc \au{Sun, X. } \& \au{Zhang, J. }} \yr{2017}  \at{Effect of the reinforced
  leading or trailing edge on the aerodynamic performance of a
  perimeter-reinforced membrane wing}.  \jt{J. Fluids Struct.}  \bvol{68},
  \pg{90--112}.

\bibitem[Tiomkin \& Raveh(2019)]{tiomkin2019on}
{\sc \au{Tiomkin, S. } \& \au{Raveh, D.~E. }} \yr{2019}  \at{On membrane-wing
  stability in laminar flow}.  \jt{J. Fluids Struct.}  \bvol{91}.

\bibitem[Tregidgo {\em et~al.\/}(2011)Tregidgo, Wang \&
  Gursul]{tregidgo2011fluid}
{\sc \au{Tregidgo, L. }, \au{Wang, Z. } \& \au{Gursul, I. }} \yr{2011}
  Fluid-structure interactions for a low aspect-ratio membrane wing at low
  reynolds numbers.  \bt{In {\em 41st AIAA Fluid Dynamics Conference and
  Exhibit\/}},  \pg{p. 3436}.

\bibitem[Tregidgo {\em et~al.\/}(2012)Tregidgo, Wang \&
  Gursul]{tregidgo2012frequency}
{\sc \au{Tregidgo, L. }, \au{Wang, Z. } \& \au{Gursul, I. }} \yr{2012}
  \at{Frequency lock-in phenomenon for self-sustained roll oscillations of
  rectangular wings undergoing a forced periodic pitching motion}.  \jt{Phys.
  Fluids.}  \bvol{24}~(11),  \pg{117101}.

\bibitem[Tregidgo {\em et~al.\/}(2013)Tregidgo, Wang \&
  Gursul]{tregidgo2013unsteady}
{\sc \au{Tregidgo, L. }, \au{Wang, Z. } \& \au{Gursul, I. }} \yr{2013}
  \at{Unsteady fluid--structure interactions of a pitching membrane wing}.
  \jt{Aerosp. Sci. Technol.}  \bvol{28}~(1),  \pg{79--90}.

\bibitem[Waldman \& Breuer(2017)]{waldman2017camber}
{\sc \au{Waldman, R.~M. } \& \au{Breuer, K.~S. }} \yr{2017}  \at{Camber and
  aerodynamic performance of compliant membrane wings}.  \jt{J. Fluids Struct.}
   \bvol{68},  \pg{390--402}.

\bibitem[Webster \& Griffin(1962)]{webster1962role}
{\sc \au{Webster, F.~A. } \& \au{Griffin, D.~R. }} \yr{1962}  \at{The role of
  the flight membranes in insect capture by bats}.  \jt{Anim. Behav.}
  \bvol{10}~(3-4),  \pg{332--340}.

\bibitem[Yang {\em et~al.\/}(2018)Yang, Dudley \&
  Harris]{yang2018aeroelasticity}
{\sc \au{Yang, H. }, \au{Dudley, J. } \& \au{Harris, R. }} \yr{2018}
  \at{Aeroelasticity validation study for a three-dimensional membrane wing}.
  \jt{AIAA J.}  \bvol{56}~(6),  \pg{2361--2371}.

\end{thebibliography}

\section*{Appendix A. Convergence study}
\setcounter{equation}{0}
\setcounter{figure}{0}
\setcounter{table}{0}
\renewcommand{\theequation}{A.\arabic{equation}}
\renewcommand{\thefigure}{A.\arabic{figure}}
\renewcommand{\thetable}{A.\arabic{table}}

We first perform the mesh convergence study to determine a proper mesh resolution for the numerical simulation of the 3D flexible membrane wing. Three different meshes namely M1, M2 and M3 are constructed for the numerical simulation. The 3D computational fluid domain is discretized by unstructured finite element mesh which consists of 341 821, 823 864 and 1 304 282 eight-node brick elements, respectively. The stretching ratio ($\Delta y_{j+1}/\Delta y_{j}$) for the mesh in the boundary layer is set as 1.25 to keep $y^+$ less than 1.0. We choose a non-dimensional time step size of $\Delta t U_{\infty}/c$=0.0364 in all simulations. The structural model is discretized by 160, 228 and 352 structured four-node rectangular finite elements, respectively.

In our simulation, the same parameters obtained from the wind tunnel experiments \citep{rojratsirikul2010unsteady,rojratsirikul2011flow} are set for both the fluid and structure domains. The critical parameters considered for the mesh convergence study are $Re$=24300, $U_{\infty}$=5 m/s and $\alpha$=$15^\circ$. The chosen physical parameters set the non-dimensional parameters to $Ae$=423.14 and $m^*$=2.4. The flexible membrane is placed in a uniform flow with a freestream velocity of $U_{\infty}$=5 m/s. The flexible membrane is glued to the rigid frame in a flat shape without pretension. The initial values of membrane displacement and velocity are set to zero. \refTab{3dmeshforce} summarizes the aerodynamic force characteristics, the non-dimensional dominant vortex shedding frequency and the structural responses for the meshes M1, M2 and M3. The percentage discrepancies are calculated with respect to the finest mesh M3. The maximum difference between the numerical solutions computed for the meshes M2 and M3 is less than $3\%$. Therefore, mesh M2 is adequate for the numerical validation study.

To perform the time step size convergence study, we select three non-dimensional time step sizes of $\Delta t U_{\infty}/c$=0.0728, 0.0364 and 0.0182. Mesh M2 is chosen in the numerical simulation. Physical parameters are set to the same as those in the mesh convergence study. The computed results of $\Delta t U_{\infty}/c$=0.0182 are the reference to calculate the percentage discrepancies. The membrane responses at three time step sizes are summarized in \reftab{3dtimeforce}. The percentage difference for $\Delta t U_{\infty}/c$=0.0364 is smaller than $3\%$. Thus, the time step size of 0.0364 is selected in the numerical simulations.

\begin{table}
	\begin{center}
		\begin{tabular}{ccccccc}
			Mesh & M1    &    M2    &   M3 \\[3pt]
			Structural elements & 160 & 228 & 352 \\
			Fluid elements & 341 821  & 823 864 & 1 304 282  \\
			Mean lift $\overline{C}_L$ & 0.9022 (0.49$\%$) & 0.8902 (0.85$\%$)& 0.8978\\
			Mean drag $\overline{C}_D$ & 0.2289 (2.97$\%$)& 0.2346 (0.55$\%$)& 0.2359\\
			r.m.s. lift fluctuation ${C_L^{\prime}}^{rms}$ & 0.0788 (16.17$\%$) &0.0928 (1.28$\%$)& 0.0940   \\
			r.m.s. drag fluctuation ${C_D^{\prime}}^{rms}$ &0.0289 (10.25$\%$)&0.0313 (2.80$\%$)& 0.0322   \\
			Dominant shedding frequency $f^{vs}c/U_{\infty}$ & 0.9668  (2.70$\%$)& 0.9937 (0$\%$)& 0.9937  \\
			Maximum mean deflection $\overline{\delta}_n^{max}/c$ & 0.03391 (1.14$\%$) & 0.03415 (0.44$\%$) & 0.03430 \\
			Maximum r.m.s. deflection fluctuation ${\delta_n^{\prime}}^{rms}$  & 0.000979 (2.88$\%$) & 0.000994 (1.39$\%$) & 0.001008  \\
			Dominant vibration frequency $f^{s}c/U_{\infty}$ & 0.9668  (2.70$\%$)& 0.9937 (0$\%$)& 0.9937 \\
		\end{tabular}
		\caption{Mesh convergence of a 3D rectangular flexible membrane wing at $Re$=24300, $Ae$=423.14, $m^*$=2.4 and $\alpha$=$15^\circ$. The percentage discrepancies in the brackets are calculated with respect to the finest mesh M3.}{\label{3dmeshforce}}
	\end{center}
\end{table}

\begin{table}
	\begin{center}
		\begin{tabular}{ccccccc}
			Non-dimensional time step size $\Delta t U_{\infty}/c$ & $0.0728$    &    $0.0364$    &     $ 0.0182$ \\[3pt]
			Mean lift $\overline{C}_L$  & 0.8336 ($7.8386 \%$) & 0.8902 ($1.58 \%$)&  0.9045  \\
			Mean drag $\overline{C}_D$  & 0.2223 ($4.6332 \%$)& 0.2346 ($0.6435 \%$)& 0.2331 \\
			r.m.s. lift fluctuation ${C_L^{\prime}}^{rms}$ & 0.0736  ($ 22.20 \%$) &0.0928 ($1.90 \%$)& 0.0946 \\
			r.m.s. drag fluctuation ${C_D^{\prime}}^{rms}$ & 0.0252 ($20.75 \%$)&0.0313 ($1.57 \%$)& 0.0318  \\
			Dominant shedding frequency $f^{vs}c/U_{\infty}$ & 0.9399 ($5.41 \%$)& 0.9937 ($0 \%$)&  0.9937 \\
			Maximum mean deflection $\overline{\delta}_n^{max}/c$ & 0.0351  ($1.45\%$) & 0.03415 ($1.30 \%$) & 0.0346  \\
			Maximum r.m.s. deflection fluctuation ${\delta_n^{\prime}}^{rms}$  & 0.000830 ($18.79\%$) & 0.000994 ($2.74 \%$) & 0.001022  \\
			Dominant vibration frequency $f^{s}c/U_{\infty}$ & 0.9399 ($5.41 \%$) & 0.9937 ($0 \%$)& 0.9937  \\
		\end{tabular}
		\caption{Time step size convergence of a 3D rectangular flexible membrane wing at $Re$=24300 and $\alpha=15^\circ$ based on mesh M2. The percentage discrepancies are calculated by using the results of $\Delta t U_{\infty}/c = 0.0182$ as the reference.}{\label{3dtimeforce}}
	\end{center}
\end{table}

\section*{Appendix B. Validation of coupled fluid-structure solver}
\setcounter{equation}{0}
\setcounter{figure}{0}
\setcounter{table}{0}
\renewcommand{\theequation}{B.\arabic{equation}}
\renewcommand{\thefigure}{B.\arabic{figure}}
\renewcommand{\thetable}{B.\arabic{table}}
To validate our coupled fluid-structure solver, we present a comparison of the membrane dynamic responses at $U_{\infty}$=5 m/s between the experimental measurements \citep{rojratsirikul2010unsteady,rojratsirikul2011flow} and the numerical simulation results in \reffig{averagediscn}. The maximum magnitude of the normalized time-averaged membrane deformation is presented in \reffig{averagediscn} \subref{averagediscna}. A difference of the time-averaged normal force coefficient between the flexible membrane wing and a rigid wing counterpart is considered for the plotting purpose in \reffig{averagediscn} \subref{averagediscnb}. It can be observed that the overall trends of the membrane deformation and the aerodynamic forces are reasonably well predicted. We further compare the circulation of tip vortices on a cross-flow plane near the trailing edge in \reffig{averagediscn} \subref{circulation}. Consistent with the experimental measurements \citep{rojratsirikul2010unsteady}, the circulation of the tip vortices is calculated by integrating the vorticity distributions over the measuring surface $\Omega^{S^m}$ covering the tip vortices on the cross-flow plane. In our numerical simulation, the integration of the vorticity is evaluated using the Gaussian quadrature
\begin{equation}
\Gamma = \iint\nolimits_{\Omega^{S^m}} \boldsymbol{\nabla} \times \boldsymbol{u}^f \cdot {\rm{d}} \boldsymbol{S}^m = \sum\limits_{e=1}^{n_{el}}\sum\limits_{p=1}^{n_p} \omega_x(\eta_p,\xi_p) {\rm{det}}J_e(\eta_p,\xi_p)W_p,
\end{equation}
where $n_{el}$ and $n_p$ are the number of the finite elements of the measuring surface and the number of Gauss points, respectively. $\eta_p$ and $\xi_p$ denote the coordinates of the Gauss nodes. $\omega_x$ is the vorticity in the freestream direction. $J_e$ is the jacobian of the $e$-th element and $W_p$ represents the Gauss weight. It can be seen from \reffig{averagediscn} \subref{circulation} that the normalized circulation obtained from our numerical simulation shows good agreement with the experimental results. The frequency spectrum of the membrane vibration at the point with maximum standard deviation in our numerical simulation is analysed via the fast Fourier transform technique. The frequency values at the frequency peaks are plotted in \reffig{averagediscn} \subref{frequency} together with the frequency spectra obtained from experiments for the validation purpose. The dominant frequency contents of the membrane vibration at various AOAs show similar distributions with the experimental results. The magnitude of the turbulent intensity on the mid-span plane at $\alpha=16^\circ$ obtained from the numerical simulation is presented in \reffig{tkecom} and the patterns are contrasted against the experimental measurements. It can be observed that the high flow fluctuation region is reasonably well predicted in our simulation.

\begin{figure}
	\centering
	\subfloat[][]{\includegraphics[width=0.5\textwidth]{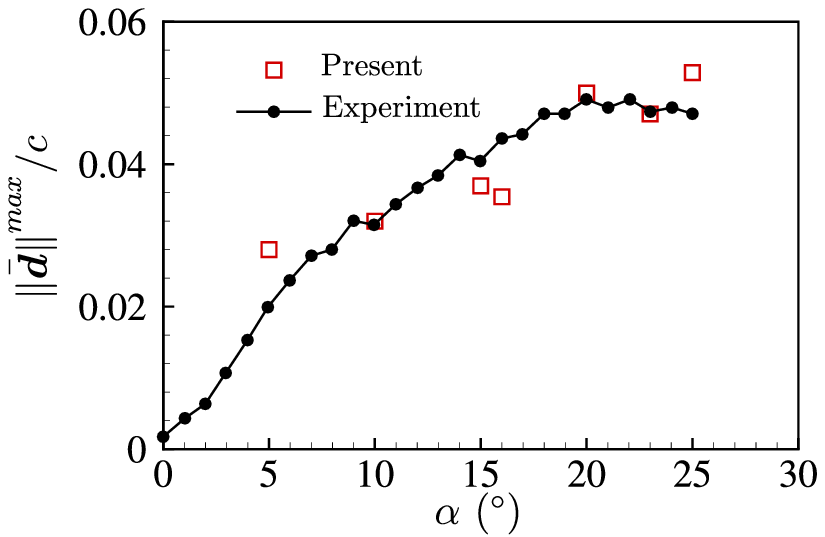}\label{averagediscna}}
	\subfloat[][]{\includegraphics[width=0.5\textwidth]{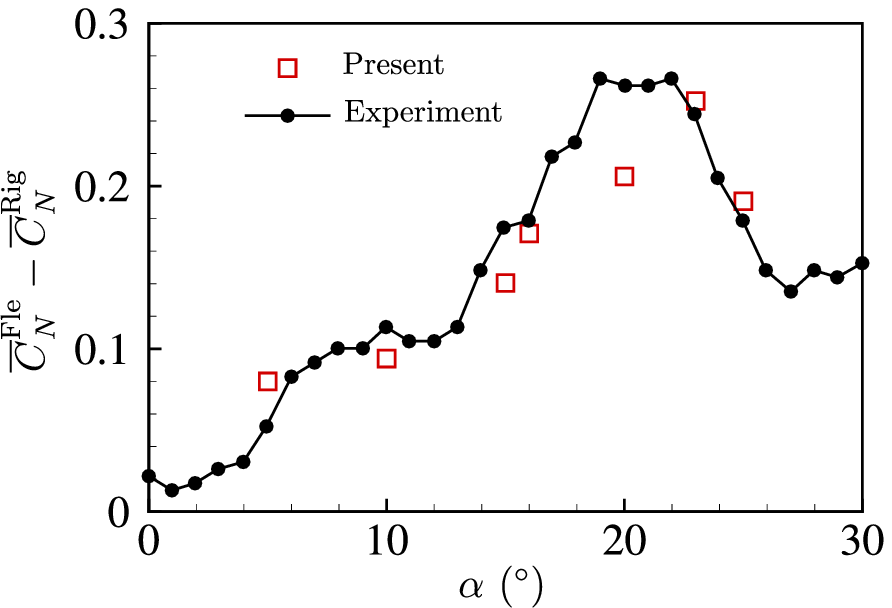}\label{averagediscnb}}
	\\
	\subfloat[][]{\includegraphics[width=0.49\textwidth]{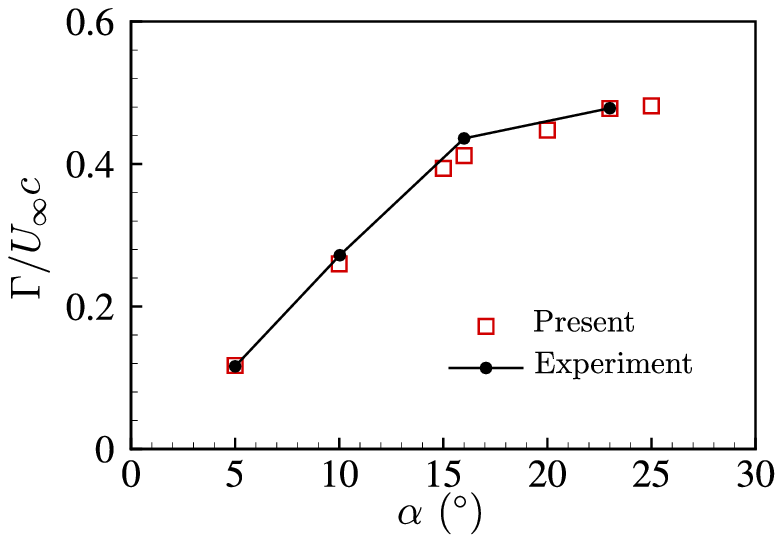}\label{circulation}}
	\subfloat[][]{\includegraphics[width=0.49\textwidth]{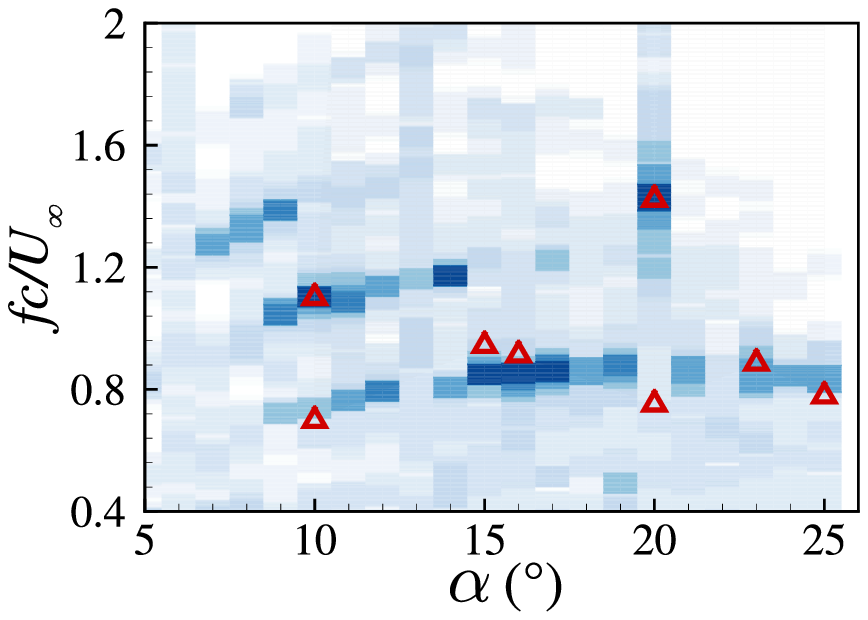}\label{frequency}}
	\caption{Comparison of membrane response characteristics between the present  simulations and experiments \citep{rojratsirikul2010unsteady,rojratsirikul2011flow} at $U_{\infty}$=5 m/s for: (a) maximum magnitude of the normalized time-averaged membrane deformation ($\bar{\left\|  \bm d \right\|}^{{max}}/c$), (b) time-averaged normal force coefficient difference ($\overline{C}_N^{\rm{Fle}}-\overline{C}_N^{\rm{Rig}}$) between the flexible membrane wing and rigid wing, (c) normalized circulation ($\Gamma/U_{\infty} c$) of tip vortices on a plane perpendicular to the freestream near the trailing edge and (d) membrane vibration frequency at the point with maximum standard deviation. Contours in (d) are coloured by the frequency spectra of the experimental results \citep{rojratsirikul2011flow}. {\color{red}$\triangle$} in (d) denotes the dominant frequency contents measured from our numerical simulations.}
	\label{averagediscn}
\end{figure}

\begin{figure}
	\centering
	\includegraphics[width=1.0\textwidth]{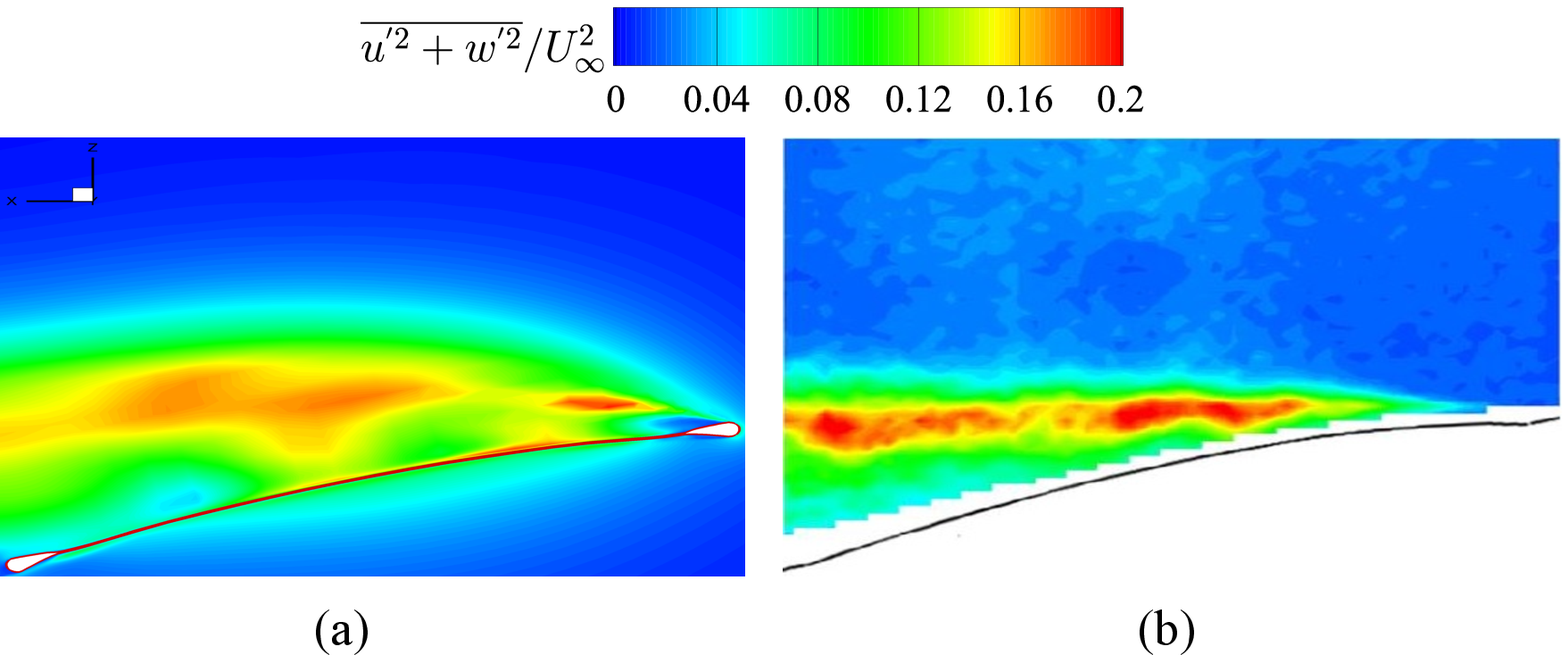}
	\caption{Magnitude of turbulent intensity on the mid-span plane obtained from: (a) present simulation at $\alpha=16^\circ$, (b) experiment \citep{rojratsirikul2011flow} at $\alpha=16^\circ$.}
	\label{tkecom}
\end{figure}

\section*{Appendix C. Natural frequency of a vibrating rectangular membrane in a uniform flow}
\setcounter{equation}{0}
\renewcommand{\theequation}{C. \arabic{equation}}
A flexible thin membrane can be excited by the unsteady separated flow which can give rise to self-excited vibration through the fluid-structure coupling effect. The natural frequency estimated from the linear model in \refeq{linear_frequency} ignores the geometric nonlinearity of the vibrating membrane, which overestimates or underestimates the actual natural frequency of a membrane immersed in a uniform flow in the DBS regime. The dynamic tension and the added mass caused by the transverse displacement of the membrane should be taken into account in the analysis of the natural frequency. In this appendix, an analytical formula of the nonlinear natural frequency for a rectangular membrane considering the added mass effect is derived. 


\refFig{membrane_strain} presents a schematic of a rectangular membrane with all fixed edges undergoing vibration in a uniform flow. The governing equation of a rectangular membrane vibrating transversely based on the large deflection theory is written as \citep{amabili2008nonlinear}
\begin{equation}
\rho^s h \frac{\partial^2 \delta_n}{\partial t^2} - (N_x^s+N_{x0}^s)\frac{\partial^2 \delta_n}{\partial x^2} - (N_y^s+N_{y0}^s)\frac{\partial^2 \delta_n}{\partial y^2} - 2N_{xy}^s \frac{\partial^2 \delta_n}{\partial x \partial y} = \Delta p(x,y,t),
\label{mem1}
\end{equation}
where $\rho^s$ and $h$ denote the membrane density and the membrane thickness. $N_{x0}^s$ and $N_{y0}^s$ are the initial tensions applied to the membrane. $N_x^s$ and $N_y^s$ represent the dynamic tensions in the $X$ and $Y$ directions caused by the transverse displacement $\delta_n(x,y,t)$, which are given as
\begin{equation}
N_x^s=h \sigma^s_x, \quad \quad N_y^s=h \sigma^s_y, \quad \quad N_{xy}^s=h \tau_{xy}^s,
\label{mem2}
\end{equation}
where $\sigma^s_x$ and $\sigma^s_y$ denote the dynamic stresses along the $X$-axis and $Y$-axis. $\tau_{xy}^s$ is the shear stress of the membrane. The Kirchhoff stresses for the membrane with isotropic and homogeneous materials are expressed as
\begin{equation}
\begin{bmatrix}
\sigma^s_x \\
\sigma^s_y \\
\tau_{xy}^s
\end{bmatrix}
=
\frac{E^s}{1-(\nu^s)^2}
\begin{bmatrix}
1 & \nu^s & 0 \\
\nu^s & 1 & 0 \\
0  &  0  &  1-\nu^s
\end{bmatrix}
\begin{bmatrix}
\varepsilon_x^s \\
\varepsilon_y^s \\
\gamma_{xy}^s
\end{bmatrix},
\label{mem3}
\end{equation}
where $\varepsilon_x^s$ and $\varepsilon_y^s$ represent the strains along $X$ and $Y$ directions. $\gamma_{xy}^s$ is the shear strain of the membrane. $E^s$ is the Young's modulus and $\nu^s$ denotes the Poisson ratio. Only the displacement along the $Z$-axis is considered and the in-plane displacements along $X$-axis and $Y$-axis are ignored in the nonlinear membrane vibration model. Based on the von Karman hypothesis, the Green's strains are written as
\begin{equation}
\varepsilon_x^s=\frac{1}{2} \left( \frac{\partial \delta_n}{\partial x} \right)^2, \quad \quad \varepsilon_y^s=\frac{1}{2} \left( \frac{\partial \delta_n}{\partial y} \right)^2, \quad \quad \gamma_{xy}^s = \frac{\partial \delta_n}{\partial x} \frac{\partial \delta_n}{\partial y}.
\label{mem4}
\end{equation}

The pressure difference between the upper and lower surfaces $\Delta p(x,y,t)$ applied to the membrane is caused by the membrane motion in a uniform flow. A simplified inviscid incompressible aerodynamic model is derived by \cite{chen2015research} to predict the pressure distribution on the membrane surface. The aerodynamic acoustic theory is employed to evaluate the aerodynamic acoustic pressure $\Delta p_a$ due to air compression caused by the membrane vibration. A quasi-static model is used to calculate the quasi-static wind pressure $\Delta p_w$ induced by the shape-change of the membrane. The total pressure $\Delta p$ is a sum of the aerodynamic acoustic pressure $\Delta p_a$ and the quasi-static wind pressure $\Delta p_w$, which is expressed as
\begin{equation}
\Delta p = \Delta p_a + \Delta p_w,
\label{mem5}
\end{equation}
\begin{equation}
\Delta p_a= -\frac{2 c \rho^f}{i \pi}  \frac{\partial^2 \delta_n}{\partial t^2},
\label{mem6}
\end{equation}
\begin{equation}
\Delta p_w = - \frac{1}{2} \rho^f U_{\infty}^2 \left[ \frac{a_{ij}}{\delta_{n0} (\omega_{ij}^s)^2 } \frac{\partial^2 \delta_n}{\partial t^2} + \frac{b_{ij}}{\delta_{n0} \omega_{ij}^s } \frac{\partial \delta_n}{\partial t} - a_0 \sin \left( \frac{i \pi x}{c} \right) \sin \left( \frac{j \pi y}{b} \right) \right],
\label{mem7}
\end{equation}
where $\rho^f$ is the air density. $\omega_{ij}^s$ denotes the vibration circular frequency for the chord-wise $i$ and span-wise $j$ mode. Here $a_0$, $a_{ij}$ and $b_{ij}$ represent the shape-change coefficients obtained from the amplitude of the pressure difference coefficient $C_{pd0}(t)$ of the numerical simulations or wind tunnel experiments, which is given as
\begin{equation}
C_{pd0}(t) = a_0 + a_{ij} \cos (\omega_{ij}^s t) + b_{ij} \sin(\omega_{ij}^s t),
\label{mem8}
\end{equation}
where $C_{pd0}(t)$ is calculated based on the pressure difference coefficient distribution on the membrane surface
\begin{equation}
C_{pd}(t) = C_{pd0}(t) \sin \left( \frac{i \pi x}{c} \right) \sin \left( \frac{j \pi y}{b} \right),
\label{mem9}
\end{equation}

\refEq{mem1} can be written in a new form by substituting \refeqs{mem5}-(\ref{mem7}) into \refeq{mem1}
\begin{equation}
\begin{split}
&\left( \rho^s h + m^{am}  \right) \frac{\partial^2 \delta_n}{\partial t^2} - (N_x^s+N_{x0}^s)\frac{\partial^2 \delta_n}{\partial x^2} - (N_y^s+N_{y0}^s)\frac{\partial^2 \delta_n}{\partial y^2} \\
& - 2N_{xy}^s \frac{\partial^2 \delta_n}{\partial x \partial y}
+c^{am} \frac{\partial \delta_n}{\partial t} = \Delta p_0(x,y),
\label{mem10}
\end{split}
\end{equation}
where $m^{am}=\frac{2 c \rho^f}{i \pi} + \frac{\rho^f U_{\infty}^2 a_{ij}}{2 \delta_{n0} (\omega_{ij}^s)^2 }$ and $c^{am}=\frac{\rho^f U_{\infty}^2 b_{ij}}{2 \delta_{n0} \omega_{ij}^s }$ are the added mass and the aerodynamic damping, respectively. $\Delta p_0(x,y)=\frac{1}{2} \rho^f U_{\infty}^2 a_0 \sin \left( \frac{i \pi x}{c} \right) \sin \left( \frac{j \pi y}{b} \right)$ is a constant value, which is related to the membrane shape. 

Substituting \refeqs{mem2}-(\ref{mem4}) into \refeq{mem10} and neglecting the damping term gives
\begin{equation}
\begin{split}
&\left( \rho^s h + m^{am}  \right) \frac{\partial^2 \delta_n}{\partial t^2} - \left(N_{x0}^s \frac{\partial^2 \delta_n}{\partial x^2} + N_{y0}^s \frac{\partial^2 \delta_n}{\partial y^2} \right) \\
&-\frac{E^s h}{2(1-(\nu^s)^2)} \left[ \left( \frac{\partial \delta_n}{\partial x} \right)^2 \frac{\partial^2 \delta_n}{\partial x^2} + \left( \frac{\partial \delta_n}{\partial y} \right)^2 \frac{\partial^2 \delta_n}{\partial y^2} \right] \\ 
&- \frac{E^s h \nu^s}{2(1-(\nu^s)^2)} \left[ \left( \frac{\partial \delta_n}{\partial x} \right)^2 \frac{\partial^2 \delta_n}{\partial y^2} + \left( \frac{\partial \delta_n}{\partial y} \right)^2 \frac{\partial^2 \delta_n}{\partial x^2} \right] \\
&- \frac{E^s h}{1+\nu^s} \frac{\partial \delta_n}{\partial x} \frac{\partial \delta_n}{\partial y} \frac{\partial^2 \delta_n}{\partial x \partial y} = \Delta p_0(x,y).
\label{mem11}
\end{split}
\end{equation}

To solve the membrane dynamic equation given in \refeq{mem11}, an assumed mode method is adopted to obtain the approximate analytical solution. The membrane displacement in the $Z$ direction can be written as
\begin{equation}
\delta_n(x,y,t) = \sum \limits_{i=1}^{\infty} \sum \limits_{j=1}^{\infty} \sin \left( \frac{i \pi x}{c} \right) \sin \left( \frac{j \pi y}{b} \right) U_{ij}(t),
\label{mem12}
\end{equation}
where $U_{ij}(t)$ is an unknown function. \refEq{mem11} is solved by the Galerkin method to obtain the following equations
\begin{equation}
\frac{{\rm{d}}^2 U_{ij} (t)}{{\rm{d}} t^2} + \omega_{ij0}^2 U_{ij} (t) + \beta \kappa \omega_{ij0}^2 (U_{ij}(t))^3=\Delta P,
\label{mem13}
\end{equation}
where
\begin{equation}
\omega_{ij0} = \sqrt{\frac{\pi^2}{\rho^s h + m^{am}} \left[ N_{x0}^s \left(\frac{i}{c}\right)^2 + N_{y0}^s \left(\frac{j}{b}\right)^2 \right]},
\label{mem14}
\end{equation}
\begin{equation}
\beta = \frac{h^2}{cb},
\label{mem15}
\end{equation}
\begin{equation}
\begin{split}
&\kappa = \frac{cb}{\omega_{ij0}^2 h} \left[ \frac{3 E^s \pi^4}{32(1- (\nu^s)^2)(\rho^s h + m^{am})} \left( \frac{i^4}{c^4} + \frac{j^4}{b^4} \right) \right. \\
&\left. + \frac{3 E^s \nu^s \pi^4}{16(1- (\nu^s)^2)(\rho^s h + m^{am})} \left( \frac{ij}{cb}  \right)^2 - \frac{E^s \pi^4}{16(1+\nu^s)(\rho^s h + m^{am})} \left( \frac{ij}{cb}  \right)^2 \right],
\label{mem16}
\end{split}
\end{equation}
\begin{equation}
\Delta P = \int \int \Delta p_0(x,y) \sin \left( \frac{i \pi x}{c} \right) \sin \left( \frac{j \pi y}{b} \right) {\rm{d}}x {\rm{d}}y.
\label{mem17}
\end{equation}


A singular perturbation method namely Poincar$\rm{\acute{e}}$-Lindstedt perturbation method \citep{drazin1992nonlinear} is adopted to obtain an approximate analytical solution for the partial differential equation of the vibrating membrane in \refeq{mem13} by setting $\Delta P$ to zero \citep{liu2018nonlinear}. The approximate analytical formula of the nonlinear natural frequency $f_{ij}^n$ is given as
\begin{equation}
f_{ij}^n = \frac{1}{2 \pi} \left( \omega_{ij0} + \frac{3 \beta \kappa \omega_{ij0}}{8}(g_0^2 + h_0^2) \right),
\label{mem32}
\end{equation}
where $g_0$ and $h_0$ are the coefficients corresponding to the initial conditions of the vibrating membrane given below
\begin{equation}
g_0 = \frac{4}{cb} \int_{0}^{c} \int_{0}^{b} \delta_{n0}(x,y) \sin \left( \frac{i \pi x}{c} \right) \sin \left( \frac{j \pi y}{b} \right) {\rm{d}}x {\rm{d}}y,
\label{mem25}
\end{equation}
\begin{equation}
h_0 = \frac{4}{cb} \int_{0}^{c} \int_{0}^{b} \frac{{\rm{d}} \delta_{n0}(x,y)}{{\rm{d}}t} \sin \left( \frac{i \pi x}{c} \right) \sin \left( \frac{j \pi y}{b} \right) {\rm{d}}x {\rm{d}}y,
\label{mem26}
\end{equation}
where $ \delta_{n0}(x,y)$ and $\frac{{\rm{d}} \delta_{n0}(x,y)}{{\rm{d}}t}=0$ are the initial displacement and velocity of the membrane.
When the membrane maintains a static equilibrium state without vibration, the natural frequency of the aerodynamically tensioned membrane is consistent with the linear natural frequency model in \refeq{linear_frequency}.

\end{document}